\documentclass[11pt]{article}

\usepackage{jheppub}

\newcommand{\order}[1]{{\cal O}\left(#1\right)}

\newcommand{\beq}{\begin{equation}}
\newcommand{\eeq}{\end{equation}}
\newcommand{\bea}{\begin{eqnarray}}
\newcommand{\eea}{\end{eqnarray}}
\newcommand{\bdm}{\begin{displaymath}}
\newcommand{\edm}{\end{displaymath}}
\newcommand{\eps}{\varepsilon}

\def\eps{\varepsilon}

\def\d{\partial}

\def \d{{\rm d} }
\def \d0 {D\O \;}
\def \fc {f_{\mathrm{cut}}}
\def \yc {y_{\mathrm{cut}}}

\newcommand{\ycut}{y_{\text{cut}}}
\newcommand{\zcut}{z_{\text{cut}}}
\newcommand{\fcut}{f_{\text{cut}}}
\newcommand{\Rprune}{R_{\text{prune}}}
\newcommand{\Rtrim}{R_{\text{trim}}}
\newcommand{\Rfilt}{R_{\text{filter}}}

\newcommand{\refeq}[1]{Eq.~$\bra{\ref{eq:#1}}$}
\newcommand{\reffig}[1]{Fig.~$\ref{fig:#1}$}

\newcommand{\bra}[1]{\left(#1\right)}

\newcommand{\T}[1]{\Theta\bra{#1}}

\newcommand{\herwig}{\textsf{Herwig++~2.7.0 }}

\newcommand{\sherpa}{\textsf{Sherpa 2.0.0 }}
\newcommand{\fasjet}{\textsf{FastJet}}
\newcommand{\eventtwo}{\textsf{EVENT2 }}
\title{On jet substructure methods for signal jets}

\author[a]{Mrinal Dasgupta,}
\author[b]{Alexander Powling,}
\author[c,d]{and Andrzej Siodmok}
\affiliation[a]{Lancaster-Manchester-Sheffield Consortium for Fundamental Physics, School of Physics
  \& Astronomy, University of Manchester, Manchester M13 9PL, United
  Kingdom}
\affiliation[b]{School of Physics
  \& Astronomy, University of Manchester, Manchester M13 9PL, United
  Kingdom}
\affiliation[c]{Institute of Nuclear Physics, Polish Academy of Sciences,\\
ul. Radzikowskiego 152, 31-342 Krak\'ow, Poland
}
\affiliation[d]{CERN, PH-TH, CH-1211 Geneva 23, Switzerland}
\preprint{{\flushright CERN-PH-TH-2015-032\\ MCnet-15-03\\ }}

\keywords{QCD, NLO Computations, Hadronic Colliders, Standard Model, Jets}

\abstract{We carry out simple analytical calculations and Monte Carlo studies
  to better understand the impact of QCD radiation on some well-known
  jet substructure  methods  for  jets arising from the decay of
  boosted Higgs bosons.  Understanding
  differences between taggers for these signal jets assumes particular
  significance in situations where they perform similarly on QCD background
jets. As an explicit example of this we compare the
Y-splitter method to the more recently proposed Y-pruning technique. We
demonstrate how the insight we gain can be used to significantly
improve the performance of Y-splitter by combining it with 
trimming and show that this combination outperforms the other taggers studied here, 
at high $p_T$. We also make analytical estimates for optimal
parameter values, for a range of methods and compare to results from Monte Carlo studies.
}
\begin{document}

\maketitle
\section{Introduction}
In recent years the detailed study and analysis of the
internal structure of hadron jets has become an area of very active
investigation.  The principle reason for such high interest has 
been in the context of Higgs boson and new physics searches at the
LHC and associated phenomenology. Due to the large (TeV scale)
transverse momenta that can be accessed at the LHC, electroweak scale
particles, such as the Higgs boson, can be directly produced with large boosts. Alternatively the decay
of yet undiscovered  heavy new particles to comparatively light standard model particles, such as top quarks or $W$/$Z$  bosons,
would also result in the production of boosted particles whose decay
products would consequently be collimated. This in turn means that
rather than producing multiple resolved jets, a significant fraction
of the time the decay products are encompassed in a single fat jet. Understanding the substructure of
such jets therefore becomes crucial in the context of discriminating
between jets originating from QCD background and those originating
from signal processes involving e.g. Higgs production and its hadronic decay. 

Though pioneering studies were carried out by Seymour several years
ago \cite{Seymour:1993mx}, and the \mbox{Y-splitter} method for tagging W
bosons was subsequently introduced in Ref.~\cite{Butterworth:2002tt} over a decade ago, the revival of interest in jet
substructure is relatively recent and owes essentially to seminal work by Butterworth,
Davison, Rubin and Salam \cite{Butterworth:2008iy}. These authors revealed the power of
substructure analyses by studying the discovery potential for a light
Higgs boson ($M_H \approx  120$ GeV) in the process $pp \to W/Z,H$
with Higgs decay to $b \bar{b}$. They demonstrated that exploiting the
boosted regime and applying jet substructure methods, specifically a mass-drop
and filtering analysis, was sufficient to turn what was previously regarded as an
unpromising channel into one of the best channels for Higgs
discovery at the LHC. Several other applications followed, dedicated to new
physics searches as well as top and $W$/$Z$ boson tagging, and numerous substructure
techniques are now in existence and being commonly employed in experimental
analyses both in the context of QCD measurements as well as for
searches \cite{ATLAS:2012am,Aad:2012meb,Aad:2013gja,CMS-substructure-studies,Aad:2012raa,Aad:2012dpa, ATLAS:2012dp, ATLAS:2012ds, Chatrchyan:2012ku,Chatrchyan:2012cx, Chatrchyan:2012yxa}.  
For the original articles introducing a selection of some of these methods we refer the
reader to Refs.~\cite{Brooijmans:2008,CMS:2009,Kaplan:2008ie,Plehn:2010st,Ellis:2009su,Ellis:2009me,Krohn:2009th,Almeida:2010pa,
Almeida:2011aa,Ellis:2012sn,Soper:2011cr} while comprehensive
reviews of the field and further studies are available in Refs.~\cite{Salam:2009jx,Abdesselam:2010pt,Altheimer:2012mn,Altheimer:2013yza}.

Most recently research has started to emerge \cite{Larkoski:2013eya,Dasgupta:2013ihk,
Dasgupta:2013via, Larkoski:2014wba}  which aims at enhancing
our understanding of jet substructure methods via the use of
analytical calculations that, where possible,  lend greater insight
and provide powerful complementary information to that available
purely from traditional Monte Carlo (MC) based investigations of jet
substructure.  In Ref.~\cite{Dasgupta:2013ihk} in particular, resummed results were
provided for jet mass distributions for QCD background jets after the
application of a variety of jet substructure methods (that we shall collectively
refer to as `taggers') and detailed comparisons to MC studies
were carried out. Jet mass distributions were examined
for the case of trimmed \cite{Krohn:2009th} and  pruned jets
\cite{Ellis:2009su,Ellis:2009me} as well as for jets obtained after the application of
the mass-drop tagger \cite{Butterworth:2008iy}.  

One of the main aims of Ref.~\cite{Dasgupta:2013ihk}  was to better understand
how aspects of tagger definition and design may interplay with QCD
dynamics to dictate the performance of taggers as reflected by their
action on background jets. The  improved analytical understanding
that was achieved led to a better appreciation of the role of tagger
parameters (including the discovery of the apparent  redundancy  of
the mass-drop parameter $\mu$ in the mass-drop tagger
\cite{Butterworth:2008iy}). The analytical studies also paved the way for
improvement of theoretical properties of taggers. Examples of
improvements that were suggested or made in
Ref.~\cite{Dasgupta:2013ihk} included  the design of
taggers with a perturbative expansion more amenable to resummation as
for the modified mass-drop (mMDT) as well as
removing undesirable tagger features as for the case of pruning via
the Y-pruning modification. Subsequent work has also
demonstrated how an analytical understanding of the action of taggers
on QCD background can be exploited to construct valuable new tools such as the {\it{soft drop}} technique introduced in 
Ref.~\cite{Larkoski:2014wba}.

While thus far there has been heavy focus on taggers applied to QCD
background, until now radiative effects for signal jets have not been investigated in the
same level of analytical detail for many commonly used substructure
methods. Detailed analytical calculations have however been performed to study the action of filtering
for $H \to b\bar{b}$ \cite{Rubin:2010fc} and for N-subjettiness
\cite{Feige:2012vc}, while the role of QCD radiation in the context of
template tagging was discussed in Ref.~\cite{Almeida:2011aa}. 

We first observe that it is common to study high $p_T$ signal jets in
some relatively narrow mass window of width $\sim \delta M$  around
the mass, $M$ of some boosted decaying heavy resonance of
interest, this mass cut being a first step in tagging signal jets. One then has a situation where there are various disparate
scales involved in the problem such as the (potentially) TeV scale
transverse momenta of the fat jets, the mass $M$ of the resonance
 (which for our studies we can consider to be around the electroweak
 scale) and the width of the window $\delta M$ which for most purposes
 we can consider as a parameter $\sim 10$ GeV. These scales are in addition, of course,
 to the various parameters corresponding to angular distances
and energy cuts introduced by tagging and jet finding.  

It is well known that in such multi-scale problems radiative
corrections have the potential to produce large logarithms involving
ratios of disparate scales. In the example of filtering studied in
Ref.~\cite{Rubin:2010fc} large logarithms in $M_H/\delta M$ arose from
considering soft emissions, which were accompanied by collinear
logarithmic enhancements in $R_{b\bar{b}}/R_{\text{filt}}$, the ratio of the $b
\bar{b}$ opening angle to the filtering radius. On the other hand
Ref.~\cite{Dasgupta:2013ihk} observed via MC studies of the signal that for the taggers studied
there (mMDT, pruning, trimming,Y-pruning) the tagger performance was
primarily driven by the action of taggers on QCD background, with
signals not appearing to display very sizeable radiative corrections for the default parameters chosen there.

In order to better understand these apparently contrasting observations it is desirable to
acquire a higher level of analytical insight into the action of
taggers on signal jets. When comparing the performance of taggers one
may also meet a situation where two taggers shall act essentially
similarly on background jets and hence their action on signal becomes
of critical significance. We shall in fact provide an explicit example
of this situation later in this article. It is also of importance to understand and assess the impact of QCD
radiative corrections and non-perturbative effects on tagger efficiency for signals, to ascertain what
theoretical tools (fixed-order calculations, MC methods,
resummed calculations or combinations thereof), should ideally be deployed to
get the most reliable picture for the signal efficiency for a given tagger.  

With all the above aims in mind, here we embark on a more detailed
study dedicated to signal jets. We shall focus our attention on the case of
a jet arising from a  boosted Higgs boson for a process
such as Higgs production in association with a vector boson $pp \to W/Z, \, H$,
with $H \to b \bar{b}$, where we will work in a narrow width
approximation. For our analytical approximations we shall also
typically consider highly boosted configurations i.e. those where the
Higgs has transverse momentum $p_T \gg M_H$  and shall further take a
fat jet with radius $R \gg\frac{M_H}{p_T}$. We shall work in a
small-angle approximation throughout, though we will often consider $R
\sim 1$.

 We stress here that we do not intend to provide precise
 high-order calculations for radiative corrections to any given process but seek
mainly  to understand and compare the behaviour of  taggers via a combination of
approximate analytics and MC cross-checks. For an example of
exact fixed-order calculations involving jet substructure and
signal processes we refer the reader to Ref.~\cite{Banfi:2012jh}.

We start in the next section by analysing the case of a plain jet
mass cut focussing on a mass window around the signal mass and deeming a
jet to be tagged as signal if the jet mass falls within this window. We consider
the impact on signal efficiency of initial state radiation (ISR), final
state radiative corrections (FSR) both analytically and in MC studies. 
We also study the impact of non-perturbative effects (NP) with MC. The results so obtained
can then provide a point of reference and comparison to judge the
improvements that are offered by use of substructure taggers, which
impose requirements in addition to a simple cut on mass. 

Next we move to analysing jets with application of various taggers. In section 3 we
study trimming at lowest order and with ISR and FSR  corrections.  
We investigate the logarithmic structure that emerges in
$M$, $\delta M$ and $p_T$ as well as in tagger parameters and compare
to MC results where appropriate. We also study non-perturbative
corrections, though purely with MC results.
In section 4 we analyse along similar lines
pruning and the modified mass-drop tagger (mMDT). FSR is
analysed further for these taggers in appendices A and B where we also
compare parton shower results to those from full leading-order
calculations i.e. those that go beyond the soft/collinear approximation.

In section 5 we study Y-pruning and the Y-splitter tagger
\cite{Butterworth:2002tt}. We  observe that while the action of
Y-splitter on QCD background is similar to Y-pruning, the signal jet
with \mbox{Y-splitter} is subject to severe loss of resolution due to ISR and underlying event (UE) effects. 
We show with MC studies that combining Y-splitter with trimming dramatically improves
the signal behaviour while leaving the background largely
unmodified. As a consequence we show that Y-splitter with trimming
outperforms the other taggers we study here, especially at high
$p_T$. To our mind this example further illustrates how even a
relatively basic analytical understanding of all aspects of taggers
(for both signal and background) can be exploited to achieve important performance
gains.\footnote{One other recent example of simple analytical arguments, based on power counting, being used to good effect for
  top tagging can be found in Ref.~\cite{Larkoski:2014zma}.}

 Finally we carry out analytical studies of optimal values
for tagger parameters, obtained by maximising signal significance, and
compare to MC results. We conclude with a summary and mention
prospects for future work.

\section{Results for plain jet mass}
Here we shall consider the plain jet-mass distribution for fat signal
jets without the application of substructure methods, other than the
imposition of a mass window $\delta M$, as previously stated.
As also mentioned before, we shall consider  the case of
Higgs boson production in association with an electroweak vector boson
$pp \to W/Z, H$ with Higgs decay to a $b\bar{b}$ quark pair  and shall
work in a narrow width approximation throughout. For the purposes
of examining the jet substructure we shall not need to  write down 
matrix elements for the production of the high $p_T$ Higgs boson and
shall  instead be  concerned purely with the details of the Higgs
decay and the resulting fat jet, as well as the impact of ISR, FSR and 
non-perturbative effects.
Let us take a boosted Higgs boson produced with
transverse momentum $p_T \gg M_H $ and purely for convenience set it
to be at zero rapidity with respect to the beam direction, so that the
corresponding energy is $\sqrt{p_T^2+M_H^2}$. We further consider
Higgs decay into $b \bar{b}$ so that in terms of four-momenta one has 
$p_H = p_b +p_{\bar{b}}$. 
Thus the invariant mass of the Higgs can be expressed as 
\begin{equation}
\label{eq:masssqrd}
 p_H^2 = M_H^2= 2 p_b\cdot p_{\bar{b}} = 2 z(1-z) (p_T^2+M_H^2) \left(
    1-\cos \theta_{b \bar{b}} \right),
\end{equation}
with $z$ and $1-z$ the energy fractions of the decay products and
$\theta_{b \bar{b}}$ the $b \bar{b}$ opening angle, where we neglected
the $b$ quark masses.
Furthermore we shall consider the highly  boosted regime taking  
$\Delta = \frac{M_H^2}{p_T^2} \ll R^2$ and shall systematically
neglect power corrections in $\Delta$. Then from
Eq.~\eqref{eq:masssqrd}, taking a small-angle approximation, we obtain the standard result:
\begin{equation} \label{eq:bbopening}
\theta_{b \bar{b}}^2 \approx \frac{\Delta}{z(1-z)}.
\end{equation}

Requiring  that the Higgs decay products are contained in a
single fat jet, $\theta_{b \bar{b}}^2 < R^2$ thus translates into a constraint on $z$:
\begin{equation}
\label{eq:constraint}
z(1-z) > \frac{\Delta}{R^2}.
\end{equation}

Let us start by providing the results for the signal efficiency, 
$\eps_\text{S}^{(0)} $ to  lowest order i.e. taking just the
$H \to b \bar{b}$ decay without any radiative corrections. This can be
considered as the fraction of decays that are reconstructed inside a
fat jet of radius $R$. Here one has to consider the relevant Feynman
amplitude for $H \to b \bar{b}$ and the full decay phase-space with an
integral over the final state parton momenta. However for the
\emph{fraction} of decays inside the fat jet,  $\eps_\text{S}^{(0)}$, we just obtain 
\begin{align}
\eps_\text{S}^{(0)} = \int_0^1 dz \, \Theta \left(R^2 -\frac{\Delta}{z(1-z)}
\right)=\sqrt{1-\frac{4\Delta}{R^2}} \Theta \left(R^2-4 \Delta \right)
\approx 1-\frac{2 \Delta}{R^2}+\mathcal{O} \left(\frac{\Delta}{R^2} \right)^2,
\end{align}
which is trivially in good agreement with corresponding MC event generator results with
all ISR, FSR and non-perturbative effects turned off. The above result simply
suggests, as one can easily anticipate, that with increasing boosts,
i.e. smaller $\Delta$, the efficiency of reconstruction inside a single
jet increases. At this lowest-order level there is of course no role
for the mass-window $\delta M$ since the jet mass $M_j$ coincides with
the Higgs mass $M_H$. 

\subsection{Initial state radiation}
Let us now account for the impact of initial state radiation on
the jet mass distribution.  We can anticipate that the impact of soft
radiation may be significant here because of the fact that we require
the invariant mass of the fat jet to be within $\delta M$ of the Higgs
mass, with $\delta M \ll p_T$. This requirement imposes a constraint on
real emissions, arising from ISR, that enter the jet since these
contribute directly to the deviation of the jet mass from $M_H$. Hence
one can expect that large logarithmic corrections arise as a
consequence. In order to understand the structure of the logarithmic
corrections that arise,  we consider the process
$pp \to ZH$ with the additional production of soft gluons radiated
by the incoming hard partons ($q \bar{q}$ pair).  Let us start by
taking a single ISR gluon which is soft i.e. has energy $\omega \ll p_T$.
In the soft limit we can work with the eikonal approximation in which
production of the ISR factorises from the Born-level hard process $pp
\to ZH$. To compute the signal efficiency we shall require the jet invariant
mass to be within a relatively narrow mass-window $\delta M$ of the Higgs mass:
\begin{equation}
M_H -\delta M< M_j < M_H+\delta M.
\end{equation}

As we observed previously at lowest order (Born level) this inequality
is always true since $M_j = M_H$, however with ISR it amounts to a constraint on the ISR gluon energy.
Neglecting corrections of order $\delta M^2$ we can write:
\begin{equation}
M_j^2-M_H^2 = 2 p_H \cdot k < 2 M_H \delta M,
\end{equation}
where $p_H$ is the four-momentum of the Higgs (or equivalently the sum
of the four-momenta of its decay products) and $k$ that of the ISR
particle. Defining $\theta$ as the angle between the soft emission and
the Higgs direction we can write:
\begin{equation}
2 p_H\cdot k = 2 \omega \left (\sqrt{p_T^2+M_H^2}-p_T \cos \theta \right).
\end{equation}

Since we take $\Delta = M_H^2/p_T^2 \ll 1$ and also use the small
$\theta$ approximation, we can expand in small quantities and write the following constraint on
gluon energy:
\begin{equation}
\label{eq:constraintenergy}
\omega < \frac{2 M_H \delta M}{p_T \,\left( \theta^2+\Delta\right)}.
\end{equation}
One can express this equation in terms of standard hadron collider
variables $k_t$, $\eta$ and $\phi$ defined wrt the beam direction by noting simply that $\omega =k_t
\cosh \eta$ and $\theta^2 \approx \eta^2+\phi^2$.

We wish to examine only the leading logarithmic structure that arises from
soft ISR emissions, starting with a single emission i.e. to leading
order in the strong coupling. Since we are considering an emission that enters the
high $p_T$ fat jet we are concerned with large-angle radiation from the
incoming hard partons. This in turn implies that there are no
collinear enhancements associated to such radiation and the resulting
leading logarithmic structure ought to be single-logarithmic, arising
purely from the infrared singularities in the gluon emission
probability.\footnote{In practice this expectation is challenged by the discovery
  of superleading logs \cite{Forshaw:2006fk}. Since these appear at order $\alpha_s^4$ and
  are suppressed as $1/N_C^2$ we do not expect them to make a
  significant impact  on the essential arguments we make here.}

The gluon emission probability is in turn given by the standard two particle antenna  for soft emissions off the
incoming quark/anti-quark:
\begin{equation}
\label{eq:antenna}
W_{ij} = 2 C_F \frac{\alpha_s}{\pi} \frac{(p_i\cdot p_j)}{(p_i\cdot k)(p_j\cdot k)}.
\end{equation}
where $C_F=4/3$.
Analogous to the case of jet mass distributions for hadron collider QCD jets
(see for example calculations in Ref.~\cite{Dasgupta:2012hg}) the ISR contribution can be
written by integrating Eq.~\eqref{eq:antenna} over the gluon emission
phase space. The result can be expressed in terms of $k_t$, $\eta$ and
$\phi$ and reads:
\begin{multline}
\label{eq:ISRcalc}
\eps^{(1)}_{\text{S},\text{ISR}} = \int dz \Theta
\left(z(1-z)-\frac{\Delta}{R^2} \right)  \times \\
\times 2 C_F \int \frac{dk_t}{k_t} d \eta
\frac{d\phi}{2\pi} \frac{\alpha_s(k_t)}{\pi}
\left(\Theta\left(\frac{2M_H \delta M}{p_T\left(\theta^2 +\Delta\right)}-k_t \cosh \eta\right)-1 \right) \Theta_{\text{jet}}.
\end{multline}

The above equation contains an integral over the energy fraction of
the partonic offspring involved in the Higgs decay (i.e. that over $z$), with a constraint
that is identical to the zeroth order requirement that the hard quarks
be contained in the fat jet, which is unmodified by the presence of
soft ISR  at leading logarithmic level, i.e. in the limit $\omega = k_t
\cosh \eta \ll
p_T$.  The step function involving a restriction on the transverse
momentum $k_t$  follows directly from Eq.~\eqref{eq:constraintenergy} and the
subsequent arguments. Virtual corrections are incorporated via
unitarity through the $-1$ term also in parenthesis. Lastly we have a factor
$\Theta_{\text{jet}}$ that is the condition that the soft ISR is
within the fat jet. 

The clustering condition $\Theta_{\text{jet}}$ is in principle quite
complicated since it involves recombination of three particles within
the fat jet, namely the $b,\bar{b}$ and the ISR gluon. In our approximation of $\Delta \ll R^2$ i.e. in the limit
of large boosts, we are considering a highly collimated quark pair,
relative to the radius of the fat jet. One can thus ignore the effect
of the finite $b \bar{b}$ opening angle as these effects contribute
only terms that are relatively suppressed by powers of $\Delta$
compared to the leading term we compute. Then one only has to consider
the fact that the soft ISR gluon is in the interior of the fat jet 
which amounts to the condition  $\Theta_{\mathrm{jet}}= \Theta
\left(R^2 -\left(\eta^2+\phi^2 \right)\right)$, since we had taken
the Higgs rapidity as zero. 

Within the context of the current purely order $\alpha_s$ calculation we shall
also consider the coupling as fixed at scale $p_T$ and ignore its
running. Running coupling effects are of course important to include
for leading logarithmic resummation and we shall do so for our final
answers. We define $\eps_\text{S,ISR} =\eps_{\text{S}}^{(0)}+\eps_{\text{S,ISR}}^{(1)}$
and carrying out the
relevant integrations with fixed coupling we get from Eq.~\eqref{eq:ISRcalc} 
the leading logarithmic result:
\begin{equation}
\label{eq:lo}
\frac{\eps_{\text{S,ISR}}}{\eps_\text{S}^{(0)}} \simeq
1-\frac{C_F \alpha_s}{\pi} R^2\ln \left(\frac{p_T^2 R^2}{2 M_H\delta M} \right).
\end{equation}

In order to obtain the above result starting from Eq.~\eqref{eq:ISRcalc} one can first integrate over
$k_t$, discarding the $\cosh \eta$ accompanying factor as this will
only generate subleading terms. The integral over $k_t$ produces the
large logarithm we seek. One can then express the $\eta, \phi$
integral as one wrt $\theta^2 =\eta^2+\phi^2$ and then integrate over
$\theta^2$ with the condition $\theta^2 < R^2$. With neglect of subleading terms, including those which
vanish with $\Delta$, one then obtains the result reported above
for the quantity $\eps_\text{S,ISR}/\eps_\text{S}^{(0)}$.
We have however chosen to retain the formally subleading logarithmic dependence on $R$ 
which can become important at smaller values of $R$, e.g. $R\sim 0.4$.

Given the large single logarithms that emerge from the above
approximate fixed-order calculation, it is natural to wish to attempt
to resum at least the leading logarithms to all perturbative
orders. This is far from a straightforward exercise. One of
the main obstacles to performing  a soft single log resummation, in the
present context, is the presence of non-global logarithms, associated clustering
logarithms \cite{DassalNG1,Appleby:2002ke, Banfi:2005gj,Delenda:2006nf} as well as
superleading logarithms referred to previously. Such
calculations pose a serious challenge to the current state of the
art and are beyond the scope of our work.

In the absence of a complete resummed calculation one can still obtain
a working estimate that can be compared to MC results, simply by exponentiating
the order $\alpha_s$ result obtained above and by including running
coupling effects. 

The exponentiated result   including the running of the QCD coupling  is given by
 \begin{equation}\label{eq:PlainISR}
\frac{\eps^{P}_{\text{S,ISR}}}{\eps^{(0)}_{\text{S}}} \approx \exp \left (-2 C_F R^2 t \right)
\end{equation}
where we defined the single-log evolution variable
\begin{eqnarray} \label{eq:Singlelogevol}
t &=& \frac{1}{2\pi}\int_{\frac{2 M_H \delta M}{p_T R^2}}^{p_T} \frac{dk_t}{k_t} 
\alpha_s(k_t) \\ \nonumber
&=& \frac{1}{4 \pi \beta_0}\ln \frac{1}{1-2 \lambda},  \, \, \lambda = \beta_0 \alpha_s(p_T) \ln \frac{p_T^2 R^2}{2M_H \delta M},
\end{eqnarray}
where $\beta_0 = \frac{1}{12 \pi} \left(11 C_A -2 n_f \right)$, and we shall use $n_f=5$.
Note that we have indicated the exponentiated result by the superfix $P$,
which indicates the resummed contribution from primary emissions alone
i.e. excluding secondary emissions which lead to non-global logarithms. 
We observe here  that the perturbative calculations break down at
$\lambda=1/2$ which corresponds to an ISR emission with $k_t \sim
\Lambda_\text{QCD}$, the QCD scale. 
This translates into a value of $\delta M$ which is
\begin{equation}
\delta M_{NP} = \frac{\Lambda_{\text{QCD}} p_T}{2 M_H}
\end{equation} 
where $\delta M_{NP}$ is the point of breakdown for perturbative calculations.
Taking a value of $\Lambda_{\text{QCD}} = 1 \, \mathrm{GeV}$ for $p_T = 3 \,
\mathrm{TeV}$ we can deduce that we should not use perturbative
results below $\delta M \sim 12 \, \mathrm{GeV}$. 

Although we have emphasised that our estimate of the ISR corrections
to the signal efficiency are incomplete, even to leading logarithmic
accuracy, it is nevertheless of interest to compare to MC
event generators. This is at least in part because MC generators
themselves do not attain full single logarithmic accuracy and certainly
exclude superleading logarithms. They do however
contain a number of effects that would be formally subleading from the
viewpoint of our calculation but could be of non-negligible
significance numerically. Hence while we do not intend to make a
detailed quantitative comparison we do expect to find qualitative
similarities with MC results.

To make this comparison, we generate $pp \to Z H$ events at 14 TeV using 
\herwig
with the ${\text{UE-EE-5-MRST}}$ tune~\cite{Seymour:2013qka}
and constrain the Higgs and Z boson to decay hadronically and leptonically respectively.

\begin{figure}[h]
\begin{center}
\includegraphics[width=0.49 \textwidth]{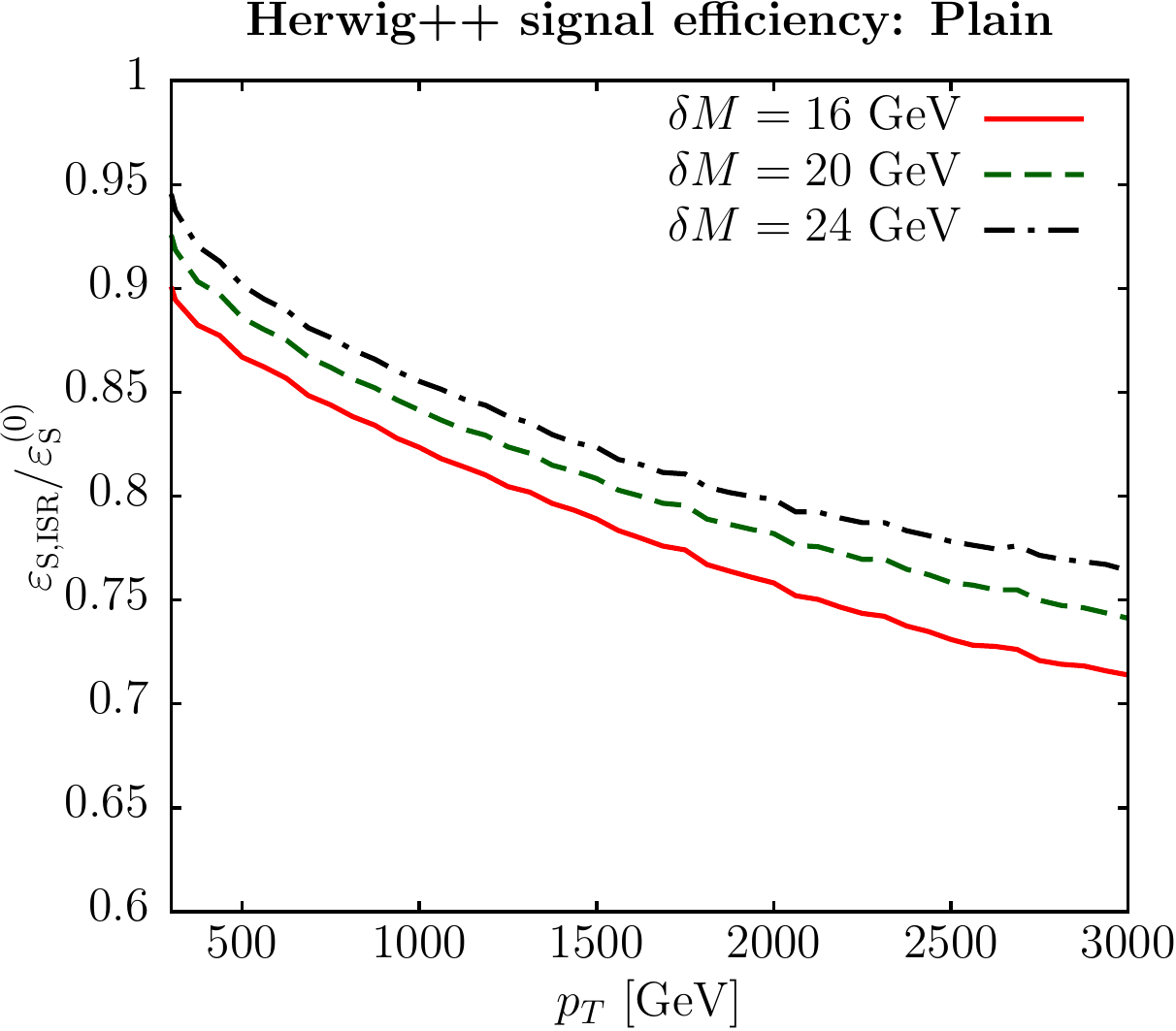}
\includegraphics[width=0.49 \textwidth]{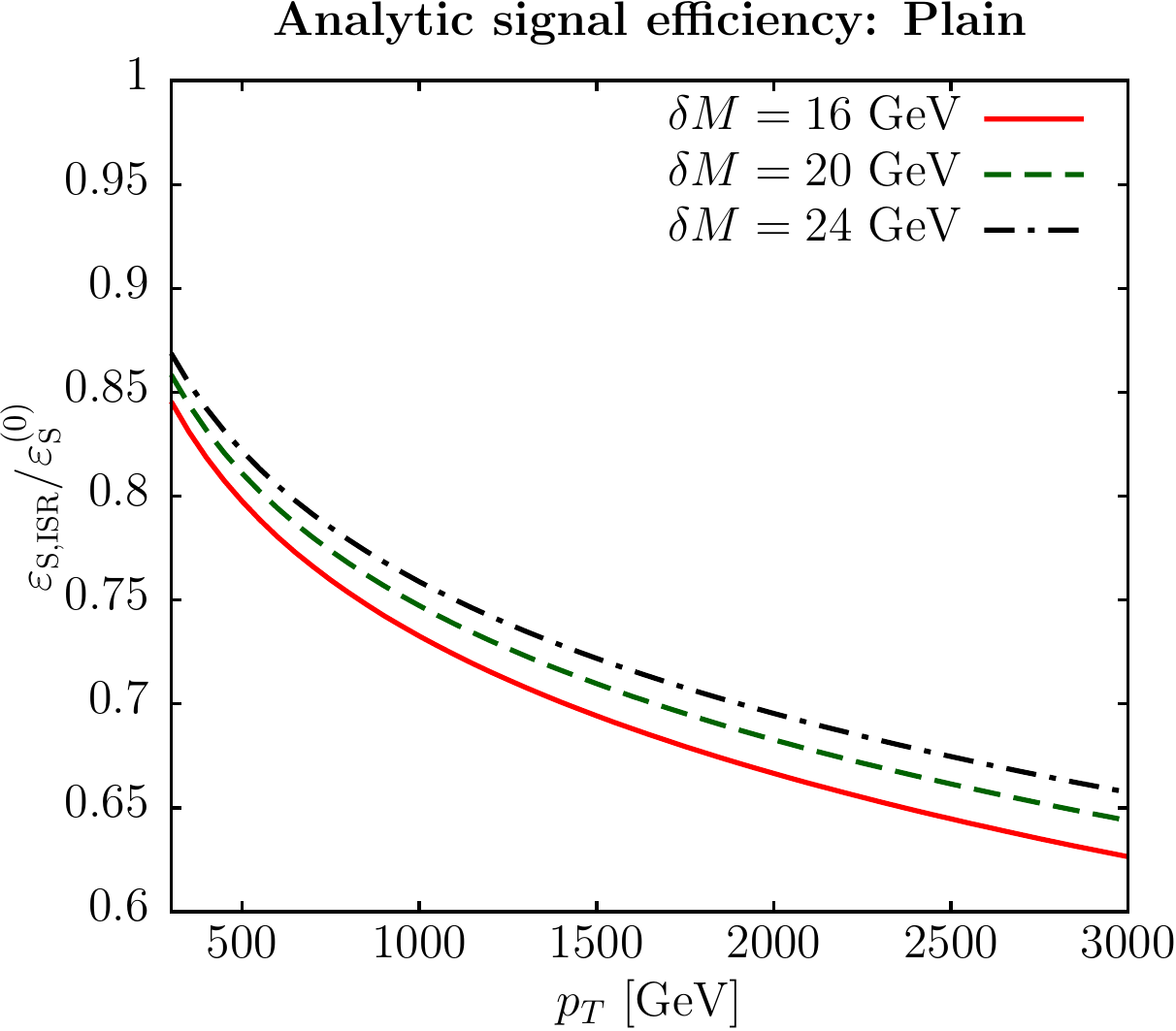}
\end{center}
\caption{Comparison of MC (left) and analytic (right) \refeq{PlainISR} tagging efficiencies for a 
range of mass windows as a function of a generator level cut on minimum jet transverse momentum $p_{T}$.
This result has been generated using \herwig for $pp \to ZH$ 
at 14 TeV with the Z decaying leptonically and $H \rightarrow b\bar{b}$, setting $M_H=125$ GeV.
We have tagged the signal jet as the highest $p_T$ Cambridge/Aachen jet with $R=1$.
In this figure we have generated events at parton level with ISR only
and divided out the contribution due to the lowest order result in both
panels for clarity.}
\label{fig:ISRplain}
\end{figure}

Each generated event is directly handed over to the Rivet package~\cite{Buckley:2010ar}, which implements our analyses. 
We tag the signal jet as the highest $p_T$ Cambridge/Aachen \cite{Dokshitzer:1997in,Wobisch:1998wt} jet with $R=1$
as implemented in \fasjet{} package~\cite{Cacciari:2011ma} and
plot the fraction of jets which lie in mass in the window
$M_H\pm \delta M$ as a function of a generator level cut on jet transverse momentum for
 three separate  values of $\delta M$.
To make a comparison with our ISR results we omit FSR and
non-perturbative corrections including hadronisation and underlying event (UE) corrections,
switching them off for the MC results. The resulting comparison is
shown in \reffig{ISRplain}, where results are displayed for the
ratio of the signal efficiency to the lowest order result. We observe
that the signal efficiency, from MC,  decreases with transverse
momentum for all values of $\delta M$ as indicated by our exponentiated
result \refeq{PlainISR}. By making the mass window wider,
i.e. choosing a larger $\delta M$, one of course obtains a smaller Sudakov
suppression and hence a larger efficiency but starts to lose the
association with a well defined signal peak. 

\subsection{Final state radiation}

For the case of plain jet-mass we would expect that the correction due to final state
radiation can be neglected in our region of interest where $p_T \gg
M_H$ or equivalently $\Delta \ll 1$.  Physically FSR  is associated to the $b \bar{b}$
dipole originating from Higgs decay. It is captured within the fat jet
as long as the FSR gluons are not radiated at angles beyond those
corresponding to the jet radius $R$. Due to angular ordering
however, we would expect that most of the FSR radiation from the $b$
quarks is emitted at angles smaller than $\theta^2_{b \bar{b}}$. In
this limit the final state emission is always recombined inside the
fat jet.  To be more precise, large-angle radiation beyond the
jet-radius $R$ is cut-off by the ratio of the dipole size ($b \bar{b}$
opening angle, given by $\Delta/(z(1-z))$, to the jet radius squared.
Upon integration over $z$ such corrections translate into terms
varying at most as $\Delta \ln \Delta$, which we shall neglect as they vanish with $\Delta$.
We have verified this with MC and find that for sufficiently
large transverse momenta, the correction due to final state radiation
is of negligible magnitude $\order{0.5\%}$ when compared to ISR
$\order{20\%}$ for $R=1$ and $p_T \gtrsim 500$ GeV.
We shall need to consider FSR more carefully when it comes to
analysing the taggers in future sections.
\subsection{Non-perturbative contributions}
In order to get a complete picture of the physical effects that
dictate the signal efficiency we also need to study how the signal
efficiency changes after including non-perturbative effects such as
hadronisation and underlying event (UE). In order to estimate those effects
we used Herwig++~2.7.0 with improved modeling of underlying event~\cite{Gieseke:2012ft} 
and the most recent UE-EE-5-MRST tune~\cite{Seymour:2013qka} which is able to describe the 
double-parton scattering cross section~\cite{Bahr:2013gkj} and underlying event data from $\sqrt{s} = 300$~GeV to $\sqrt{s} = 7$~TeV.
It can readily be anticipated that the underlying event effect in particular will 
significantly degrade the mass peak and hence lead to a loss of signal.

For this study, we consider all final state hadrons to be stable, therefore we switch off the decay handler module in Herwig++.
In doing so, we eliminate the chance of $b$ flavour hadrons decaying into invisible particles such as neutrinos.
If one were to include hadronic decay via invisible particles, one notices a universal reduction in signal efficiency for each tagger due to a loss of signal mass resolution.
This is particularly important for the jets formed from the decay $H\rightarrow b\bar{b}$ as compared to $W$/$Z$ jets because these electroweak bosons instead couple strongly to light quarks.
For further information on experimental techniques to mitigate the impact of these particular sources of missing transverse energy, see for example \cite{Aaltonen:2011bp}.
We also assume a $b$-tagging efficiency of $100\%$ which is sufficient for a relative comparison of tagger performance and behaviour. 
The reader is referred to Ref. \cite{Butterworth:2008iy} for a discussion on the impact of $b$-tagging efficiency on signal significance.

\begin{figure}[t]
  \centering
  \begin{minipage}{0.49\linewidth}
    \includegraphics[width=\textwidth]{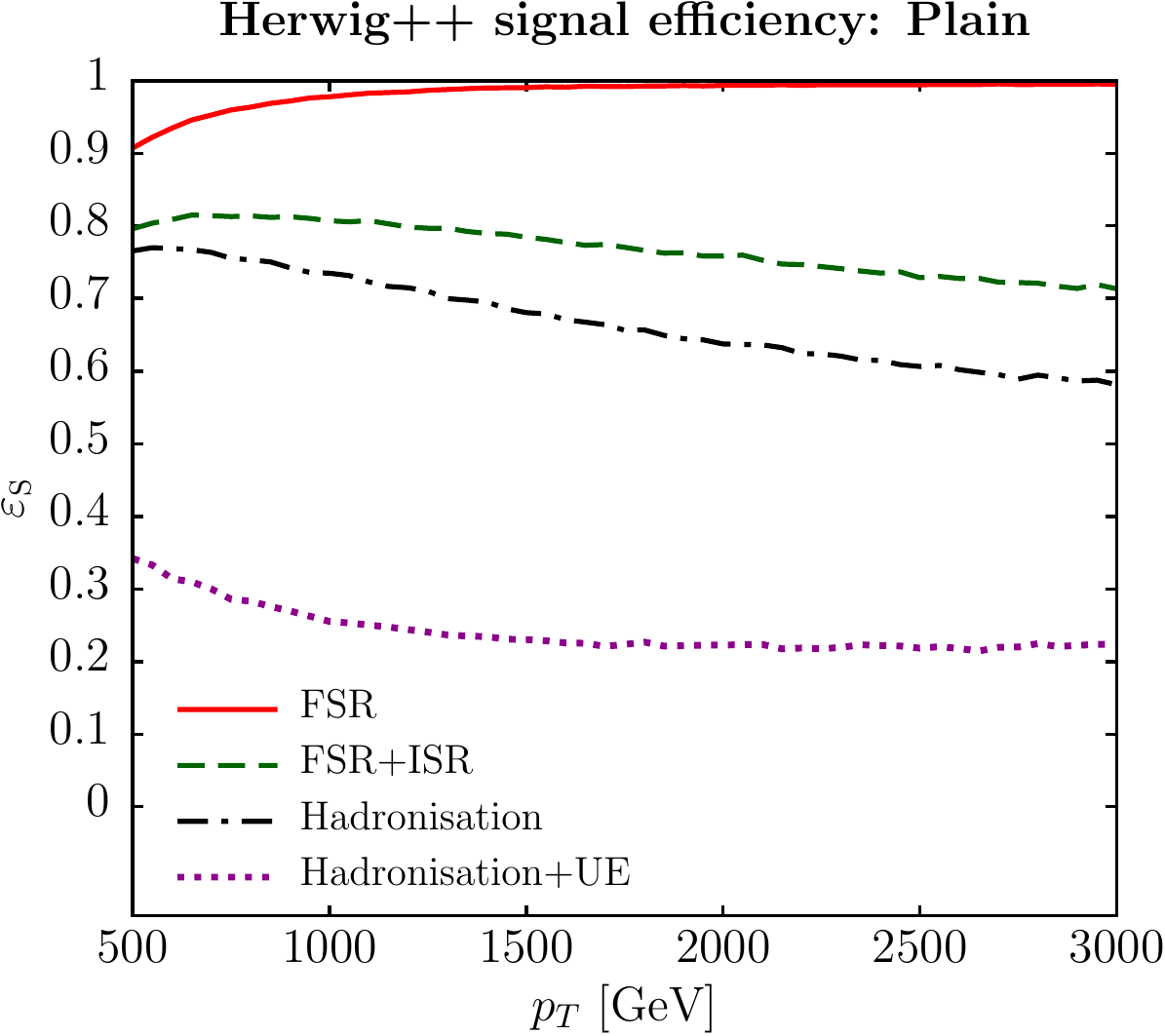}
  \end{minipage}\hfill
  \begin{minipage}{0.49\linewidth}
    \caption{An MC study of the impact of hadronisation and underlying
      event on the signal efficiency for a plain jet mass cut as a function of the minimum jet transverse momentum.
    One can see the sizeable impact of  both hadronisation and especially underlying event on 
    the signal efficiency in the window $\delta M = 16$ GeV.
    Details of the generation are as  in Fig.~\ref{fig:ISRplain} but now we also include FSR at parton level.
      \label{fig:NPPlain}
    }
  \end{minipage}
\end{figure}

In \reffig{NPPlain} we see how non-perturbative effects such as
hadronisation and underlying event affect the signal efficiency when
using plain jets tagged with $\delta M = 16$ GeV. One immediately
notices that whilst hadronisation has a more moderate effect on the
signal efficiency, which however increases with $p_T$ (more precisely
like $\sqrt{p_T}$, see Ref.~\cite{Dasgupta:2007wa}), the dominant
contribution comes from underlying event contamination which reduces
the efficiency at $p_T=3$ TeV from about $60$ percent to around $20$
percent. This implies simply that one needs to consider removal of the
UE for efficient tagging, which we shall discuss when we come to the boosted object
taggers. We have also presented results here for $R=1$ and the
averaged UE contribution to the squared jet mass varies as $R^4$
\cite{Dasgupta:2007wa}. Thus working with smaller $R$ jets one may
expect this contribution to be less significant. One should of course
consider also the presence of considerable pile-up contamination,
which we do not treat in this paper (see \cite{Soyez:2012hv,Krohn:2013lba,Cacciari:2007fd} 
for discussion of pileup subtraction techniques), but to which the plain jet mass
will also be very susceptible.
 
For now it is evident (as is well known) that the plain jet, with a mass window cut, is
not a useful option from the viewpoint of tagging signal jets due principally
to effects such as ISR, UE and pile-up contamination. It however provides a reference
point for the discussions to follow. 

\section{Trimming}
Trimming \cite{Krohn:2009th} takes all the particles in a jet defined
with radius $R$ and reclusters them into subjets using a new jet
definition with radius $\Rtrim < R$. It  retains only the subjets
which carry a minimum fraction $\fcut$ of the original jet transverse momentum
$p_T^{\mathrm{(subjet)}}>\fcut \times p_T^{\mathrm{(jet)}}$ and
discards the others. The final subjets are merged to form the trimmed jet.

It is standard to use the
Cambridge-Aachen (C/A) jet algorithm
\cite{Dokshitzer:1997in,Wobisch:1998wt} for substructure studies with trimming (and other taggers) and this is what we shall employ here.

\subsection{Lowest order result}
Compared to the plain jet mass,  trimming already has a more
interesting structure even without considering any additional radiation.
If the opening angle between the $b \bar{b}$ pair is less then
$\Rtrim$ then trimming is inactive. However, if the angle is greater
than $\Rtrim$, one removes the softer particle if its energy fraction is below $\fc$.
The result for signal efficiency is therefore given by an integral
over $z$ which can be expressed as,
\begin{equation}
\label{eq:trim}
\eps^{(0)}_{\text{S}} = \int_0^1 dz \left( 1-\Theta \left (\fc-\mathrm{min}\left [z,1-z \right] \right)\Theta \left(\frac{\Delta}{\Rtrim^2}-z(1-z) \right) \right).
\end{equation}
Strictly we should also have written above the condition for the hard
prongs to be inside the fat jet as we did for the plain jet
case. However since this condition only results in terms varying as
$\Delta/R^2$ we shall neglect it here, consistent with our approximation
$\Delta/R^2 \ll 1$.

The subtracted term in the above equation represents the removal of any
prong that has energy fraction below $\fc$, in the region where
trimming is active.

Evaluating the integral in Eq.~\eqref{eq:trim} gives the result 
\begin{multline} \label{eq:TrimBorn}
\eps^{(0)}_{\text{S}} = \left(1-2 \fc \right) \Theta \left(1-2 \fc \right)+\sqrt{1-\frac{4 \Delta}{\Rtrim^2}} \, \Theta \left (\frac{1}{4}-\frac{\Delta}{\Rtrim^2} \right)\Theta \left(\fc-\frac{1}{2} \right) + \\ + \left(2 \fc -1+\sqrt{1-\frac{4 \Delta}{\Rtrim^2}} \right)\Theta \left(\frac{1}{4}-\frac{\Delta}{\Rtrim^2} \right) \Theta \left(\frac{1}{2}-\fc \right) \Theta \left(\fc-\frac{1}{2} \left(1-\sqrt{1-\frac{4\Delta}{\Rtrim^2}}\right)\right)
\end{multline}

For now we shall consider values of $\fc$ that are standard in trimming
analyses and therefore are considerably smaller than $0.5$. For such
choices of $\fc$ the second term in Eq.~\eqref{eq:TrimBorn}, which
requires $\fc >1/2$, clearly does not contribute. While the result in \refeq{TrimBorn}  is general, let us for illustrative purposes consider
values of $R_\text{trim}$ not too small, such that
$\Delta/R_{\text{trim}}^2 \ll 1$. Then Eq.~\eqref{eq:TrimBorn} implies
a transition point at $\Delta \simeq \fc R_{\text{trim}}^2$, which
translates to a transition point at $p_T =M_H/(\sqrt{\fc}
R_{\text{trim}})$.
We remind the reader that the mass distribution for
    background jets also had transition points at $M_j^2/p_T^2=\fc R^2$
    and $M_j^2/p_T^2 =\fc \Rtrim^2$
    \cite{Dasgupta:2013via,Dasgupta:2013ihk}. The latter  
transition point is coincident with that reported above for the signal
 and corresponds to the minimal jet mass that can be obtained with trimming for a
    splitting with opening angle $\Rtrim$. As one
    increases $p_T^2$ beyond $M_j^2/(\fc \Rtrim^2$) the background
    distribution starts to grow due to the onset of a double
    logarithmic behaviour so the mistag rate increases.

 Below this value of $p_T$ the signal efficiency is
given by $1-2\fc$ and is therefore $p_T$ independent while above it one
obtains $\sqrt{1-\frac{4 \Delta}{R_{\text{trim}}^2} }$ and hence acquires a
  $p_T$ dependence. We remind the reader that these results apply
  specifically to the Higgs decay and for processes involving $W$/$Z$
  tagging different results will be obtained.
  This is due to the different splitting functions involved in hadronic $W$/$Z$ decay.

\begin{figure}[h]
\begin{center}
\includegraphics[width=0.49 \textwidth]{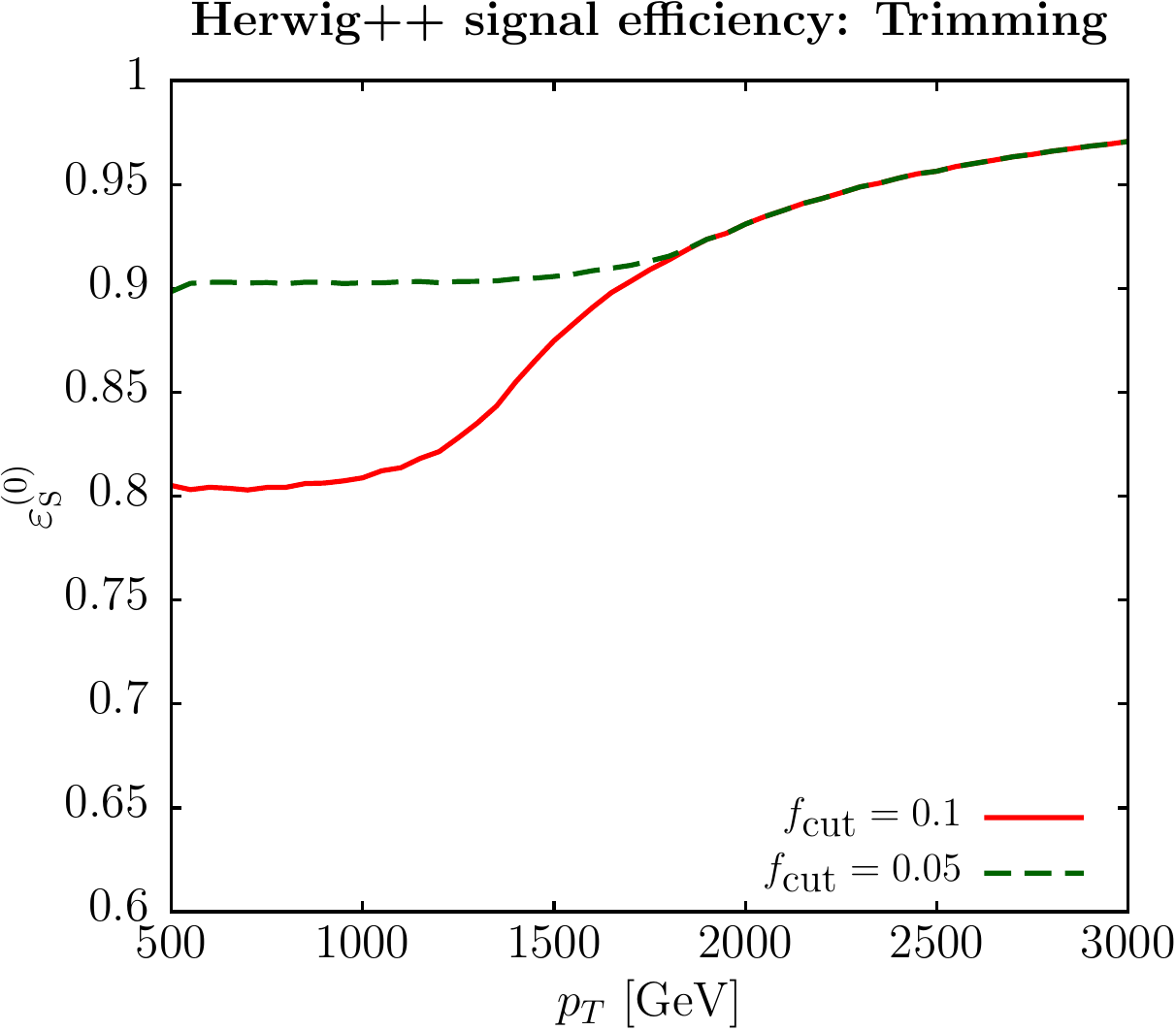}
\includegraphics[width=0.49 \textwidth]{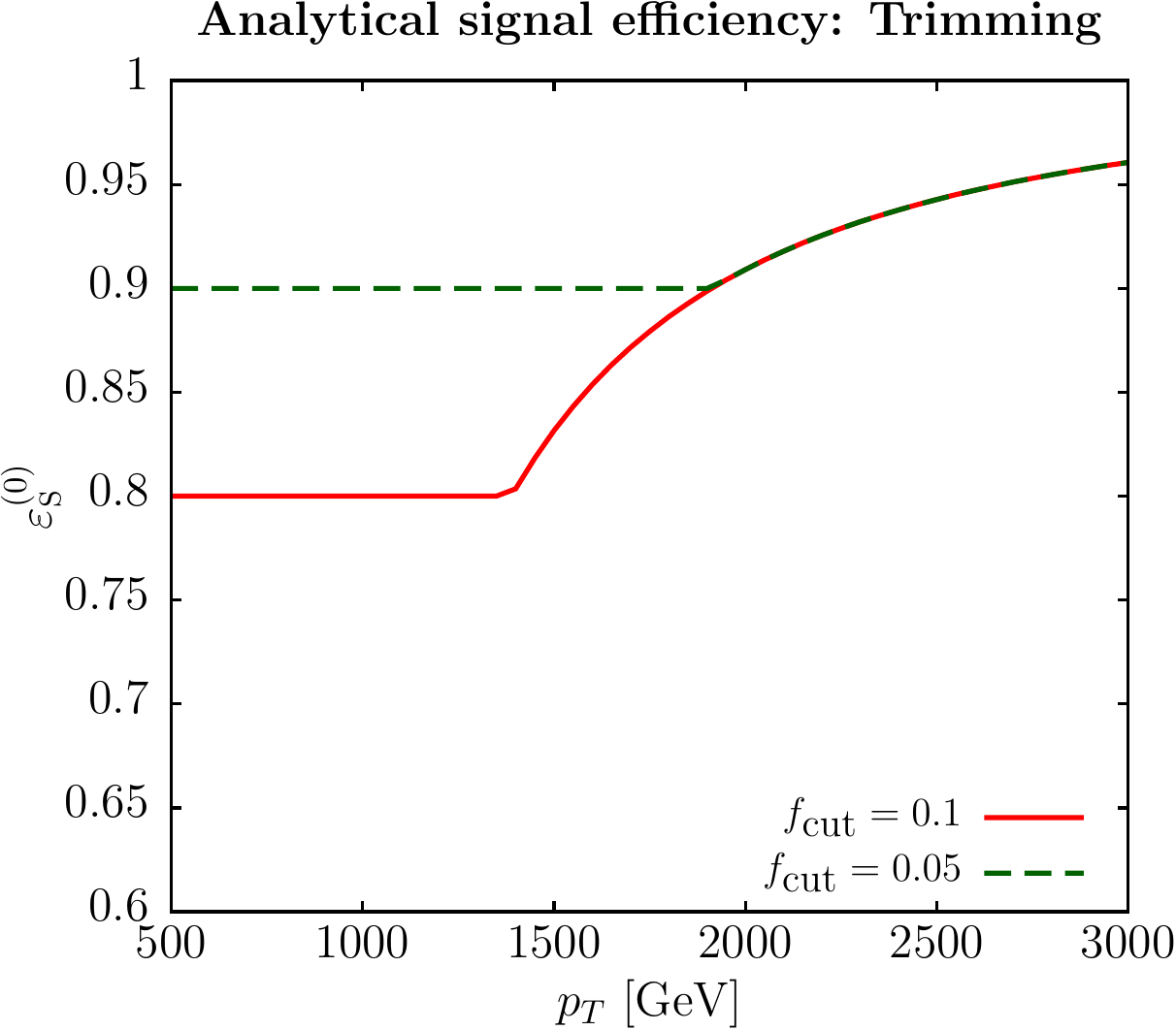}
\end{center}

\caption{Comparison of MC (left) with the analytic result \refeq{TrimBorn} (right) with $\Rtrim=0.3$ for 
the tagging efficiency for two values of $\fc$ as a function of generator level of jet transverse momentum.
This result has been generated using \herwig
at parton level with no additional radiation for $H\rightarrow
b\bar{b}$ jets.
We note that the location of the transition points are reproduced by MC.
\label{fig:TrimBORN}
}
\end{figure}

In \reffig{TrimBORN} we compare the signal efficiency using \herwig for trimming applied to boosted Higgs jets with no ISR, FSR or
non-perturbative effects to the analytical calculation above \refeq{TrimBorn}.
We generate the tagging efficiency with two different $\fcut$ values
and for $R_{\text{trim}}=0.3$ as a function of $p_T$ for both MC (left) and analytics (right).
One observes, as we would expect, that the MC clearly reproduces the analytic
behaviour of the tagger at lowest order and, for our choice of
parameters, the expected transition points at around 1320 GeV and 1860 GeV
for $\fc =0.1$ and $\fc=0.05$ respectively.

\subsection{Initial state radiation}
Let us consider the action of trimming on ISR and compare to the case
of the plain jet. For the plain jet we found a large logarithmic term
that results in loss of signal with increasing $p_T$. On the other
hand we would expect trimming to substantially remove ISR radiation and hence
wish to check the impact on the logarithmically enhanced terms that emerge
from considering soft ISR.
The key difference with the plain jet case is that when the angle
between the ISR gluon  and the jet axis exceeds $R_{\text{trim}}$ the soft
gluon is retained only if it has $k_t/p_T$ greater than $\fc$, where
$k_t$ is the transverse momentum of the soft gluon. If the $k_t$
fraction is below $\fc$ the ISR emission is removed by trimming, thus not contributing
to the jet mass, and hence in this region there is a complete cancellation with virtual corrections.
Alternatively, if the ISR falls into the trimming radius, we always
retain the emission, much like the plain jet case. These constraints
on real emission can be expressed as:
\begin{align}
\label{eq:constrainttrim}
  \Theta_\text{ISR,trim} &= \Theta\left(\theta^2-R_\text{trim}^2\right)
  \left(\Theta\left(x-\fc\right) \Theta \left(\frac{2M_H \delta M}{p_T^2\left(\theta^2+\Delta\right)}-x
      \right)+\Theta \left(\fc-x \right) 
    \right)\\ \nonumber &+\Theta\left(R_\text{trim}^2-\theta^2\right) \Theta
    \left(\frac{2M_H \delta M}{p_T^2\left(\theta^2+\Delta \right)}-x
      \right),
\end{align}
where we defined $x$ as $k_t/p_T$ and $\theta^2 =\eta^2+\phi^2$ is the
angle between the ISR gluon and the fat jet axis.

One can then repeat the calculation carried out for the plain jet mass
in the previous section using the above constraint. Taking the ISR
emission probability in the eikonal approximation as before, and
incorporating virtual corrections  we get (in a fixed-coupling approximation)

\begin{equation} \label{eq:TrimISR}
\frac{\eps_\text{S,ISR}}{\eps_\text{S}^{(0)}} = 1+C_F
\frac{\alpha_s}{\pi} \int_0^1 \frac{dx}{x} d\theta^2 \left [
  \Theta_\text{ISR,trim} -1\right ].
\end{equation}

We can evaluate the integrals straightforwardly and again shall discard
terms that are power suppressed in $\Delta$. The result obtained has
two distinct regimes:
For $\fc >\frac{2 M_H \delta M}{R^2 p_T^2} $ one gets an answer of the
form 
\begin{equation}
\label{eq:TrimISReval1}
\frac{\eps_\text{S,ISR}}{\eps_\text{S}^{(0)}} \approx
1-C_F \frac{\alpha_s}{\pi}  \left(
  R^2 \ln{\frac{1}{\fcut}}+\Rtrim^2 \ln \left( \frac{f_{\mathrm{cut}}
      p_T^2 \Rtrim^2}{2 M_H \delta M} \right)\T{\fcut-\frac{2M_H\delta
      M}{p_T^2 \Rtrim^2}} \right), 
\end{equation}
while for $\fc < \frac{2 M_H \delta M}{R^2 p_T^2}$, we get
\begin{equation}
\label{eq:TrimISReval2}
\frac{\eps_\text{S,ISR}}{\eps_\text{S}^{(0)}} \approx  1-C_F \frac{\alpha_s}{\pi} R^2 \ln \frac{R^2 p_T^2}{2 M_H
  \delta M}.
\end{equation}

Eqs.~\eqref{eq:TrimISReval1} and \eqref{eq:TrimISReval2} basically tell us that for sufficiently large
$\fc$ i.e. above $\frac{2 M_H \delta M}{R^2 p_T^2}$ one eliminates the
logarithm we obtained for the plain jet mass replacing it by a less
harmful $\ln 1/\fc$, provided one chooses $\fc$ not too small. On the
other hand for smaller $\fc$ we see a transition to the logarithmic
dependence seen for the plain mass. There is an additional correction
term in  equation Eq.~\eqref{eq:TrimISReval1} that represents
the region of integration with $\theta^2 < R_{\text{trim}}^2$. This
term vanishes as $R_{\mathrm{trim}} \to 0$ and suggests that choosing
smaller $R_{\mathrm{trim}}$ values will result in less contamination
from ISR as one may readily expect. We shall however see later, when
studying FSR radiative corrections, that we cannot choose $R_{\mathrm{trim}}^2$ too small,
i.e. $\ll \Delta$, due to degradation of the jet due to FSR loss.   If one
chooses $R_\text{trim}^2 \ll1 $ but of order $\Delta$, then within our
small $\Delta$ approximation we can simply ignore this term. If on
the other hand one chooses $R_{\text{trim}}$ to not be too small then
at very high $p_T$ one should also consider the presence of this term,
which appears only for $\fc> \frac{2M_H\delta M}{p_T^2 \Rtrim^2}
$. For most practical purposes, with commonly used parameter values,
this term can safely be ignored. For example with $\fc=0.1$, $\Rtrim =0.3$ and $\delta M=16$ GeV, even at
$p_T = 3$ TeV it only contributes order 10 percent corrections
relative to the main $\ln  1/\fc$ piece.

In principle, we should also resum the
logarithms of $\fc$ that are obtained with trimming. Such a
resummation is however also beset by non-global and clustering
logarithms and therefore highly involved. Moreover the $\ln 1/\fc$ terms
also play only a modest role numerically, for typical choices of
$\fcut \sim 0.1$ and thus their resummation is not
particularly motivated on phenomenological grounds. We note that $\ln
1/\fcut$ enhanced terms are also produced in corresponding
calculations for QCD background \cite{Dasgupta:2013ihk} and were not
resummed in that case either.
Consequently, unlike the plain jet case, we do not exponentiate the radiative 
corrections to the signal efficiency for trimming, or any of the other taggers studied in this paper.

\begin{figure}[h]
\begin{center}
\includegraphics[width=0.49 \textwidth]{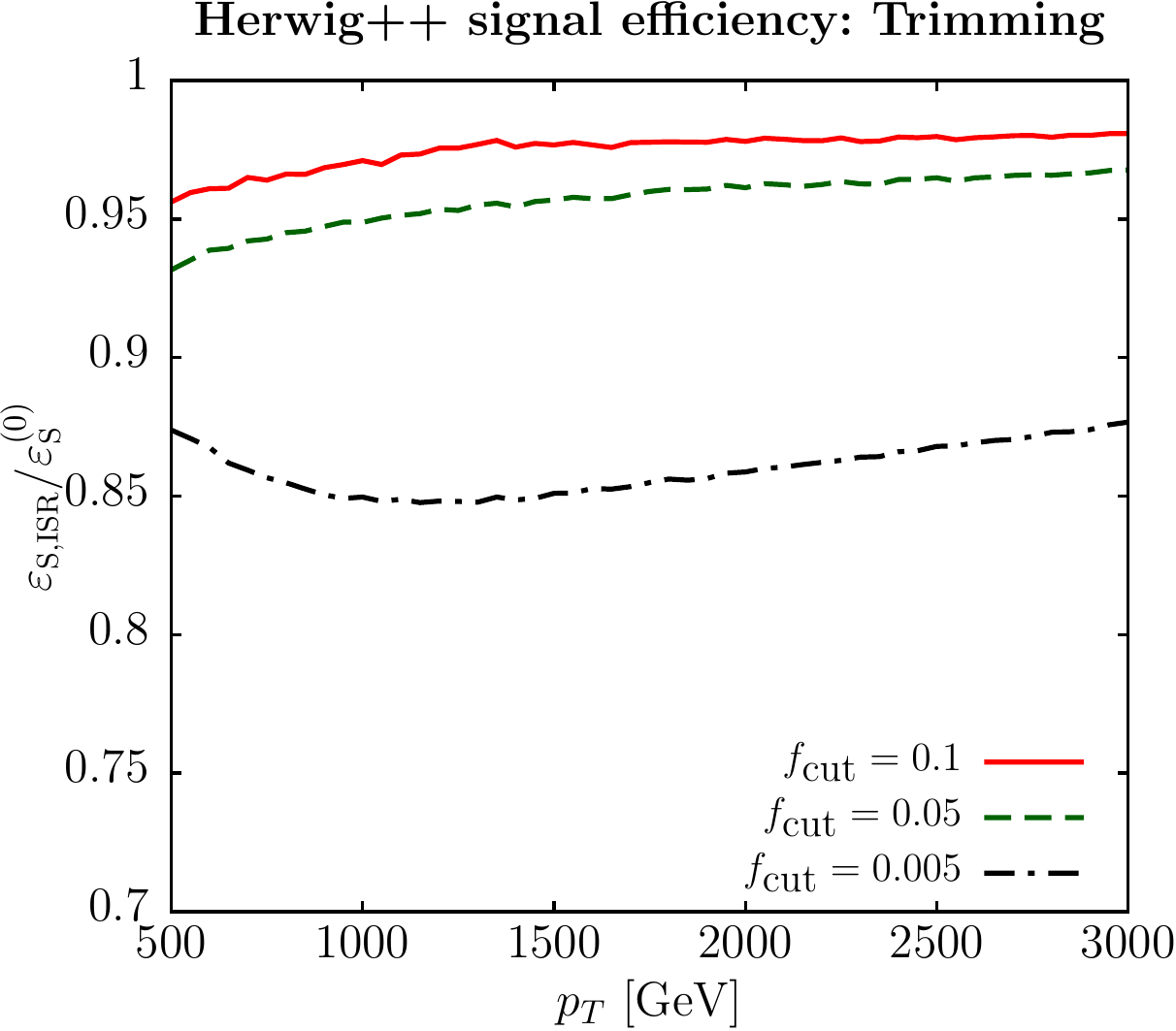}
\includegraphics[width=0.49 \textwidth]{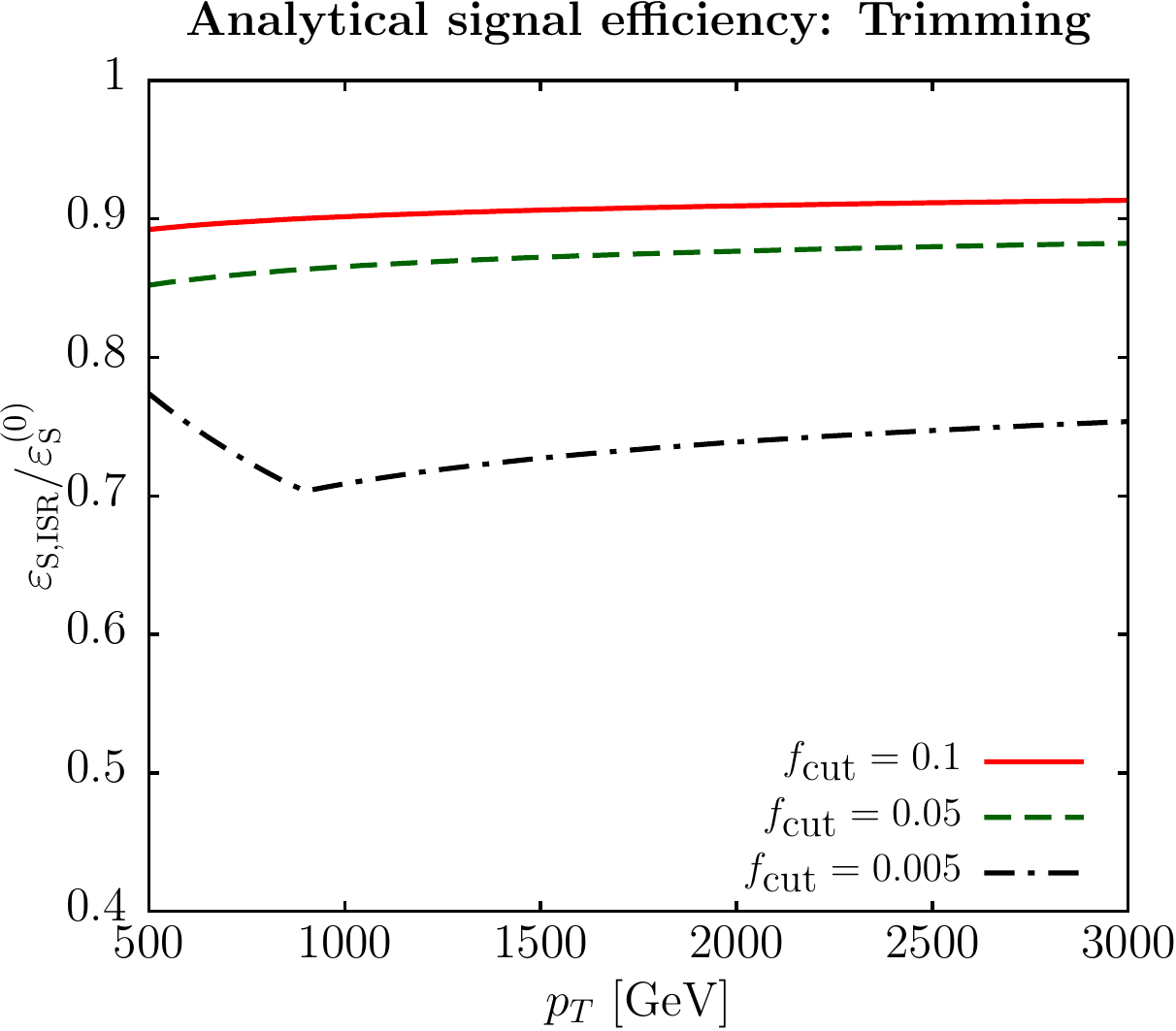}
\end{center}

\caption{Comparison of MC (left) and analytic \refeq{TrimISR} (right) trimming ($R_{\mathrm{trim}}=0.3$) 
tagging efficiencies for a range of $\fcut$ values as a function of a generator level cut on jet transverse momentum.
This result has been generated using \herwig \cite{Bahr:2008pv} at 
parton level with ISR only for $H\rightarrow b\bar{b}$ jets, setting $M_H=125$ GeV and $\delta M = 16$ GeV.
The transition points correspond to the change from plain jet mass
like behaviour to a $\ln{\fc}$ term, discussed in the main text.
\label{fig:TrimISR} }
\end{figure}

Let us then compare the main features of our simple analytical NLO
approximation, augmented to include running coupling effects, to what is seen in MC event generators.
In \reffig{TrimISR} we again compare our analytical approximations,
with running coupling effects as in the plain mass case \refeq{Singlelogevol}, to \textsf{Herwig++~2.7.0}.
For the MC studies we turn on ISR effects with boosted $H \to b\bar{b}$ jets for a range
of $\fcut$ values, as a function of jet $p_T$, keeping $\delta M$ fixed at
$16$ GeV.

Plotting the ratio of the  ISR corrected
signal efficiency to the lowest order result, we can see that the approximate NLO analytic result
reproduces the MC trends reasonably well. For values of $\fc
=0.1$ and $0.05$ we do not obtain a transition over the range of $p_T$
values shown (transition points are expected at roughly 200 GeV and
280 GeV, which are beyond the range shown) and none is seen in the MC plots. The
behaviour over the entire plotted $p_T$ range is quite
flat with $p_T$ since it depends mainly on $\ln 1/\fc$, with
running coupling and uncalculated subleading effects (in the case of
the MC results) providing the mild $p_T$ dependence that is seen.
Instead for $\fc=0.005$  we would
anticipate a plain mass like degradation of the signal efficiency
until the transition at about 890 GeV and
then for higher $p_T$ a flatter behaviour with $p_T$, consistent with
MC results. This relative flatness over a large range of $p_T$ is of course in contrast
to the pure mass cut case.

\subsection{Final state radiation}
Let us consider the response of trimming to final state radiation. 
In principle there are a number of parameters to be considered,
in particular $\fc$, $\Delta$ and $\Rtrim$ as well as the mass window
$\delta M$ and transverse momentum $p_T$.
Final state radiation, when not recombined into the fat jet, results in
a shift in mass which can cause the resulting jet to fall outside the
mass window $\delta M$. Imposing a veto on soft FSR that degrades
the jet mass results in the appearance of large logarithms whose
structure we examine here. Additionally a relatively hard FSR gluon can
also result in one of the primary $b$ quarks falling below the
asymmetry cuts that are used in taggers, and hence loss of the
signal. Such hard configurations can still come with collinear
enhancements and so their role should also be considered.

In order for an FSR gluon to be removed by trimming it has to be
emitted at an angle larger than $\Rtrim$ wrt both the hard primary
partons. In addition its energy, expressed as a fraction of the fat
jet energy, must fall below the $\fc$ cut-off. Lastly for the
resulting jet to be retained, the consequent loss in mass must be less than
$\delta M$. 

One can therefore write the following result for real emission
contributions, valid in the soft limit where the gluon energy $\omega \ll p_T$:

\begin{equation}
\label{eq:anttrim}
\eps^{(1)}_\text{S,FSR,REAL} =C_F \frac{\alpha_s}{\pi}\int_{\fc}^{1-\fc} dz
\int\frac{d\omega}{\omega} \frac{d\Omega}{2 \pi}
\frac{\left(b\bar{b}\right)}{\left(bk\right) \left(\bar{b}k\right)}
\Theta^{\text{FSR}}_{\text{trim}}  \, \, \Theta \left(\fc-\frac{\omega}{p_T}
\right) \,\,  \Theta\left (\delta M-\left(M_H-M_j\right) \right),
\end{equation}
where $d\Omega$  is the solid angle element for the emitted gluon and 
the spatial distribution of radiation has been expressed in
terms of the standard antenna pattern with the notation $\left(i j
\right)=1-\cos\theta_{ij}$. The condition
$\Theta^{\text{FSR}}_{\text{trim}}$ simply represents that the angular
integration should be carried out over the region where trimming is
active i.e. when the emitted gluon makes an angle larger than $\Rtrim$
with both $b$ and $\bar{b}$. Moreover there is an additional step
function that constrains the energy fraction to be below $\fc$ and the
factor $ \Theta\left (\delta M-\left(M_H-M_j \right) \right)$ represents the constraint on real
emissions due to the mass window $\delta M$.

There are three distinct regimes one can consider according to the
value of $\Rtrim$. Firstly when one has $\Rtrim \ll \theta_{b
  \bar{b}}$ then one can expect a collinear enhancement with a
logarithm in $\Rtrim$, that accompanies a soft logarithm arising out
of the $\delta M$ constraint. This should be the most singular
contribution one obtains for trimming so we analyse it in more detail
below. In the region where $\Rtrim \sim \theta_{b \bar{b}}$ on the
other hand there will be no collinear enhancement and one obtains a
pure soft single logarithm. In this region trimming is similar to
pruning and the mMDT as far as FSR is concerned, and we shall comment
on the results in somewhat more detail in the next section. Finally in
the region where $\Rtrim \gg \theta_{b\bar{b}}$, the FSR correction for trimming becomes more
like the plain jet where large angle corrections are strongly
suppressed.

\begin{figure}[h]
\begin{center}
\includegraphics[width=0.49 \textwidth]{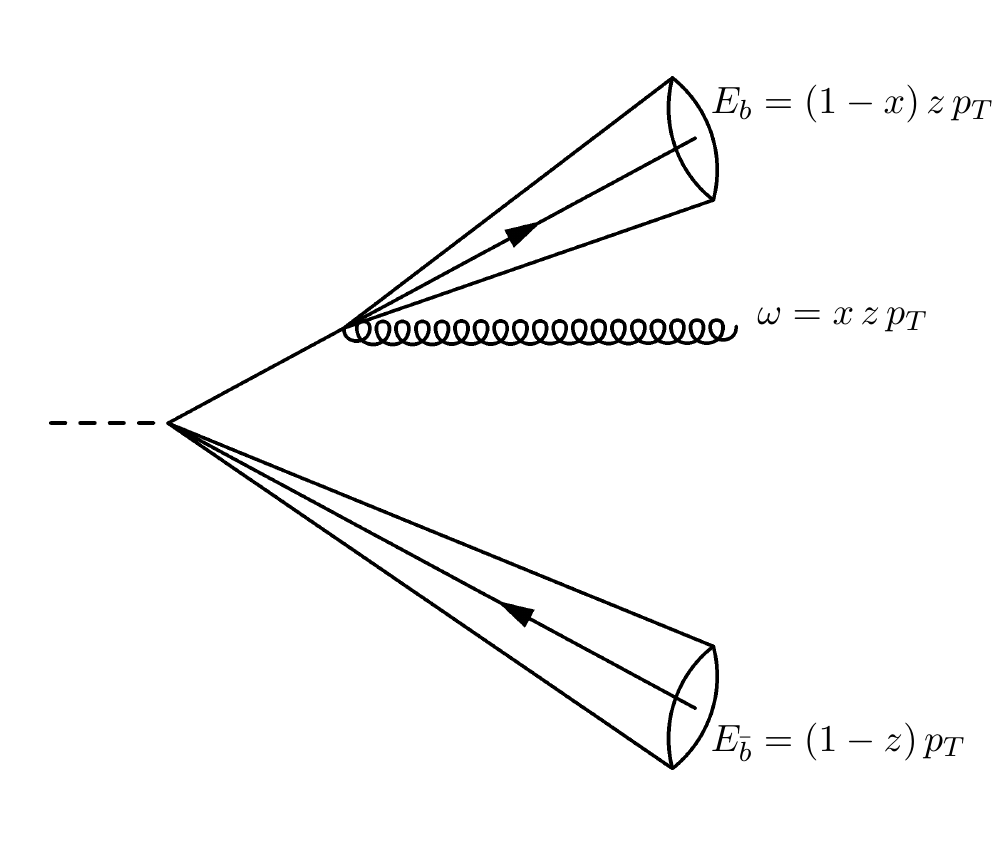}
\end{center}

\caption{A configuration which contributes to the FSR correction to trimming signal efficiency in the 
region $\Rtrim \ll \theta_{b\bar{b}}$. Here gluon $k$ is emitted collinear to the $b$ quark with momentum fraction $x$ outside a cone with radius $\Rtrim$.
\label{fig:TrimFSRDiagram} }
\end{figure}

For the soft and collinear enhanced region $\Rtrim \ll \theta_{b \bar{b}}$, let
  us perform a more detailed calculation. First let us examine the
  loss in mass in more detail. One has:
\begin{equation}
M_H^2-M_j^2 = 2 \left(p_b \cdot k+p_{\bar{b}} \cdot k \right) = \omega\left(E_b\theta_{bk}^2+E_{\bar{b}}\theta_{\bar{b}k}^2 \right).
\end{equation}

Consider first  that the gluon $k$ is emitted collinear to the
$b$ quark with momentum fraction $x$, as shown in \reffig{TrimFSRDiagram}. In this limit one can neglect the $E_b \theta_{bk}^2$
contribution above and set $\theta_{\bar{b}k}^2 \approx
\theta_{b\bar{b}}^2=\frac{\Delta}{z(1-z)}.$  Requiring $M_H-M_j < \delta M$ and neglecting terms of order $\delta M^2$ relative to $M_H
\delta M$ gives the mass window constraint $2 M_H \delta M > \omega p_T
\Delta/z$. Next, defining $x$ as the energy fraction of the soft gluon wrt
the energy of the hard emitting prong  i.e. $x=\omega/(z p_T)$,
one can write the mass window constraint as the condition  $x < 2 \delta
M/M_H$, where we used the fact that  $\Delta =M_H^2/p_T^2$. Likewise
the $\fc$ cut expressed as a condition on $x$ just gives $\fc/z>x$.

In the collinear limit, one can also simplify the angular integration
in Eq.~\eqref{eq:anttrim} which then assumes a simple
$d\theta^2/\theta^2$ form, resulting in a logarithmic contribution, $\ln \frac{\theta_{b \bar{b}}^2}{\Rtrim^2}$. It is also possible to
perform the angular integration exactly i.e. beyond the collinear
limit, to account for less singular soft large-angle contributions.
More details of the derivation and results on the angular integration are provided in
appendix A. 

We can therefore express the
soft-collinear contribution to the FSR corrections as:
\begin{equation} \label{eq:SoftTrimFSR}
\eps_{\text{S,FSR}}^{(1)} = 2 C_F \frac{\alpha_s}{\pi} \int_{\fc}^{1-\fc}  dz \ln \frac{\theta_{b
    \bar{b}}^2}{\Rtrim^2}\int\frac{dx}{x}\Theta \left(\fc/z -x \right) \left[\Theta 
\left(\frac{2 \delta M}{M_H}-x \right)-1\right],
\end{equation}
where a factor of $2$ has been inserted to account for an identical
result from the region where $k$ is collinear to $\bar{b}$ rather than
$b$ and virtual corrections have been introduced corresponding to the
$-1$ term in square brackets.

Now let us write $\theta_{b\bar{b}}^2 =\Delta/(z(1-z))$ and carry out the integration over $x$ which gives 
\begin{equation} \label{eq:SoftTrimFSR}
\eps_\text{S,FSR}^{(1)} = -2 C_F \frac{\alpha_s}{\pi}
\int_{\fc}^{1-\fc}  dz \left(\ln
\frac{\Delta}{\Rtrim^2}-\ln(z(1-z))\right) 
   \ln \frac{\fc}{z \epsilon} \Theta
 \left(\fc-z \epsilon\right),
\end{equation}
where we introduced $\epsilon = \frac{2 \delta M}{M_H}$.
The structure of the above result is, in essence, a double logarithmic
form with a soft divergence in the limit $\delta M \to 0$ and an
accompanying collinear divergence when $\Rtrim \to 0$. Let us take
$\Rtrim^2 \ll \Delta$, \footnote{In this limit, the leading order result \refeq{TrimBorn} is simply $1-2\fcut$, i.e. the two subjets never form a single subjet and are always subject to the asymmetry condition $x>\fcut$.} for simplicity ignore the accompanying $\ln(z(1-z))$
term, and integrate over $z$ to generate the following result:
\begin{equation}
\label{eq:epsFSRlog}
\eps_{\text{S,FSR}}^{(1)}= -2 C_F \frac{\alpha_s}{\pi} \ln
\frac{\Delta}{\Rtrim^2} \left[C_1 \left(\fc,\epsilon
  \right)\Theta \left(\fc-\frac{\epsilon}{1+\epsilon}\right)+C_2 \left(\fc,\epsilon
  \right) \Theta \left(\frac{\epsilon}{1+\epsilon}-\fc\right)\right],
\end{equation}
where 
\begin{align}
\label{eq:coefficients}
C_1 & = (1-2 \fc) +(1-2 \fc) \ln \frac{\fc}{\epsilon}+\fc\ln \fc -(1-\fc) \ln(1-\fc)
\, ,
\\ \nonumber
C_2 &= \frac{\fc}{\epsilon}-\fc -\fc\ln \frac{1}{\epsilon}.
\end{align}

We note that for values of $\epsilon \ll \fc$ the signal efficiency
will be dominated by a $\ln \frac{\fc}{\epsilon}$ term in the
coefficient $C_1$. The presence of the $\fc$ constraint however means that in practice such logarithms
make only modest or negligible contributions for a wide range of
values of $\epsilon$ or equivalently $\delta M$. This can be
contrasted to the case of filtering, computed in Ref.~\cite{Rubin:2010fc},
which has an identical collinear divergence to that for trimming
above, but where additionally the absence of an $\fc$ condition
leads to a much stronger $\ln \frac{1}{\epsilon}$ enhancement, which
needs to be treated with resummation. It is also straightforward to include the effects of hard collinear
radiation by considering the full $p_{gq}$ splitting function rather
than just its divergent $1/x$ piece. In this region it is possible for
the quark to fall below the $\fc$ threshold and therefore to be
removed. Such corrections do not come with soft enhancements and
produce terms that vanish with either $\fc$ or $\epsilon$, hence do not have a sizeable numerical effect that would require resummation.
For this reason, we do not calculate these terms explicitly,
continuing to work in the soft and collinear limit. 

To make the above statements more explicit let us consider the
situation at $p_T  =300$ GeV  and
choose $\fc =0.1$. The zeroth order result for signal efficiency is
then $\eps^{(0)}_\text{S}=1-2 \fc=0.8$.  If one chooses a value of
$\Rtrim=0.1$ then $\ln \left(\Delta/\Rtrim^2\right) \sim \ln 17$ and one may expect significant (collinear enhanced) radiative
corrections. Choosing a larger $\Rtrim = 0.3$ one can instead reduce $\ln
\Delta/\Rtrim^2$ to $\sim \ln 2$, which is not enhanced and does not
require resummation, implying a much more modest FSR contribution.
However we should also examine the effect of this increased $\Rtrim$ on ISR and UE
contributions.
For our choice of parameters it is evident
from MC studies that we do not pay a significant price for the
increased $\Rtrim$ value in terms of the ISR contribution.
At the same value of $p_T$ the UE contribution for $\Rtrim=0.3$ is also
small (see \reffig{NPTrimming}). If one moves to higher $p_T$, say 3
TeV, one should correspondingly lower $\Rtrim$. Here one has $\Delta
=0.0017$ and choosing a value of $\Rtrim \sim 0.1$ would ensure a
small FSR contribution as well as reduce the impact of ISR and the
underlying event.  This illustrates that by an appropriate choice of
$\Rtrim$ one can negate large radiative losses due to FSR, without
necessarily suffering from large ISR/UE effects. In general the
optimal value of $R_{\mathrm{trim}}$ will involve a trade-off between
FSR radiative corrections and ISR/UE effects. We shall return to this
point in section \ref{sec:optimalvalues}.

We should also examine the role of soft divergences that are formally
important in the $\epsilon \to 0$ limit. Taking a value of $\delta M =
2$ GeV leads to $\epsilon =0.032$. For our choice of $\fc = 0.1$, we
have $\ln \fc/\epsilon \sim \ln 3$, which is not a genuinely large
logarithm. The overall coefficient of $-2 C_F \frac{\alpha_s}{\pi} \ln
\Delta/\Rtrim^2$, which is given by the $C_1$ and $C_2$ terms
in Eq.~\eqref{eq:coefficients}, is for $\delta M= 2$ GeV,
approximately $1.58$ and for $\delta M =10$ GeV approximately 0.34,
thus indicating that resummation of soft logarithms is not a necessity. 
Expressed as a percentage of the tree level result $1-2\fc$,
the FSR corrections, as computed above,  constitute a roughly two percent to
ten percent effect  for $\delta M$ ranging from $2$ GeV to $10$ GeV,
if one chooses $\Rtrim^2 \simeq \Delta/2$.\footnote{As we shall see
  later, this constitutes a somewhat non-optimal  choice for
  $R_{\mathrm{trim}}$ and is made here for purely illustrative
  purposes, in order to estimate the size of soft but non-collinear
  enhanced effects.}

Hence,  we find that even the leading soft-collinear enhanced
 contribution makes only modest contributions to the signal
 efficiency, at best comparable to pure order $\alpha_s$ corrections.
   The main implication of this
finding is that full fixed-order calculations or combinations of
fixed-order results with parton showers (see Refs.~\cite{Buckley:2011ms,Richardson:2012bn}
for a review of the latter methods), would give a
better description of the signal efficiency than pure soft showers. We
further explore in appendix B, in somewhat more detail, the role of fixed-order
calculations in a description of the signal efficiency. 

Eq.~ (\ref{eq:epsFSRlog}) is intended to address the formal limit
$R_{\mathrm{trim}}^2 \ll \Delta$. It indicates that choosing such
  small values of $R_{\mathrm{trim}}$ is problematic due to
  degradation of the jet from FSR loss. In the opposite limit
  i.e. $R_{\mathrm{trim}}^2 \gg \Delta$ the FSR  dependence will be
similar to the plain jet mass i.e. one may expect FSR losses to be
negligible. On the other hand Eq.~(\ref{eq:TrimISReval1}) for ISR
corrections warns us that large choices of $R_{\mathrm{trim}}$ may
not be optimal due to increased ISR (and UE) contamination. One is
thus led to think about the region $R_{\mathrm{trim}}^2 \sim
\Delta$. This is reminiscent of the choice made in pruning for
$R_{\mathrm{prune}}^2 \sim M_j^2/p_T^2$. In this limit the
behaviour of FSR corrections for trimming is therefore expected to be similar to
that for pruning for which a detailed calculation is carried out in
section \ref{sec:pruning}. We simply note here that in the region
$R_{\mathrm{trim}}^2 \sim \Delta$ FSR corrections are relatively
modest and can be thought of as pure order $\alpha_s$ corrections,
rather than carrying significant logarithmic enhancements.

\subsection{Non-perturbative contributions}
Let us now study the impact of non-perturbative corrections to the signal efficiency
using trimming on boosted Higgs jets.

\begin{figure}[h]
\begin{center}
    \includegraphics[width=0.49\textwidth]{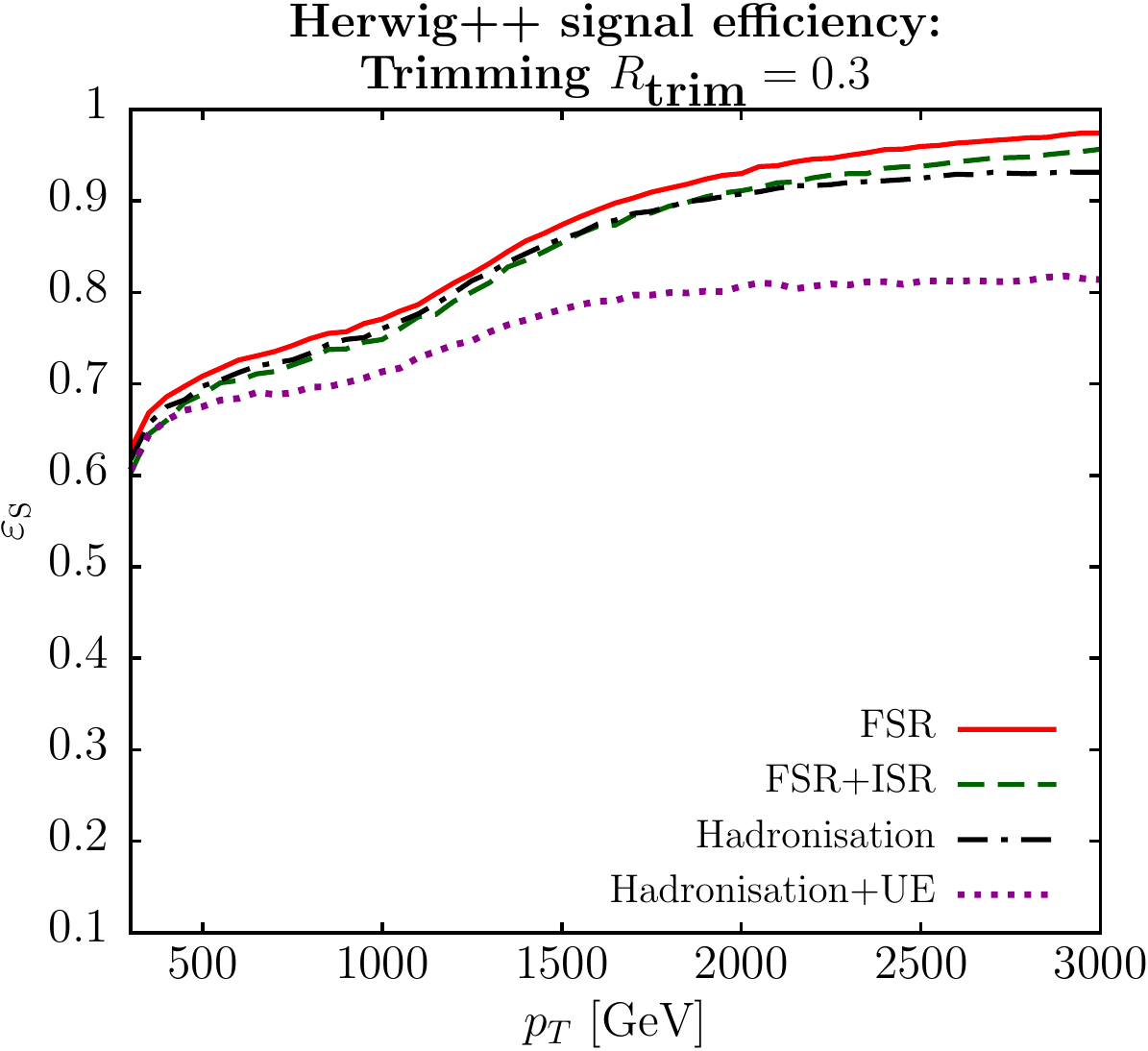}
    \includegraphics[width=0.49\textwidth]{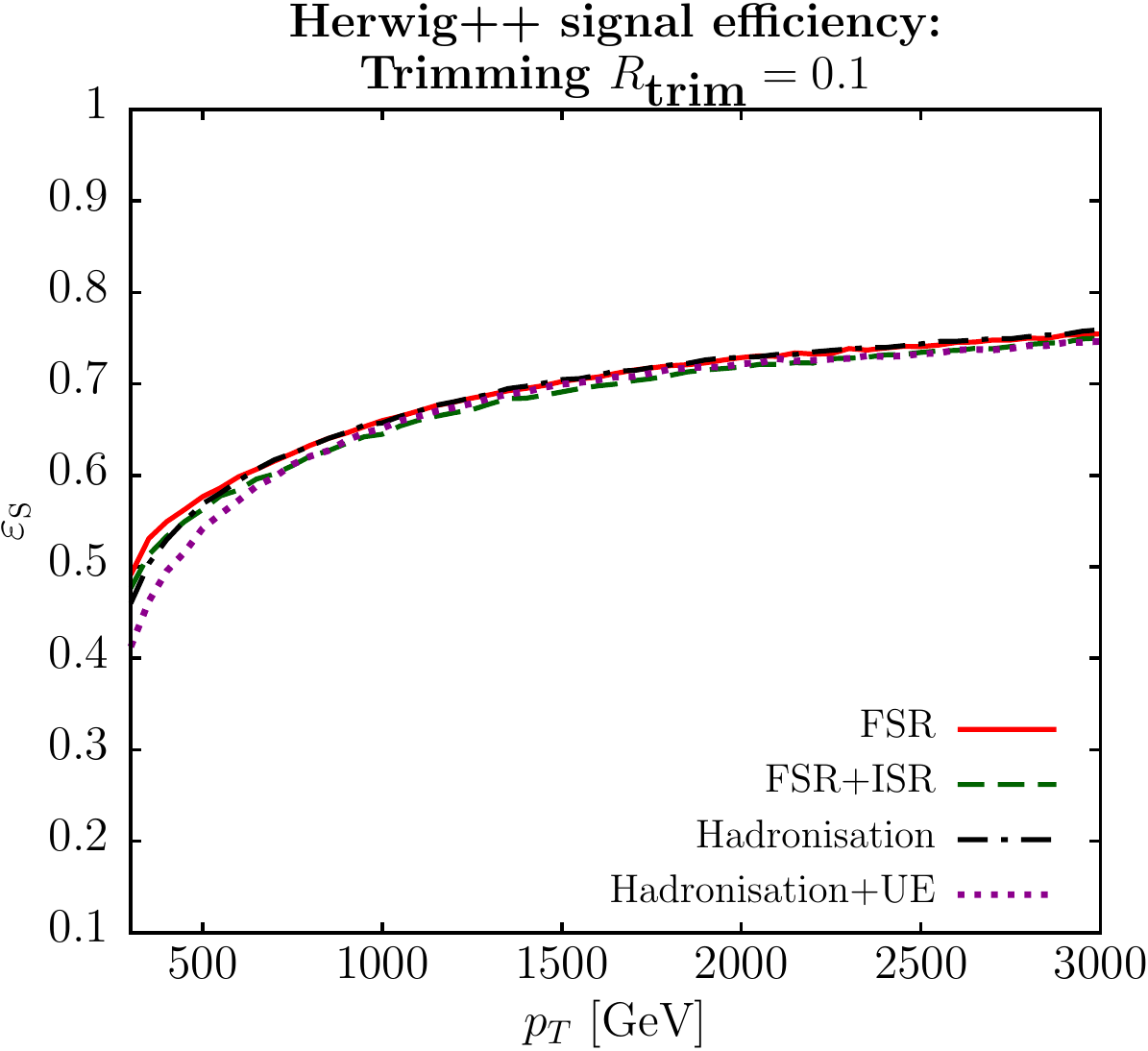}
\end{center}

\caption{An MC study of the impact of hadronisation and underlying
      event (UE) on the signal efficiency using the the trimmed jet
      ($\fcut=0.1$) as a function of the minimum jet
      transverse momentum for two different values of $\Rtrim$. One can see the impact of
      hadronisation and
      underlying event on the signal efficiency in the window $\delta
      M = 16$ GeV. Details of generation given in~\reffig{NPPlain}.
     \label{fig:NPTrimming} }
\end{figure}

In \reffig{NPTrimming} we show the signal efficiency for a boosted Higgs signal jet after application 
of trimming with  parameters $\Rtrim=0.3$ (left), $\Rtrim=0.1$ (right) with $\fcut=0.1$,
as a function of the jet transverse momentum.
One can see that hadronisation has little effect on the tagging rate of signal jets, 
due to the action of trimming on contributions which are soft and wide angle in the jet.
UE has a larger impact on the signal efficiency due to soft contamination which 
is not checked for energy asymmetry.
In other words inside the trimming radius the algorithm is inactive, and we
automatically include all contamination coming from UE,
which inside this region would contribute on average to a change in
the jet mass squared varying as $\Rtrim^4$.
The UE contribution could thus be substantially reduced by choosing a smaller $\Rtrim$. 
This is in particular required
at higher $p_T$ as evident from \reffig{NPTrimming}.
Also, in contrast to the plain jet result in \reffig{NPPlain}, one
notes a significant reduction in sensitivity to non-perturbative
effects when tagging signal jets using trimming. 
\section{Pruning and mMDT}
In this section we shall study pruning
\cite{Ellis:2009me,Ellis:2009su} and the modified mass drop
tagger \cite{Butterworth:2008iy,Dasgupta:2013ihk}. We describe these
methods together because unlike for the case of the QCD background
studied in detail in Ref.~\cite{Dasgupta:2013ihk} where the taggers can
exhibit substantial differences, for the signal one finds quite similar
behaviour.  
\subsection{Pruning}
\label{sec:pruning}
Pruning uses the initial jet to
calculate a pruning radius which is dependent on the mass of the jet
and its transverse momentum $\Rprune=R_{\mathrm{fact}} \times
\frac{2M_j}{p_T}$ where $R_{\mathrm{fact}}$ is a parameter of the 
tagger. It proceeds by reclustering the jet, at each step checking if
both the angle between the two objects $i$ and $j$ is greater than the
pruning radius $\Delta R_{ij}>\Rprune$ and the splitting is $p_T$
asymmetric i.e. $\mathrm{min}\bra{p_{T_i},p_{T_j}}<\zcut \times p_{T_{\bra{i+j}}}$.
If these conditions are both true, pruning discards the softer of $i$
and $j$, else $i,j$ are combined as usual. This is repeated for each clustering step until all particles are either discarded or combined into the final pruned jet.
For this study we use the default value
$R_{\mathrm{fact}}=\frac{1}{2}$ \cite{Ellis:2009su} and again use the C/A
algorithm to both find and recluster the jets.

At zeroth order the two signal prongs are always at an angle larger than $\Rprune$
and so the result is simply $1-2 \zcut$. For initial state radiation
one can consider pruning to be similar to trimming with $\Rtrim$
replaced by $\Rprune$. The pruning radius is given by 

\begin{equation}\label{eq:prunerad}
R_{\mathrm{prune}}^2 = \frac{(p_1+p_2+k)^2}{p_T^2} \approx \Delta + \frac{2 p_H\cdot k}{p_T^2} \approx 
\Delta + x \theta^2,
\end{equation}
where $\theta$ is the angle between the soft gluon and the Higgs
direction (or equivalently, with neglect of recoil against soft ISR,
the fat jet axis).

One then ends up comparing the gluon angle $\theta^2$ to $x\theta^2+\Delta$
and thus for sufficiently soft emissions i.e. in the limit $x\to 0$,
responsible for logarithmic corrections, one can just  replace $\Rprune^2$
by $\Delta$. The situation is therefore identical to trimming but with
$\Rtrim^2$ replaced by $\Delta$. Since we work in the limit $\Delta \ll
1$ we can neglect corrections varying as powers of $\Delta$, that
replace the $\Rtrim$ dependence in Eq.~\eqref{eq:TrimISReval1}. The result
should then be identical to trimming in that one should obtain a $\ln
\frac{1}{\zcut}$, dependence with a transition to the plain mass behaviour visible 
for smaller $\zcut$ values as in \reffig{TrimISR}. We have verified
that this is indeed the case with MC and that the efficiencies for pruning and trimming
look essentially identical in terms of the response to ISR. An MC plot
comparing ISR for all taggers is shown in the next section (\reffig{ISRall}), after we
discuss the cases of mMDT, Y-pruning and Y-splitter.

Next we discuss briefly the situation with regard to FSR, in the
context of pruning. Again one can employ the insight we gained in the
previous section for the case of trimming. For the case of pruning
there is no collinear enhancement since radiation that is lost is
emitted at an angle (wrt both hard prongs) larger than $\Rprune^2 \sim \Delta =
z(1-z) \theta_{b \bar{b}}^2$ i.e. essentially of order $\theta_{b
  \bar{b}}^2$. Thus the angular integration produces a finite
  $\mathcal{O} \left(1 \right)$ coefficient and we will thus obtain only a
  single logarithmic enhancement, that results from the loss of soft
  radiation at relative large angles i.e. those comparable to the $b \bar{b}$
  dipole size. The corresponding loss in mass can be expressed as 
\begin{equation}
M_H^2-M_j^2 = (p_1+p_2+k)^2-(p_1+p_2)^2 = 2k\cdot (p_1+p_2) \approx \omega p_T
\left(\theta^2+\Delta \right),
\end{equation}
where $\theta$ is the angle between the gluon and the jet
direction. Noting that $\theta^2$ is at most of order $\Delta$
(contributions from the region where $\theta^2 \gg \Delta$ are
negligible due to angular ordering of soft radiation) one can replace
$M_H^2-M_j^2$ by $\omega p_T \Delta$, ignoring any multiplicative factors of order
one, that lead to only subleading logarithmic terms. The condition on
gluon energy due to the mass window constraint is then $2 \frac{\delta
M}{M_H} > \frac{\omega}{p_T}$. One also requires a constraint on the
gluon energy such that it fails the $\zcut$ condition \footnote{Strictly, 
with the precise definition of the $\zcut$ condition we would
have to consider a cut on $\omega$ normalised to the energy of the
corresponding declustered prong i.e. $z p_T$ or $(1-z) p_T$ but at single-log
accuracy one can just always take a cut on $\omega/p_T$.}. In the soft limit
and to leading logarithmic accuracy this condition is just that
$\frac{\omega}{p_T} < \zcut$. Denoting $\frac{\omega}{p_T}$ by $x$, and
accounting for virtual corrections as for the case of trimming, we
have the expression for FSR corrections to pruning:
\begin{equation}
\eps_{\text{S,FSR}}^{(1)} \simeq- C_F \frac{\alpha_s}{\pi} \int_{\zcut}^{1-\zcut} dz \, \int
\frac{dx}{x} \Theta \left(\zcut-x \right) \Theta \left(x-\epsilon \right) \int
\frac{(b \bar{b})}{(bk) (\bar{b} k)} \frac{d\Omega}{2 \pi} \Theta^{\text{FSR}}_{\text{prune}},
\end{equation}
where the condition $\Theta^{\text{FSR}}_{\text{prune}}$ is simply the
condition that the gluon is emitted outside an angle $\Rprune$ wrt
both hard prongs. The angular integration and $z$ integration is
performed in appendix A and carrying out also the integrals over $x$ we obtain the
result:
\begin{equation}
\label{eq:pruningfsr}
\eps_{\text{S,FSR}}^{(1)} =- C_F \frac{\alpha_s}{\pi} \frac{2 
  \pi}{\sqrt{3}} \ln \frac{\zcut}{\epsilon}, \, \zcut > \epsilon.
\end{equation}

The above results suggest that logarithmic enhancements for pruning are
in principle present for $\epsilon \ll \zcut$. However even with a
choice of $\delta M$ as low as 2 GeV, with a choice of $\zcut=0.1$ we
obtain a modest logarithm $\sim \ln 3$, implying that soft enhanced
effects can be neglected. To therefore assess FSR corrections in more
detail, as we found for trimming, it is necessary to go beyond the soft approximation and study
hard corrections, which we explore further in appendix B. However it should be apparent that radiative corrections due
to FSR corrections represent essentially order $\alpha_s$ corrections
without significant log enhancements over a wide range of values of
$\delta M$, $p_T$ and $\zcut$. We can exploit this stability against
radiative corrections in optimising the tagger parameters, which we do
in a subsequent section.

To obtain a complete picture of pruning we also need to account for
non-perturbative corrections arising from hadronisation and UE
corrections. We shall comment on MC results for these aspects together
with the mMDT results below.

\subsection{Modified mass drop tagger}
Here we shall consider the modified mass drop tagger along similar
lines. We start by recalling the definition first of the regular mass
drop tagger:
The mass drop tagger (MDT) \cite{Butterworth:2008iy} is intended for
use with jets clustered using the C/A algorithm.
For each jet $j$ one applies the algorithm:
\begin{enumerate}
\item If jet $j$ contains subjets, split $j$ into two subjets $j_1$ and $j_2$ by undoing the last stage of clustering such that $m_{j_1}>m_{j_2}$.
\item If there is a significant mass drop $\mu \times m_j>m_{j_1}$ \emph{and} the splitting is not too asymmetric $\mathrm{min}\bra{p^2_{T,j_1},p^2_{T,j_2}}\Delta R^2_{j_1,j_2}/m_j^2 >\ycut$ deem $j$ to be a ``tagged'' jet and exit the loop.
\item Else relabel $j_1=j$ and repeat from step 1.
\end{enumerate}

The modified mass drop tagger corrected a flaw in the mass drop tagger
so that in the event that the mass drop and asymmetry conditions are
not satisfied one follows the more energetic (higher $p_T$ branch) rather than the
heavier branch  $j_1$ as advocated above. This is not only
physically relevant (as one ensures that one identifies hard
substructure rather than, for a small fraction of events, following
soft massive jets)  but also ameliorates significantly the perturbative structure for
calculations related to QCD background jets, rendering for instance
the QCD jet mass distribution purely single logarithmic and free from
non-global logarithms. 

One other observation that was made in Ref.~\cite{Dasgupta:2013ihk}
concerned the role of the mass drop parameter $\mu$ itself. There it
was noted that the mass drop condition had a negligible impact on the
result obtained for the jet mass distribution for QCD background jets
and hence that it was possible to entirely ignore the mass drop
requirement. For the current paper we shall consider this variant of
the mMDT, where we do not impose the mass-drop condition but just the
asymmetry requirement via a $\ycut$ cut-off.

At zeroth order we obtain a signal efficiency
$\eps_\text{S}^{(0)}=1-2\ycut$ coming from the asymmetry cut, which is the
same result as for pruning.
As far as the response to ISR is concerned, one can straightforwardly
see that the general behaviour will be similar to the taggers we have
considered before. Consider a fat jet consisting of a $b \bar{b}$ pair
and an ISR gluon. If the gluon makes an angle less than
$\theta_{b \bar{b}}$ with either of the hard prongs of the fat
jet then on declustering, we will break the jet into a massless prong and
a prong with a small mass consisting of a quark
and the soft gluon. In the soft limit the asymmetry condition
will pass if the hard prongs are sufficiently energetic, i.e. exactly
as at zeroth order, and the soft ISR will contaminate the jet. Such
corrections will vanish with $M_H/p_T$ just like for the case of pruning
and hence we can ignore them here, at high $p_T$. For relatively large angle ISR, the
soft gluon emerges first on declustering the jet.  If it fails the asymmetry condition it is
removed and its effects cancel against virtual corrections. If it
passes the asymmetry cut one obtains, in the small $\Delta$ limit,
essentially a logarithm in $\ycut$ as for pruning and trimming.
One should note that, if one uses the asymmetry condition, with the
$\ycut$ measure exactly as defined above, the condition for the gluon
to pass the asymmetry can be expressed as $x^2 \theta^2/(\Delta+x
\theta^2)>\ycut$, where $x=\omega/p_T$. Combining this with the
condition $\theta^2<R^2$, on the angular integration, we get the constraint:
\begin{equation}
\label{eq:MDTconstr}
x > \frac{\ycut}{2} \left(1+\sqrt{1+\frac{4 \Delta}{\yc R^2} }\right),
\end{equation}
which for $\ycut \gg \Delta/R^2$ reduces to the same constraint as for the
case of pruning and trimming i.e. $x > \ycut$. The main effect of this slightly  different
relationship between the gluon energy and $\ycut$ manifests itself, 
only for rather small $\yc$ values and at low $p_T$, in terms of a
change in the transition point to the plain jet mass like
behaviour. The position of the transition can be computed as before by setting
the RHS of Eq.~\eqref{eq:MDTconstr} equal to $\frac{2 M_H \delta
  M}{p_T^2}$ recalling that we take the fat jet radius $R=1$. Let us
consider $\fc=0.005$ and $\delta M=16$ GeV, for which one obtains a
transition point with trimming at $p_T \sim 900$ GeV,  below which one
sees a plain jet like behaviour. For the same value of $\yc$ the
corresponding transition point for mMDT occurs at roughly $p_T \sim 400$ GeV,
i.e. is absent over the range of $p_T$ considered here.  In any case very small values of $\yc$ would mean that large
logarithms in $\yc$ become important and hence should in general be
avoided. For reasonable values of $\yc \sim 0.1$, mMDT behaves essentially identical to pruning and trimming. We have verified all of
the above points with MC studies and shall provide a plot comparing
the response of taggers to ISR in the next section, after discussing
Y-pruning and Y-splitter (see \reffig{ISRall}).
Lastly we shall mention that for FSR corrections in the soft
approximation, we do not observe any significant differences between
mMDT and pruning. To understand this it is enough to realise that a
soft FSR gluon emitted at an angle smaller than $\theta_{b \bar{b}}$
  wrt either the $b$ or $\bar{b}$ direction is not examined for the
  asymmetry condition and hence does not contribute to a loss in jet
  mass, implying also the absence of any collinear
  enhancements. Emissions at an angle larger than
  $\theta^2_{b\bar{b}}=\frac{\Delta}{z(1-z)}$ contribute to a loss in
mass and give a soft single logarithmic contribution identical to that
for pruning. The result obtained is identical to Eq.~\eqref{eq:pruningfsr}
with the coefficient $2 \pi/\sqrt{3}\approx  3.63$ replaced by a coefficient that
we have determined numerically. For $p_T$ =
3 TeV the coefficient we obtain is $\approx 0.646$. \footnote{This is the same result, $J\bra{1}$, found by Rubin in Ref. \cite{Rubin:2010fc} for the coefficient of the FSR soft logarithm for filtering when $\Rfilt=\theta_{b\bar{b}}$.} The key point however is that no large
logarithmic corrections arise due to soft FSR emissions, owing to the
presence of the $\ycut$ cut-off. Once again it would therefore be of
interest to study the role of genuinely hard radiative corrections
beyond the eikonal approximation, a study we carry out in appendix
B.

\subsection{Non-perturbative effects and MC results}
We have analysed the effects of ISR and FSR for both pruning and mMDT
and concluded that the taggers have an essentially similar behaviour
for the case of signal jets. Of course our studies have focussed thus
far on the perturbative component and hence it is prudent to
examine non-perturbative effects before reaching any firm
conclusions. 

\begin{figure}[h]
\begin{center}
\includegraphics[width=0.49 \textwidth]{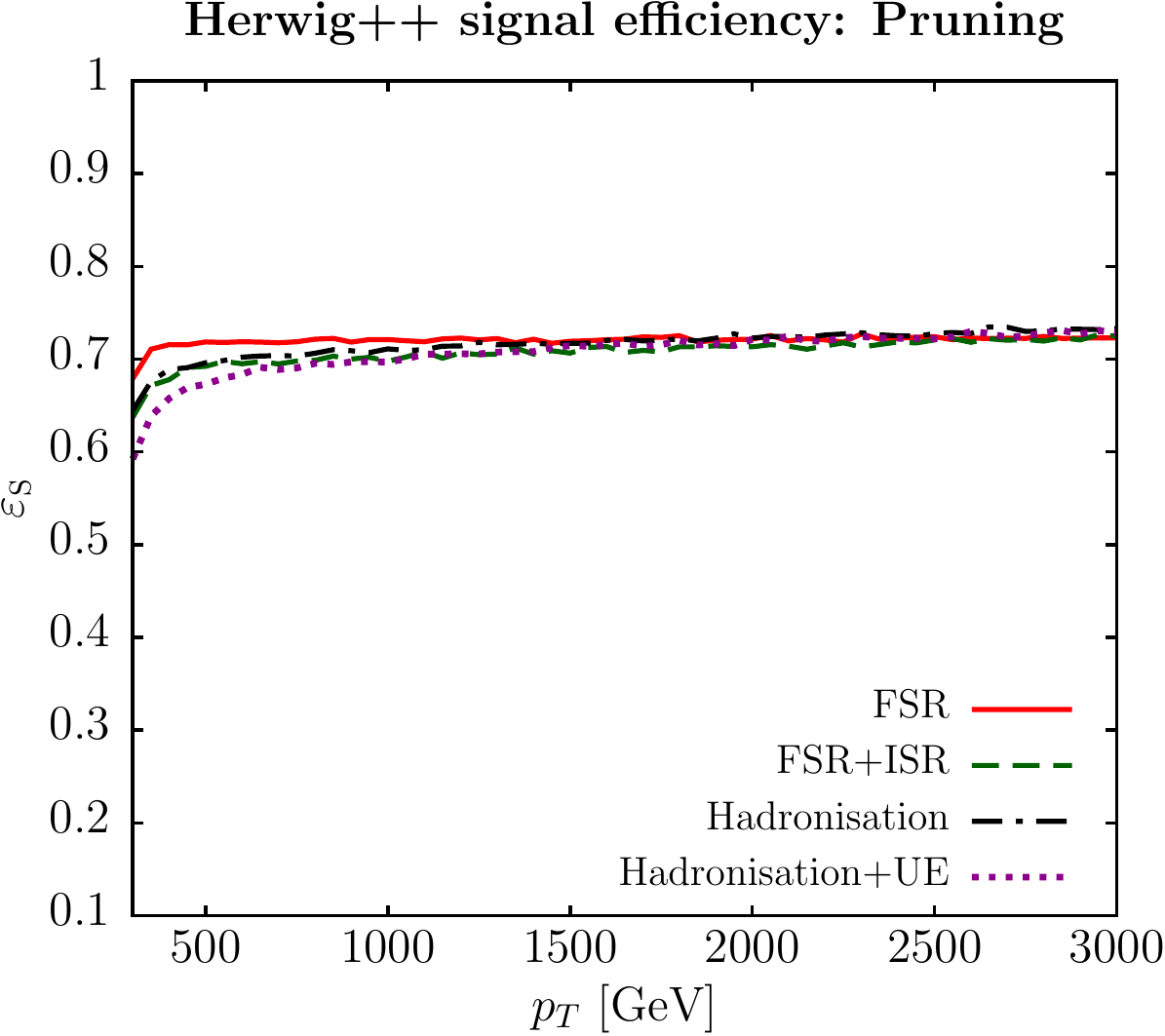}
\includegraphics[width=0.49 \textwidth]{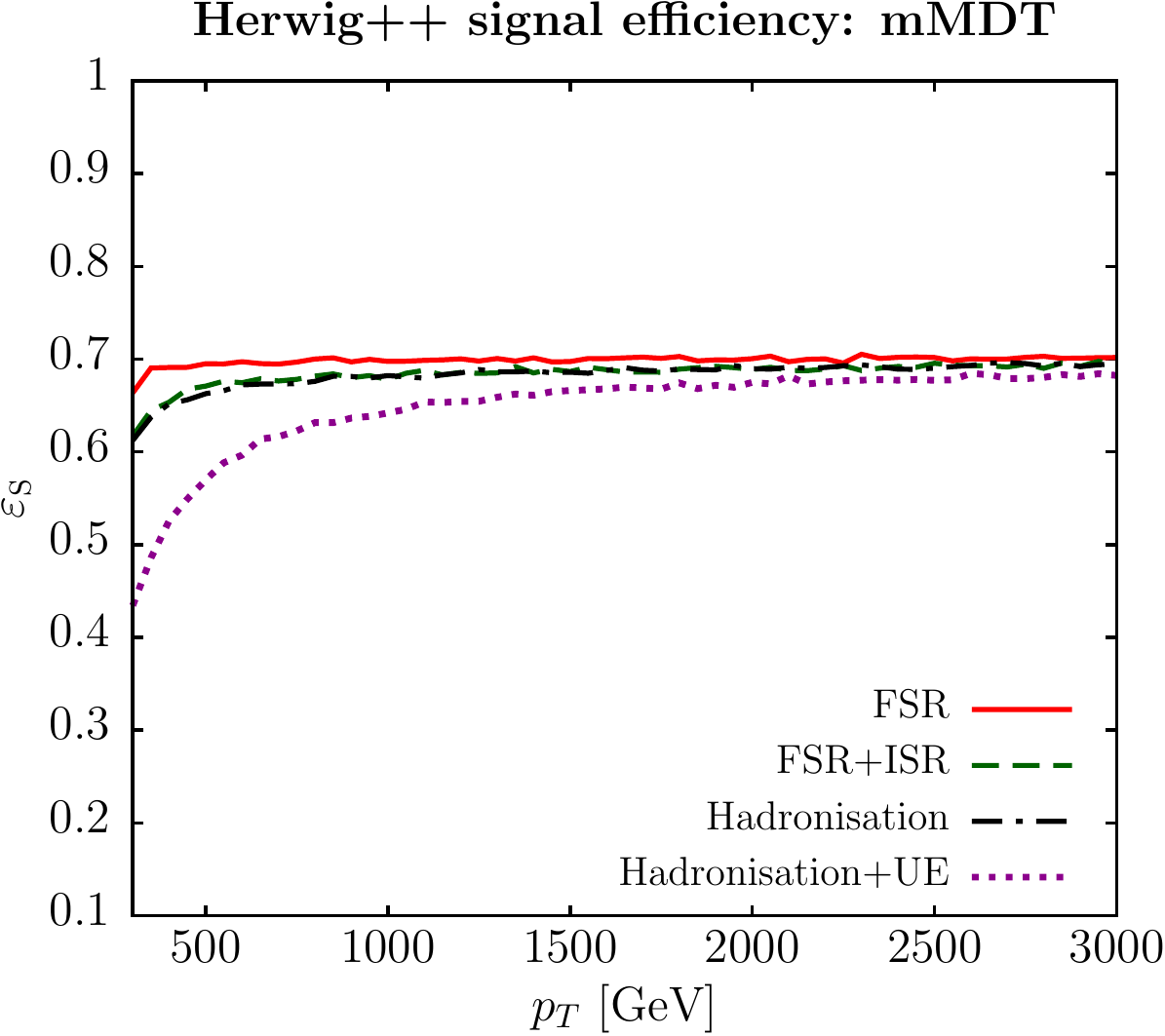}
 \caption{An MC study of the impact of hadronisation and
   underlying event (UE) on the signal efficiency for pruning (left)
   and mMDT (right)  ($\zcut,\ycut=0.1$) as a function of  jet transverse
   momentum with $\delta M =16 \, \text{GeV}$. 
   Details of generation are given in~\reffig{NPPlain}.\label{fig:mMDTpruneall}}
\end{center}
\end{figure}

We show in Fig.~\ref{fig:mMDTpruneall} the MC results for pruning and
mMDT. One observes that the signal efficiency has only a
weak dependence on $p_T$ and that relative to the lowest order
expectation of $\eps^{(0)}_\text{S}=0.8$, at high $p_T$ one sees a
roughly 10 percent difference for the full parton level result with
radiative corrections. One also sees a remarkable similarity between
the two taggers over the entire $p_T$  range as far as parton level
results and those including hadronisation are concerned. The
UE contamination is however more clearly visible in the mMDT case towards
lower $p_T$ values which owes to the larger effective radius
$\theta_{b\bar{b}}=\frac{M_H}{p_T \sqrt{z(1-z)}}$ as compared to
$\Rprune \approx M_H/p_T$ for pruning as well as differences in the definitions of the asymmetry
parameters $\ycut$ vs $\zcut$.\footnote{It is of course possible to
  use mMDT with a $\zcut$ constraint defined as for pruning instead of
  $\ycut$, as was studied in Ref.~\cite{Dasgupta:2013ihk}. This choice
  would further enhance the similarity we observe for signal jets and is the default in the current public implementation of mMDT in \fasjet~\cite{Cacciari:2011ma}.}

 At lower $p_T$ therefore it has been
standard practice to use the mass drop tagger in conjunction with filtering as
suggested in the original reference Ref.~\cite{Butterworth:2008iy}.  One
should also bear in mind the results of Ref.~\cite{Dasgupta:2013ihk}
where for QCD background jets much more pronounced non-perturbative
effects were observed for pruning than for mMDT, and in the final
analysis one expects the impact on the background to dictate the
ultimate performance of the taggers, rather than the comparatively
small corrections one sees here  for the signal, over most of the
$p_T$ range studied.

A final point to make about Fig.~\ref{fig:mMDTpruneall} is about the
contrast between the FSR corrections observed for mMDT and pruning to
those seen in Fig.~\ref{fig:NPTrimming} for trimming. To make the comparison we
note the fact that for Fig.~\ref{fig:NPTrimming} we have chosen $\fcut=0.1$ and
consider $R_{\mathrm{trim}}=0.1$. Then the zeroth order result for
trimming is simply $1-2\fcut$ as for mMDT and pruning, within the $p_T$ range we are studying. 
It is evident from Figs.~\ref{fig:NPTrimming} and \ref{fig:mMDTpruneall} that while
the  FSR results for mMDT and pruning show hardly any dependence on
$p_T$ over the range studied, the corresponding results for trimming show a more pronounced $p_T$
dependence. This feature already emerges from our simplified
fixed-coupling analytics where the FSR corrections for trimming depend on $p_T$ via a
dependence on $\ln \Delta/R_{\mathrm{trim}}^2$ while for pruning  and
mMDT we have shown that the FSR corrections are $p_T$ independent (see
e.g. Eq.~\eqref{eq:pruningfsr} for pruning). We shall return to this
point in section \ref{sec:optimalvalues}.

\section{Y-pruning and Y-splitter}
In this section we shall study the Y-pruning modification of pruning
suggested recently \cite{Dasgupta:2013ihk}, along with the older
Y-splitter method \cite{Butterworth:2002tt}. We shall study these two
methods together because they have a remarkably similar action on QCD
background jets, and are particularly effective in cutting out QCD
background in the vicinity of signal peaks, for boosted Higgs and
electroweak gauge bosons, which makes them potentially valuable
tools. They however differ very significantly from
each other and from other taggers in their response to initial state
radiation (and even more significantly to UE), for reasons we
highlight below, and which were also mentioned in the case of
Y-pruning, in Ref.~\cite{Dasgupta:2013ihk}. In the next subsection we
shall use the insight we gain in the present section to suggest
improvements to Y-splitter in particular.
\subsection{Y-pruning}
We begin by examining the case of Y-pruning. Let us recall that this
is a modification of pruning where one requires that at least one
clustering is explicitly checked for and passes the pruning criteria
else one discards the jet. In this way one removes spurious configurations
where all emissions that are left after pruning is applied, are within
an angular distance $\Rprune$ of one another and hence never get
examined for an asymmetry condition, resulting in the tagging of
structures with only a single hard prong.  

A known issue with Y-pruning for the case of signal jets, already
discovered in Ref.~\cite{Dasgupta:2013ihk}, concerns its response to soft
wide-angle emissions from ISR or UE. Here one can have a situation
where a soft ISR emission contributes to setting the pruning radius
but is itself removed by pruning. If the pruning radius set by the ISR emission is
larger than $\theta_{b \bar{b}}$ then one would discard the resulting
jet as it does not satisfy Y-pruning, causing a loss of signal. In
the same kinematic region virtual corrections would lead to a jet
accepted by Y-pruning (assuming the hard prongs arising from Higgs
decay satisfy the $\zcut$ criterion), and hence contribute to the
signal tagging efficiency, as we shall demonstrate below.

We first consider that for an ISR emission with energy, or
equivalently $k_t$, fraction $x \ll 1$ which makes an angle $\theta$ with the fat jet axis (again we neglect
recoil against soft ISR), we have that $\Rprune^2 = M_j^2/p_T^2 \approx
\Delta+x \theta^2$. We wish to compute the virtual contribution in the
region where the real ISR is removed i.e. in the kinematic region where
$\theta^2 > \Rprune^2$ and where $x < \zcut$. Moreover we require that
$\Rprune^2 > \theta_{b \bar{b}}^2$ so that the hard prongs are inside
the pruning radius. 

Thus we have, for the contribution of uncancelled virtual gluons, the equation
\begin{equation}
\Delta \eps_{\text{S}}=-C_F \frac{\alpha_s}{\pi} \int_{\zcut}^{1-\zcut}dz \int \frac{dx}{x}
d\theta^2 \, \Theta \left( \theta^2 - \Delta \right) \Theta\left(\theta^2-\frac{\Delta}{x} f(z) \right) \Theta(R^2-\theta^2) \Theta(\zcut-x),
\end{equation}
where we have denoted the extra contribution for Y-pruning, relative
to pruning, by $\Delta\eps_{\text{S}}$, have defined the function
$f(z) =\frac{1}{z(1-z)}-1$ and where the overall minus sign indicates
that we are considering the virtual contribution. The first step function in the above equation comes from the
requirement that we are considering the region where the ISR
emission lies at a larger angle than $R_{\mathrm{prune}}$ relative to
the jet axis (with $x \ll1$) while the second step function expresses the constraint
that the $b\bar{b}$ opening angle is less than $\Rprune$, the pruning
radius. Lastly we have conditions corresponding to the ISR radiation being in
the fat jet $R^2>\theta^2$ and the energy condition $x<\zcut$
corresponding to removal of the corresponding real radiation.
We can straightforwardly carry out the  integrals over $\theta^2$, $x$ and $z$. In
particular the integration over $x$ produces the logarithmically
enhanced term we seek, where one obtains a logarithm in the ratio of
$\zcut$ to  $\Delta$. One may expect that this logarithm becomes large
and hence should have a visible effect at high $p_T$, for values of
$\zcut$ that are not too small. The final result, discarding all other
terms that are less singular in the high $p_T$ limit (e.g. those that vanish with $\Delta$), is
\begin{align} \label{eq:YPruneISR}
\Delta \eps_{\text{S}} \approx-C_F \frac{\alpha_s}{\pi} R^2 \ln
\frac{\zcut R^2}{\Delta} \Theta \left(\beta-3 \right) &\Bigg[ \sqrt{1-\frac{4}{1+\beta}} \Theta \left(\frac{1}{1+\beta}-\zcut \left(1-\zcut \right) \right)\nonumber\\
&+(1-2\zcut) \Theta\left(\zcut(1-\zcut)-\frac{1}{1+\beta} \right) \Bigg],
\end{align}
where we defined $\beta = \frac{\zcut R^2}{\Delta}$.

At high $p_T$ therefore, $\Delta \eps_S$  potentially dominates the normal logarithmic dependence of pruning on $\zcut$.
We can therefore distinguish Y-pruning from other taggers by looking at the transverse momentum dependence of the signal response to ISR in the high $p_T$ limit.

\begin{figure}[h]
\begin{center}
\includegraphics[width=0.49 \textwidth]{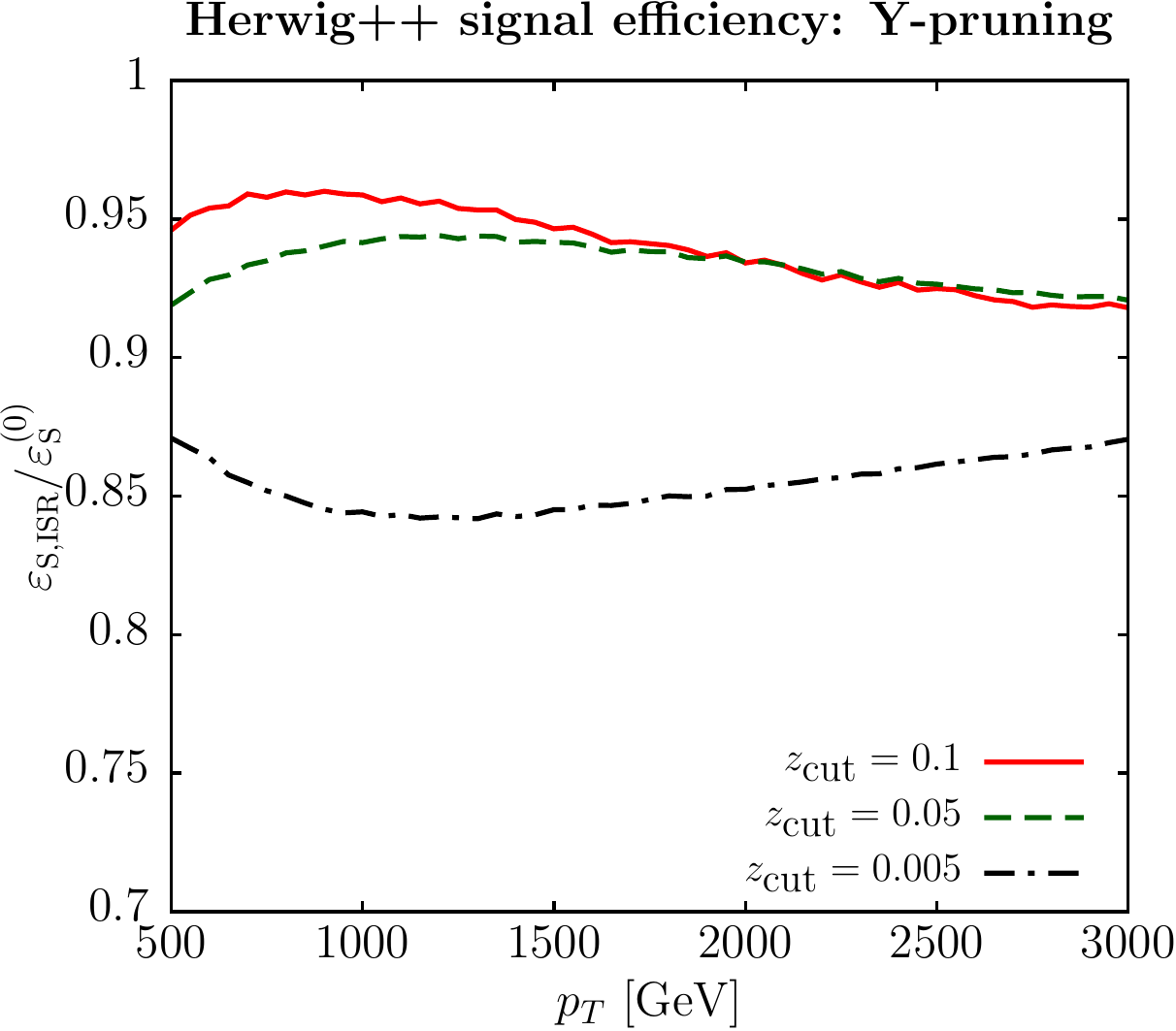}
\includegraphics[width=0.49
\textwidth]{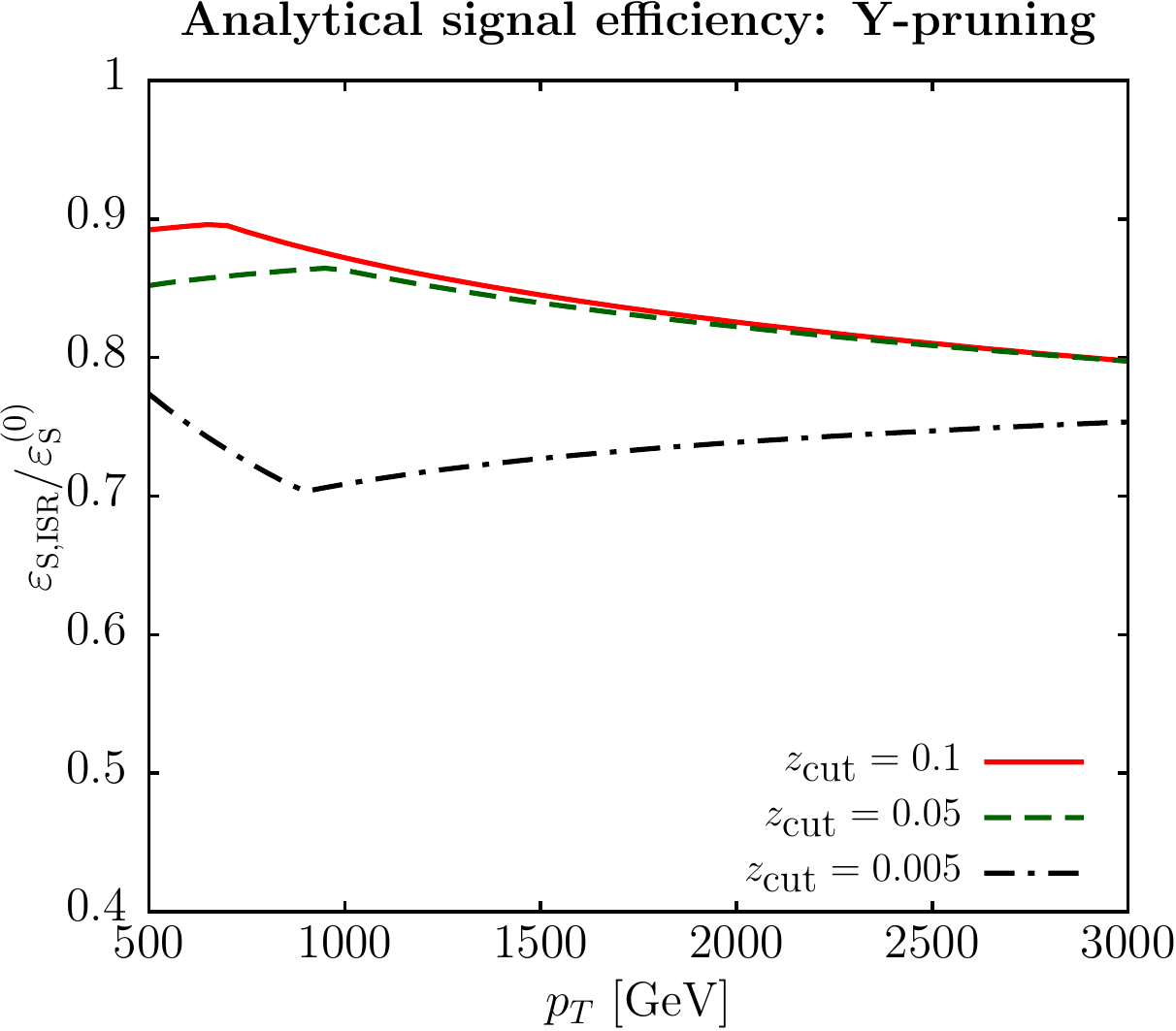}
\caption{Comparison of MC (left) and analytic (right) Y-pruning signal tagging efficiencies for a range of $\zcut$ values as a function of a generator level cut on jet transverse momentum.
This result has been generated using \herwig \cite{Bahr:2008pv} at parton level with ISR only for $H\rightarrow b\bar{b}$ jets, setting $M_H=125$ GeV and $\delta M = 16$ GeV.
We have divided out the contribution due to the born configuration in both panels for clarity.
\label{fig:YPruneISR}}
\end{center}
\end{figure}

In \reffig{YPruneISR} we compare the sum of the analytic result from
pruning and the additional contribution described in \refeq{YPruneISR} to MC with ISR  for a range of $\zcut$ values.
One immediately notices, in both analytical and MC plots,
that the $p_T$ dependence of Y-pruning is significantly different from
that of pruning for the commonly used value $\zcut=0.1$, (see \reffig{ISRall}).
In agreement with our expectations the signal
efficiency as given by MC in \reffig{YPruneISR} first increases with  $p_T$ as for the case
of pruning and then decreases beyond a certain point which we expect
to be the onset of the logarithmic behaviour we have computed
above. Our calculation indicates that the
onset of the logarithm in $\zcut/\Delta$ is for $\beta > 3$ which for
$\zcut =0.1$ corresponds to approximately 680 GeV and for $\zcut=0.05$
to approximately 970 GeV. This is consistent with what is seen in MC
though of course the  transitions are not as sharp as in the analytical result.

As far as final state radiation is concerned there is no significant difference
between Y-pruning and pruning. The soft large-angle contributions that
are responsible for the loss of signal we saw for ISR are strongly
suppressed for the case of FSR, due to the colour singlet nature of
the parent Higgs particle and angular ordering.
\begin{figure}[t]
  \centering
  \begin{minipage}{0.49\linewidth}
    \includegraphics[width=\textwidth]{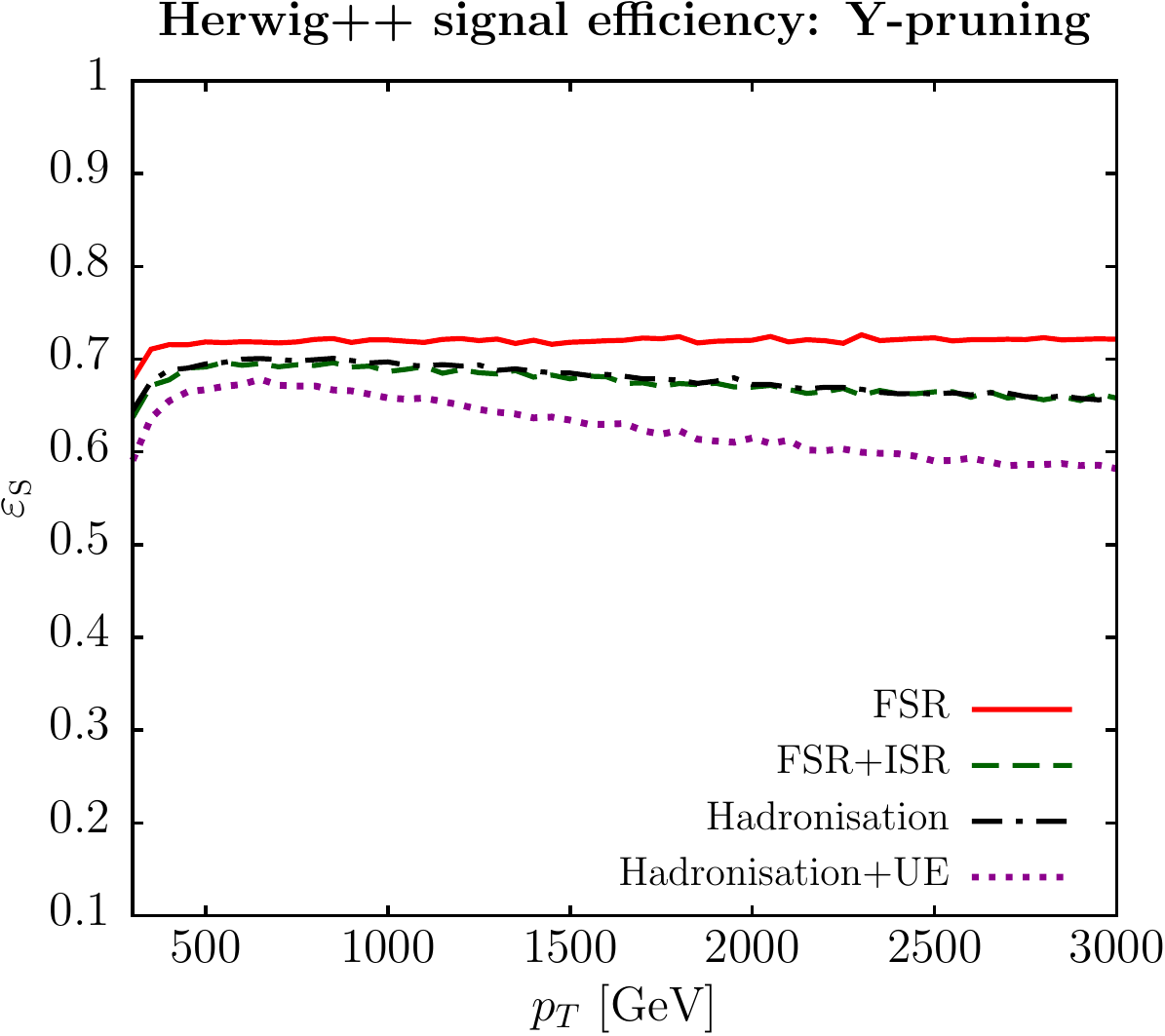}
  \end{minipage}\hfill
  \begin{minipage}{0.49\linewidth}
    \caption{An MC study of the impact of hadronisation and underlying event (UE) on the signal efficiency using the the Y-pruned jet ($\zcut=0.1$) as a function of the minimum jet transverse momentum.
    One can see negligible impact coming from hadronisation and but some degradation coming from underlying event contamination on the signal efficiency in the window $\delta M = 16$ GeV.
    Details of generation given in~\reffig{NPPlain}.
      \label{fig:NPYPruning}
    }
  \end{minipage}
\end{figure}
We conclude by showing in \reffig{NPYPruning}   an MC plot for Y-pruning
at both parton level (including both ISR and FSR) and with non-perturbative
corrections. As expected there is some significant loss of signal due
to UE contributions for precisely the same reasons as for the case of
ISR and as also observed in Ref.~\cite{Dasgupta:2013ihk}. In spite of
this deficiency it was also shown in Ref.~\cite{Dasgupta:2013ihk} that
due to its strong suppression of  background jets Y-pruning produced a signal
significance that was at least comparable and at high $p_T$ exceeded that
from the other taggers studied (mMDT, pruning and trimming),
especially for gluon jet backgrounds. In the next section we shall
study an older method, Y-splitter, that has a similar action to
Y-pruning for background jets, which makes its action on signal worth
exploring further.

\subsection{Y-splitter}
The Y-splitter technique was first introduced in Ref.~\cite{Butterworth:2002tt} in
the context of W boson tagging. The main observation was that the $k_t$
distance measure (as employed in the $k_t$ algorithm \cite{Ellis:1993tq})  between the two
partonic prongs of a $W$ decay tended to be close to the $W$ mass, which
is a consequence of a typically symmetric energy sharing between the two
prongs, in contrast to the case of QCD background where the energy
sharing is typically asymmetric. To exploit this fact one takes a fat
jet constructed with the $k_t$ algorithm and undoes the last step of
the clustering. This produces two prongs and we let them have energy
fractions $z$ and $1-z$. The $k_t$ distance $d_{ij}$ is given, at
small opening angles, by the square of the transverse momentum of the
softer prong wrt the direction of the harder prong which gives:
\begin{equation}
d_{ij} =\text{min}(z,1-z)^2 p_T^2 \theta_{ij}^2.
\end{equation}

One can either cut directly on this distance by requiring it to be of
order $M_W^2$ ($M_H^2$ in our case) or cut on the ratio of $d_{ij}$ to
the jet invariant mass squared $M_j^2=p_T^2  z(1-z) \theta_{ij}^2$ (see
e.g. Ref.~\cite{Salam:2009jx,Thaler:2008ju}). In the present case we shall choose the
latter option and hence demand that 
\begin{equation}
\label{eq:Ysplitter}
\frac{d_{ij}}{M_j^2}= \frac{\text{min}(z,1-z)}{\text{max}(z,1-z)} > \ycut.
\end{equation}

If this condition is satisfied then one tags the jet else one discards
it. Taking for instance $z <1-z$ the cut amounts to requiring that
$z>\ycut/(1+\ycut)=\ycut-\mathcal{O}\left(\ycut^2 \right)$. Likewise
for $z>1-z$ one obtains $z< 1-\ycut+\mathcal{O} \left(\ycut^2
\right)$. 

The Y-splitter method has not been as widely used in recent times as
some of the other methods we have studied here, though one relatively
recent application has been for the purposes of top tagging \cite{Brooijmans:2008}. 
Also, a detailed comparison of Y-splitter with
N-subjettiness was carried out in Ref.~\cite{Thaler:2010tr}.

Let us first consider the action of Y-splitter on QCD  background. If
one considers a quark jet with an additional soft-collinear gluon
emission, then the $\ycut$ condition is active on the gluon energy,
which means it regulates the soft divergence associated to gluon
emission. The usual double logarithmic structure of the QCD jet mass
gives way to a single logarithmic answer precisely as for the mMDT, pruning and
Y-pruning methods \cite{Dasgupta:2013ihk}:
\begin{equation}
\frac{\rho}{\sigma} \frac{d\sigma}{d\rho}^{\text{(Y-splitter,LO)}} \simeq C_F \frac{\alpha_s}{\pi} \left(\ln
  \frac{1}{\ycut}-\frac{3}{4} \right), \, \, \rho < \ycut,
  \label{eq:yspliterqcd}
\end{equation}
with $\rho =\frac{M_j^2}{p_T^2 R^2}$ and where we neglected terms
varying as powers of $\ycut$. For $\rho > \ycut$ one obtains a
transition to the normal jet mass result.  

At all orders the result for Y-splitter can be derived using methods
similar to those in Ref.~\cite{Dasgupta:2013ihk}. Since the derivation of
this result takes us away from our current focus on signals, we shall
not provide it here, but shall do so in a forthcoming paper \cite{PowDasSiod:2015}.
The basic fixed coupling result, for small $\rho$ can be expressed in the form:
\begin{equation}
\label{eq:Ysplitbkg}
\frac{\rho}{\sigma} \frac{d\sigma}{d\rho}^{\text{(Y-splitter)}} \simeq C_F\frac{
  \alpha_s}{\pi} \left(\ln \frac{1}{\ycut}-\frac{3}{4} \right) \exp
\left [-\frac{C_F \alpha_s}{2\pi} \ln^2 \frac{1}{\rho} \right],
\end{equation}
which represents a Sudakov suppression of the leading order
result. The form of this result is identical to that derived for
Y-pruning in the region $\rho < \zcut^2$ and when $\alpha_s \ln
\frac{1}{\zcut} \ln\frac{1}{\rho} \ll 1$ (see Eq.~(5.10b) of
Ref.~\cite{Dasgupta:2013ihk}), though subleading logarithmic terms will differ. 
One can verify this similarity of Y-splitter to Y-pruning, for the
case of QCD jets, by examining
the results produced by MC and we shall do so in the next subsection. 

Next we shall study the response of Y-splitter to signal jets, for our
case of Higgs decays. At zeroth order the result is similar to that
for mMDT and pruning and with neglect of order $\ycut^2$ terms one simply gets
$\eps_\text{S}^{(0)}=1-2\ycut$, which is as usual a consequence of the uniform $z$
distribution and the asymmetry cuts on $z$.

Beyond zeroth order however one should expect very significant
differences between ${\text{Y-splitter}}$ and the other taggers. These come
essentially from the response to ISR (and UE and as one can expect also
from pile-up). In order to understand the ISR response let us consider
our usual configuration of a $b \bar{b}$ pair and a large-angle soft
ISR gluon with $\theta \sim 1 \gg \theta_{b \bar{b}}$. There are two
configurations of interest. Firstly when the distance $d_{ij}$ between
the ISR gluon and both the hard prongs is larger than that of the $b
\bar{b}$ quark pair one examines  $k_t^2/M_j^2$ where $k_t$ is the
transverse momentum of the gluon wrt the jet axis, and require that
this be greater than $\ycut$, for the jet to be retained. 
Also we require that the mass window constraint is satisfied as for
the plain jet mass. 

On the other hand when the $d_{ij}$ between the gluon and the
$b\bar{b}$ pair is smaller than that between the $b$ and $\bar{b}$
prongs, the gluon is simply clustered into the jet and one just
imposes the $\ycut$ condition on the two hard prongs as at zeroth
order. The gluon thus contaminates the jet and one has to impose the
mass window constraint again as for the plain jet mass. 

In fact one can argue that the first configuration where the gluon has
a larger $k_t$ distance, along with the fact that it should not be too
energetic so as to comply with the mass window constraint, is limited
to a small corner of phase space that vanishes with $\delta M /p_T$,
where $\delta M$, as before, is the size of the window. 

To see this most straightforwardly  one notes that the gluon has a
$k_t$ distance from the $b \bar{b}$ pair (or equivalently in our soft
large-angle approximation from either the $b$ or $\bar{b}$) which is
given essentially by $d_{ij} = x^2 p_T^2 \theta^2\simeq x^2 p_T^2$
where $x=\omega/ p_T$.  On the other hand we have the $k_t$ distance between 
the $b$ and $\bar{b}$ is $\text{min}\left(z,1-z \right)^2
p_T^2 \theta_{b \bar{b}}^2$, and let us for convenience suppose that $z <1/2$.  
One requires therefore that  $x^2 p_T^2 > z^2 p_T^2 \theta_{b\bar{b}}^2$  , 
while the mass window condition again for $\theta^2 \sim 1 \gg \Delta$ gives $x
<\frac{2 M_H\delta M}{p_T^2 }$. These conditions are only simultaneously
satisfied if $z/(1-z) < \left(2 \delta M/p_T \right)^2$, which corresponds to
  a negligibly small region of phase space and given the uniform
  distribution in $z$, can be ignored.
\begin{figure}[t]
  \centering
  \begin{minipage}{0.49\linewidth}
    \includegraphics[width=\textwidth]{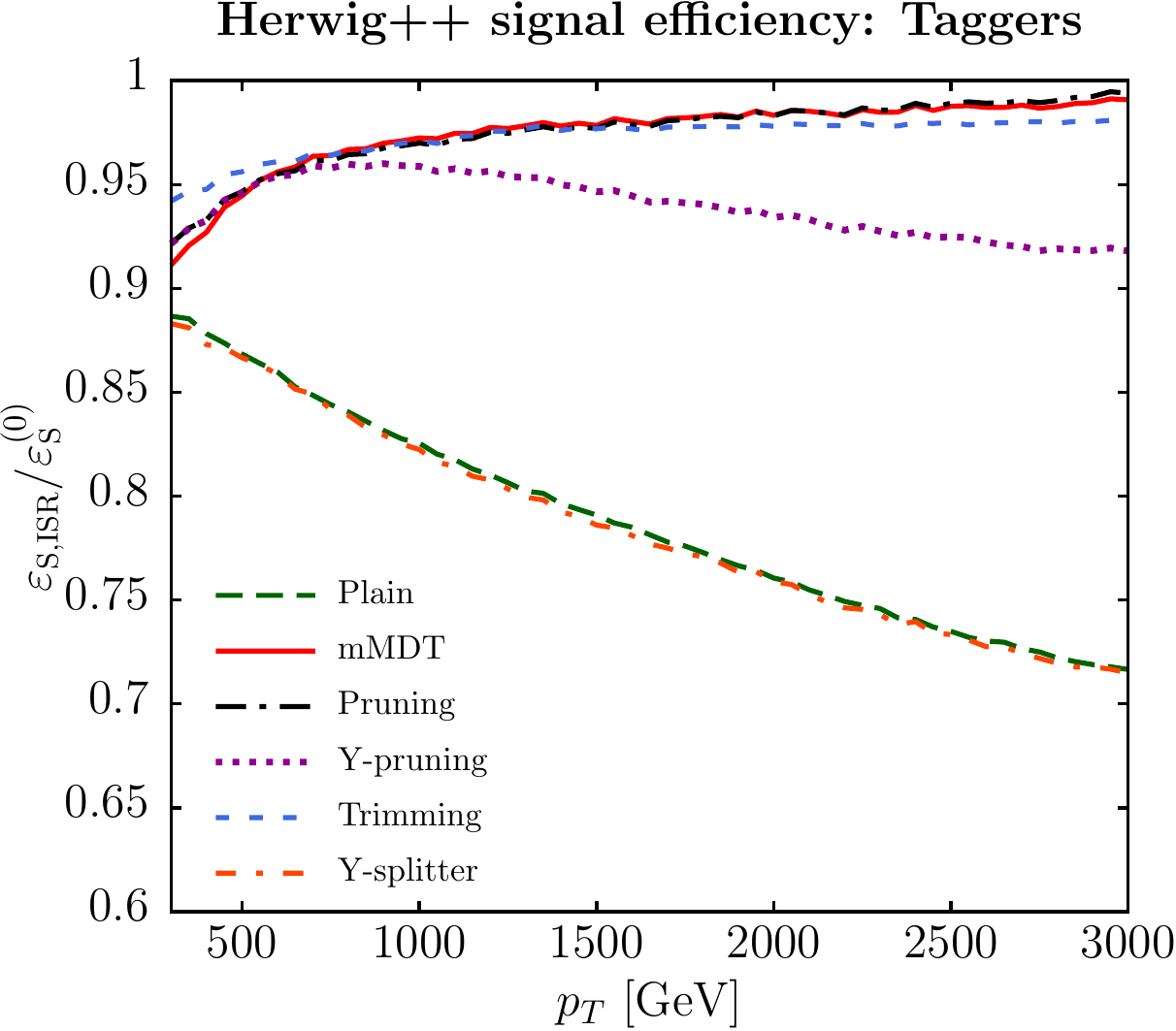}
  \end{minipage}\hfill
  \begin{minipage}{0.49\linewidth}
    \caption{A MC study of the impact of ISR on the signal
      efficiency for various taggers  ($\ycut, \zcut, \fcut=0.1, \Rtrim=0.3$) as a 
      function of the minimum jet transverse momentum.
   One can note the similarity of mMDT, pruning and trimming while
   Y-splitter and Y-pruning are different, with Y-splitter in
   particular virtually indistinguishable from the plain jet mass.
    Details of generation given in \reffig{ISRplain}.
       \label{fig:ISRall}}
  \end{minipage}
\end{figure}
Hence we are left with the situation that, modulo small corrections, the
ISR gluon contaminates the jet and gives a result that is essentially
like the plain jet mass. This implies considerable degradation in mass
due to ISR and further due to UE and, of course in the final analysis,
pile-up.

Let us compare Y-splitter with the other taggers and the
plain jet mass using MC. In Fig.~\ref{fig:ISRall} we show the response
of all taggers studied thus far to ISR, as a function of jet $p_T$.
One can immediately see that Y-splitter and the plain jet mass are
essentially identical. One also notes that mMDT, trimming and pruning
have a very similar behaviour to one another as we expected from our
analytical estimates while Y-pruning suffers at high $p_T$ as already
observed, while still remaining far better than Y-splitter.

As far as FSR is concerned, one does not expect any significant issues
with Y-splitter. In contrast to ISR a soft FSR gluon will nearly always be
clustered with the hard emitting partons, as a consequence of its
softness and angular ordering, and so end up as part of the
fat jet, thus not contributing to a loss in mass. Its effects will
cancel against soft virtual corrections leaving us to study genuinely
hard non-collinear configurations which ought to have a relatively
modest impact at the level of pure order $\alpha_s$ corrections. 

As far as NP effects are concerned we have also carried out MC studies for
Y-splitter with hadronisation and UE. The findings here are that the
effects are comparable in size to the plain jet mass.

Thus in the final analysis it  appears  that Y-splitter may not be as
useful as the other methods studied here and in particular Y-pruning,
even though it shares a very similar suppression of the QCD
background. While Y-splitter appears effective
at identifying hard substructure and removing background, it is not
effective in grooming away soft contamination, as is inbuilt to
varying extents in the other methods we have studied. This suggests using Y-splitter
along with another method more effective in grooming may  alleviate
some of the issues we see with the
signal. Therefore in the next subsection we shall consider its
combination with trimming, which we find has some noteworthy
features and produces interesting results. 

\section{Y-splitter with trimming}
Here we shall study the combination of Y-splitter with trimming, in
view of the lack of any effective grooming element in Y-splitter, as
mentioned above. We do not have to necessarily choose trimming in this
respect and it is possible to study  a combination of Y-splitter with
other methods such as mMDT and the recently introduced soft drop method
\cite{Larkoski:2014wba}. Indeed it has been known for some time that
combinations of substructure tools can often produce better results
than the individual tools themselves \cite{Soper:2010xk} and thus one
may hope to improve the performance of Y-splitter using a suitable
complementary tool.

We first study the impact of applying trimming on signal jets that are tagged
by Y-splitter. For the combination of Y-splitter with trimming we
choose $\fc=\yc=0.1$ and $\Rtrim=0.3$.

\begin{figure}[t]
  \centering
  \begin{minipage}{0.49\linewidth}
    \includegraphics[width=\textwidth]{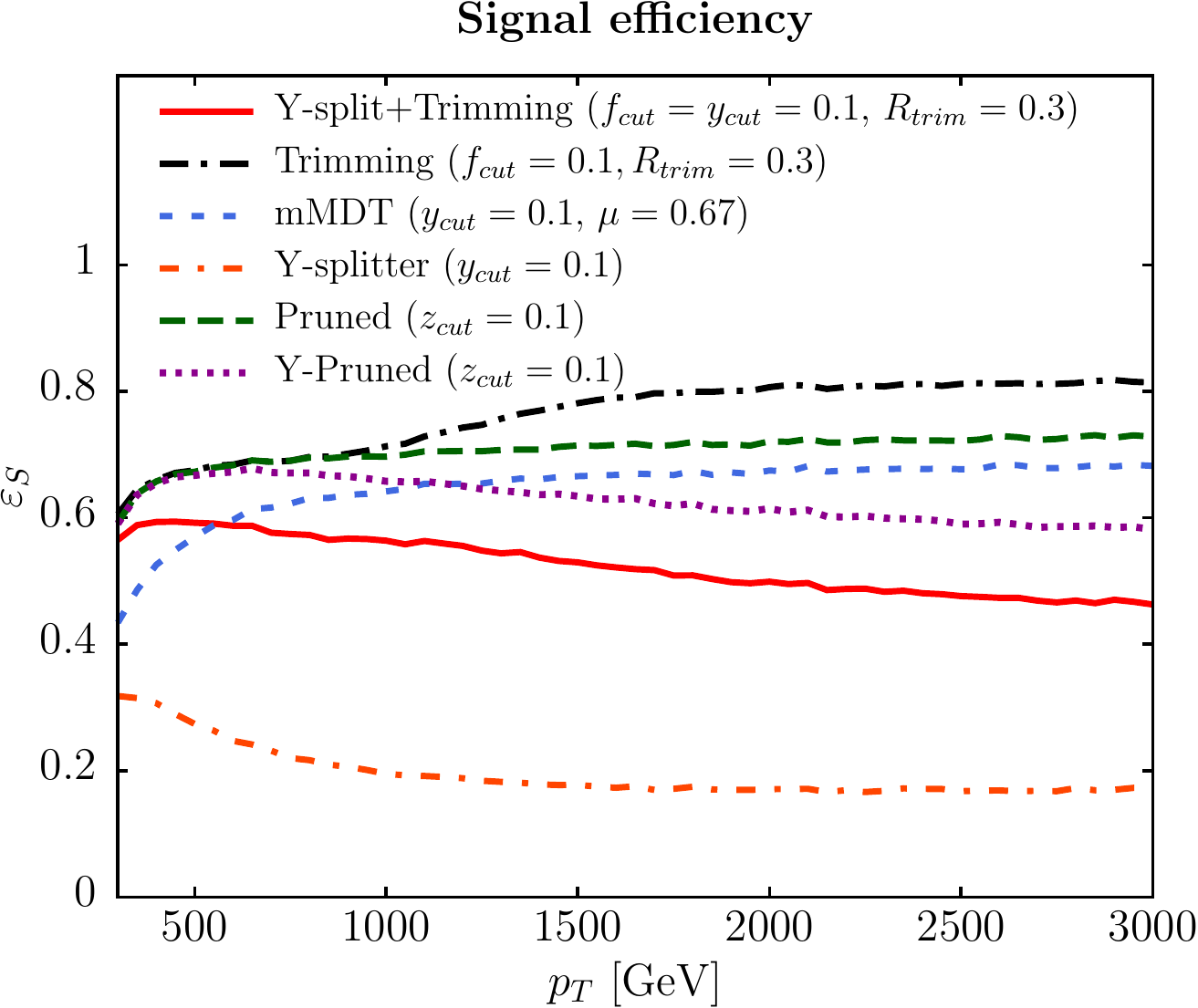}
  \end{minipage}\hfill
  \begin{minipage}{0.49\linewidth}
    \caption{Signal efficiency  for tagging hadronic $H$ jets using \herwig 
      \cite{Bahr:2008pv} with
      underlying event and hadronisation as a function of a generator
      level cut $p_T$ on transverse jet momentum.
     \label{fig:efficiency} }
  \end{minipage}
\end{figure}

The MC analysis is shown in \reffig{efficiency}
which demonstrates that the use of trimming substantially ameliorates
the loss of signal we saw with Y-splitter. It is evident from
the same figure that while Y-splitter with trimming still does not
reach the signal efficiency of some other methods, the difference is much less
pronounced than before. In fact one observes that the signal efficiency for
Y-splitter with trimming bears a qualitative similarity to
Y-pruning. The reason for this is that the use of trimming turns the
plain mass like behaviour of Y-splitter into the behaviour for trimming, except for
configurations that have been rejected by Y-splitter, on which
subsequent trimming does not act. This corresponds to the Y-splitter rejection region
whereby an ISR gluon has the largest $k_t$ distance but fails the $\ycut$
requirement. This kinematic configuration is reminiscent of that which resulted in
the extra $\Delta \eps_\text{S}$ term for Y-pruning. In the present case uncancelled
virtual corrections integrated over the Y-splitter rejection region
produce a term $\sim -C_F \frac{\alpha_s}{\pi} R^2 \ln
\frac{\yc}{\sqrt{\Delta}}$, which corrects the simple trimming
  result. However the overall efficiency remains considerably higher than
  the Y-splitter or plain mass result, due to elimination of the plain
  jet mass like logarithm.

\begin{figure}[t]
  \centering
  \begin{minipage}{0.49\linewidth}
    \includegraphics[width=\textwidth]{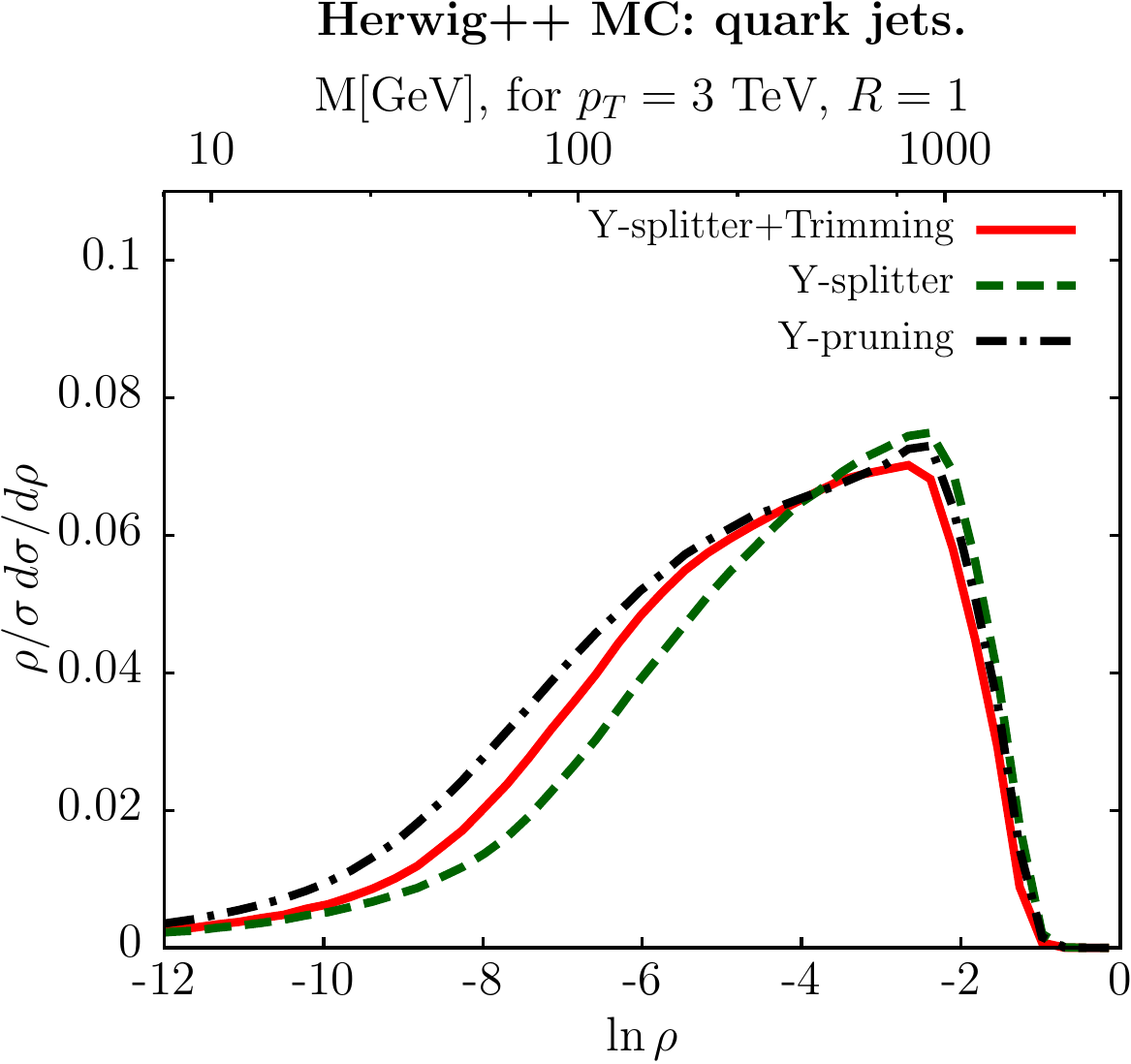}
  \end{minipage}\hfill
  \begin{minipage}{0.49\linewidth}
    \caption{MC results for the differential jet mass distribution using Y-splitter with trimming $\bra{\yc=0.1,\Rtrim=0.3}$, Y-splitter $\bra{\yc=0.1}$ and Y-pruning $\bra{\zcut=0.1}$.
This result has been generated using \herwig \cite{Bahr:2008pv} at parton level.
A minimum $p_T$ cut on generation of the hard process $qq\rightarrow qq$ was made at 3 TeV for 
14 TeV pp collisions.
We take the mass of the two hardest jets found using the C/A algorithm with $R=1.0$ and apply 
the algorithms with a fixed $\zcut/\ycut$.
  \label{fig:Y2TvsPrune_MC}
    }
\end{minipage}
\end{figure}

Next one should study also the impact of using trimming in conjunction
with Y-splitter on the QCD background. In this article, given our
focus on signal jets, we shall not attempt to provide a detailed
analytical study of the background case, which together with the
derivation of the basic Y-splitter formula Eq.~\eqref{eq:Ysplitbkg} we
shall carry out in a forthcoming article \cite{PowDasSiod:2015}. Here
we shall confine ourselves to MC studies and the result of these is
shown in \reffig{Y2TvsPrune_MC}. Once again we study the action of
trimming on jets that have been tagged by Y-splitter where we choose
$\fcut=\ycut=0.1$ for both Y-splitter and trimming and $\Rtrim=0.3$.
In this plot, we remind the reader that $\rho =\frac{M_j^2}{p_T^2  R^2}$ i.e. the
normalised jet mass as defined and used in Ref.~\cite{Dasgupta:2013ihk}.

One notes from \reffig{Y2TvsPrune_MC}  that Y-pruning, Y-splitter and
Y-splitter+trimming all have a fairly similar action on background jets and
provide, for our choice of parameters, a significant suppression of background
around the signal mass-peak $\sim 100$ GeV. These results are for quark backgrounds
but similar results are obtained for gluon jets. It is noteworthy that
the action of trimming for the chosen parameters, appears only to have
an apparently subleading effect and hence the desirable property of
Y-splitter, that of reducing background via a Sudakov suppression term  (see \refeq{Ysplitbkg}), is largely unaffected.
Such findings are certainly worthy of analytical follow-up for general
choices of parameters, which we shall provide in our forthcoming work. 

\begin{figure}[t]
  \centering
  \begin{minipage}{0.49\linewidth}
    \includegraphics[width=\textwidth]{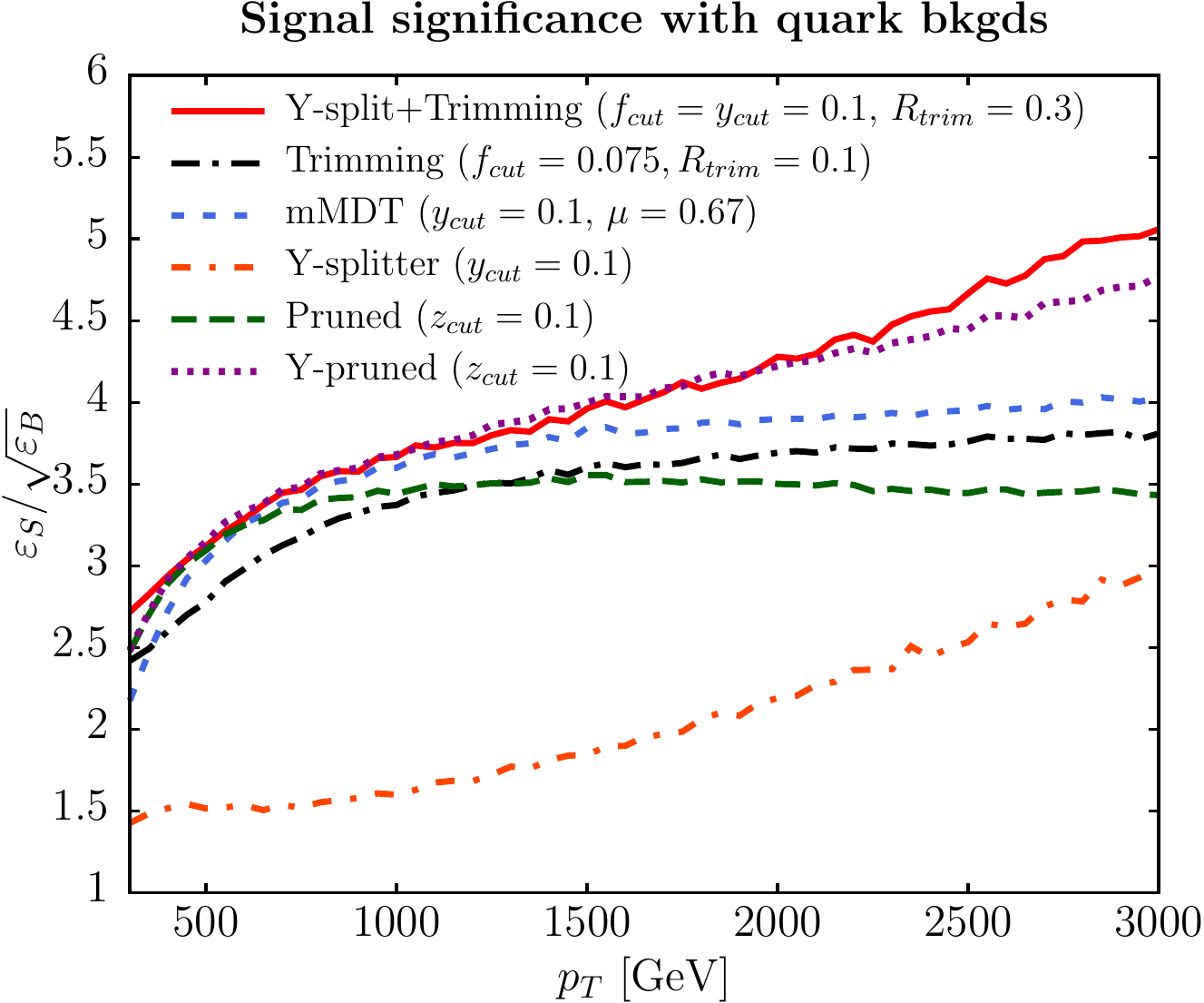}
  \end{minipage}\hfill
  \begin{minipage}{0.49\linewidth}
  \includegraphics[width=\textwidth]{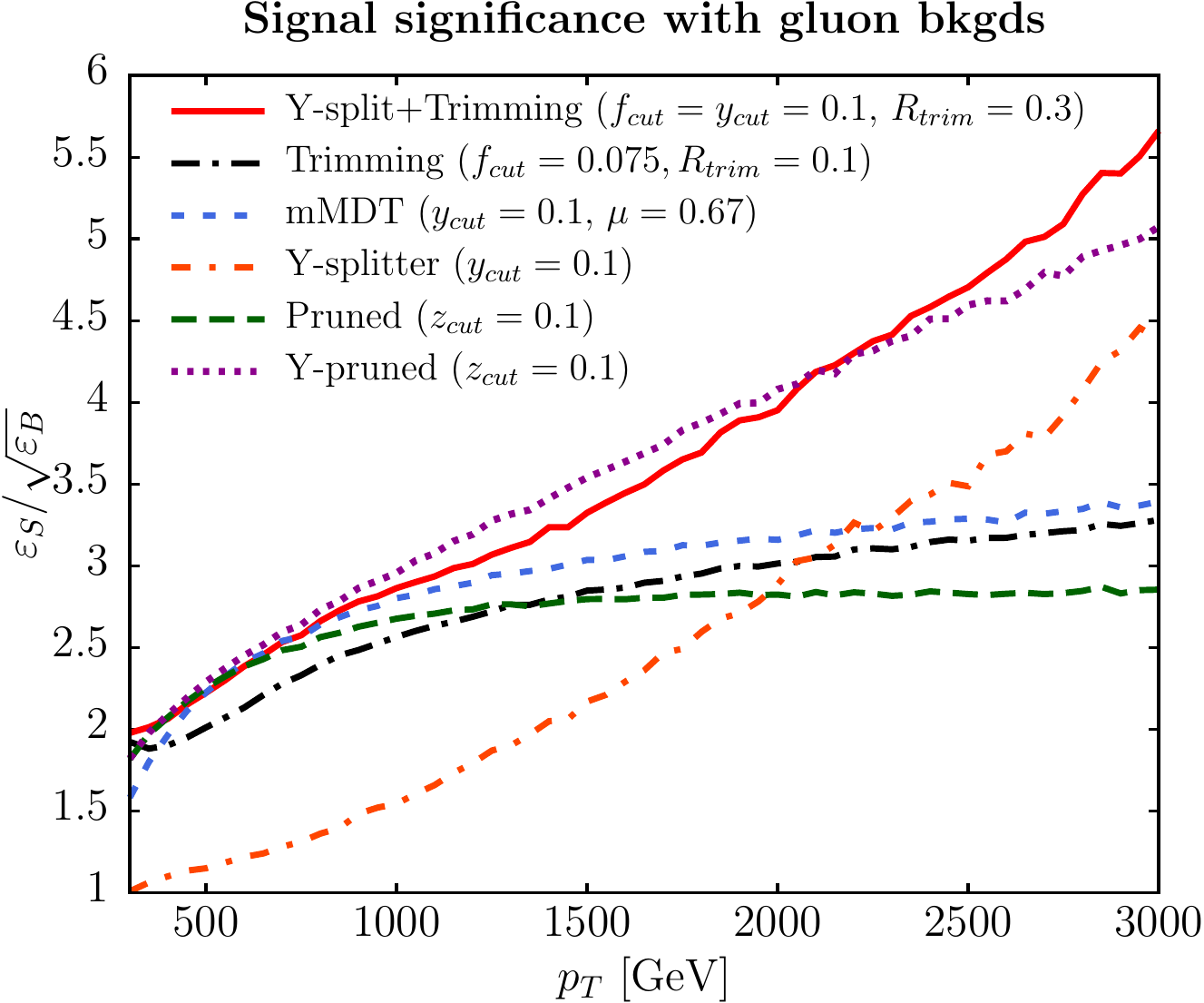}
  \end{minipage}
    \caption{Signal significance for tagging hadronic H jets with quark (left panel) and gluon (right panel) backgrounds using \herwig
   \cite{Bahr:2008pv} with underlying event and hadronisation as a function of a generator level cut $p_T$ on transverse jet momentum. 
    We compare the signal significance for different algorithms to Y-splitter+trimming and find that the latter outperforms the others at 
    high $p_T$. Here we have used $R_{\mathrm{trim}}=0.1$ for
    pure trimming since, in contrast to Y-splitter+trimming, we expect
    this value to be closer to optimal than $R_{\mathrm{trim}}=0.3$ at high transverse momenta (see Fig.~18).
\label{fig:SS_YSTrim}
}
\end{figure}

Given the improvement in signal efficiency that we have achieved with
Y-splitter with trimming, and the fact that the backgrounds are comparably
(and in fact apparently somewhat more) suppressed compared to Y-pruning in the mass
region of interest, it is worth examining the signal significances
(i.e. the ratio $\eps_S/\sqrt{\eps_B}$ of signal efficiency to the square
root of background efficiency) that can be achieved with the various
taggers, as a function of transverse momentum. These are shown in \reffig{SS_YSTrim} 
for quark and gluon backgrounds. One observes that the Y-splitter with
trimming method outperforms the taggers discussed here, particularly at
high $p_T$. Note that we have used $R_{\mathrm{trim}}=0.3$ for
Y-splitter with trimming. For trimming however this represents a
non-optimal choice at high $p_T$ (see \reffig{Trim2D} later) and hence we have
chosen to present our results for trimming with
$R_{\mathrm{trim}}=0.1$. A detailed study of optimal parameters for Y-splitter+trimming remains to be carried out and we shall aim to present the results of such a study in forthcoming work.

\begin{figure}[t]
  \centering
  \begin{minipage}{0.49\linewidth}
    \includegraphics[width=\textwidth]{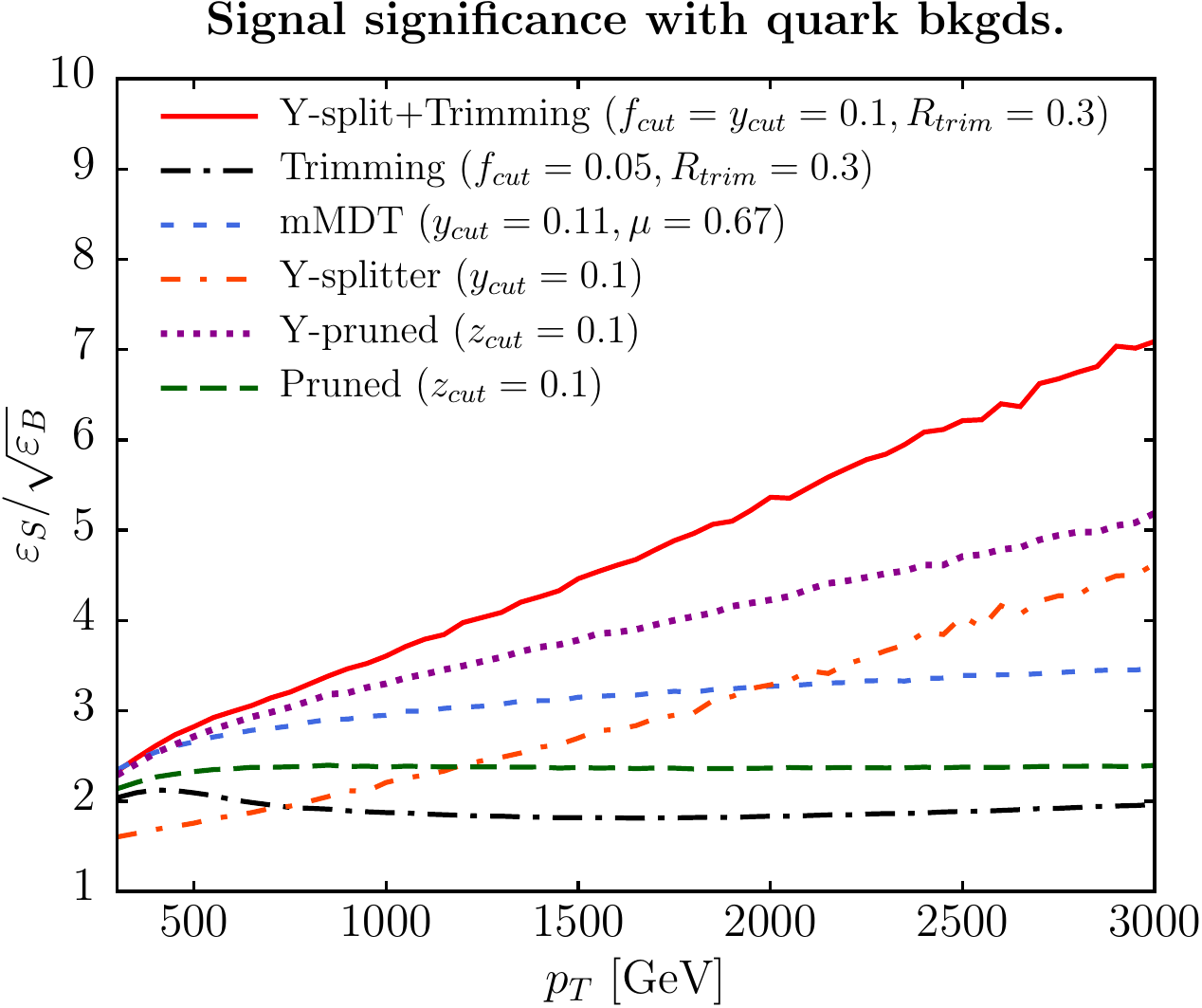}
  \end{minipage}\hfill
  \begin{minipage}{0.49\linewidth}
  \includegraphics[width=\textwidth]{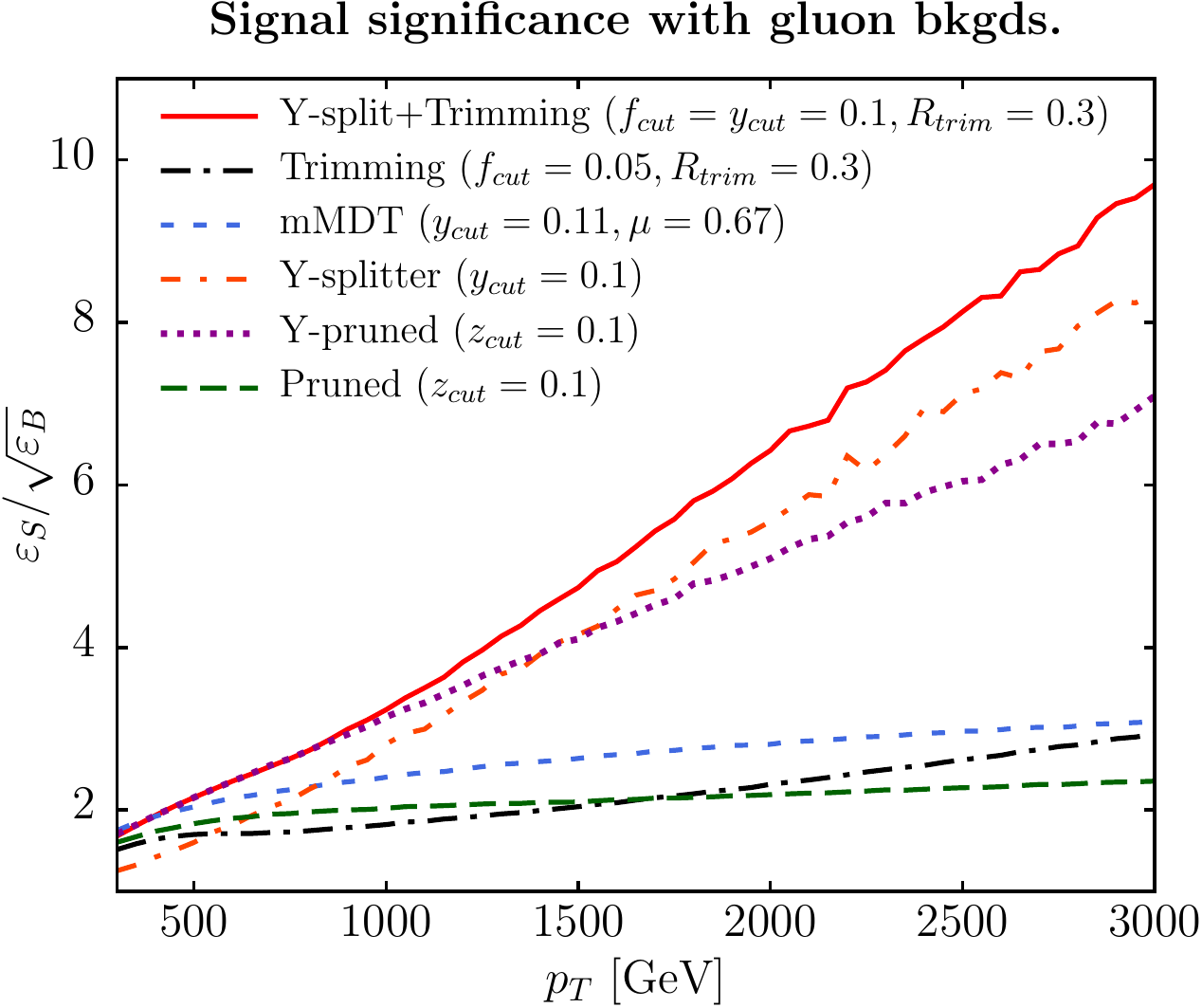}
  \end{minipage}
    \caption{Signal significance for tagging hadronic W jets with quark (left panel) and gluon (right panel) backgrounds using \herwig
   \cite{Bahr:2008pv} with underlying event and hadronisation as a function of a generator level cut $p_T$ on transverse jet momentum.  We deem a jet tagged if it has a final 
  mass in the window $64-96$ GeV.
    We compare the signal significance for different algorithms to Y-splitter+trimming and find that the latter outperforms the others at 
    high $p_T$.  In this plot, we use all tagger parameters which match those used
    for W tagging in the paper \cite{Dasgupta:2013ihk} for ease of
    comparison. 
\label{fig:SS_YSTrimWZ}
}
\end{figure}

The results shown in \reffig{SS_YSTrim}  are for our standard process, $pp\to ZH$, 
but similar results are also obtained for $W$ tagging as shown in \reffig{SS_YSTrimWZ}.
Here we observe that Y-splitter with trimming now consistently
outperforms the other taggers discussed over a range of $p_T$. This emerges from the different mass window of the W boson ($64-96$ GeV) compared to the Higgs ($109-141$ GeV). In the window $M_W\pm16$ GeV, the background mass distribution of Y-splitter+trimming is smaller relative to Y-pruning than the window around $M_H$ (see \reffig{Y2TvsPrune_MC}).
Hence, we observe a greater signal significance tagging $W$ rather than Higgs relative to the other taggers for large $p_T$. \footnote{We have performed preliminary studies for other possible combinations such as Y-splitter+mMDT/pruning/soft drop. These all have a similar qualitative effect on both the background and signal jet mass distribution as Y-splitter+trimming. Hence, one observes a comparable gain in signal significance over Y-splitter for all of these combinations. However, we find that Y-splitter+trimming has the best signal significance for tagging W bosons over background in the high $p_T$ limit.}

\section{Optimal parameter values} \label{sec:optimalvalues}
In this section we shall use analytical expressions to derive values of
parameters that maximise the signal significance
$\frac{\eps_S}{\sqrt{\eps_B}}$ for the different taggers. We
do not expect the values so derived to really be optimal in the sense
that they will not take into account non-perturbative
effects. Indeed we should emphasise that optimal parameter values have already been extracted using full MC studies for
all methods considered in the original papers and also examined in subsequent
studies such as in Ref.~\cite{Abdesselam:2010pt}. Analytical studies
of optimal parameters have also been carried out by Rubin in Ref.~\cite{Rubin:2010fc} in the context of a
filtering analysis, which we do not consider here.

 Nevertheless we can regard it as one of the tests of the robustness of
these methods that the values derived here with analytical formulae as inputs
should be reasonable approximations to what one obtains in complete MC
studies. This is because one wants ideally to have substructure methods where
statements about performance are largely independent of our detailed
knowledge about non-perturbative corrections. We are also interested
in examining to what extent general trends that emerge with analytics, such
as the dependence of optimal parameters on $p_T$, are replicated in
full MC studies. For the following studies we confine ourselves to quark
backgrounds as we have no reason to believe that gluon backgrounds
will differ significantly in terms of the conclusions we reach here.

Having observed in this paper the relatively small radiative
corrections, both for  ISR and FSR, that emerge for signal processes
over a broad range of parameter values, one feels encouraged in a first
approximation to turn off these effects and treat the signal in a
tree-level approximation, except for the case of trimming as we
discuss below in more detail. In other words we anticipate that the signal
significance ought to primarily be driven by the tree-level results
for signal while for the background we shall use the resummed formulae
first derived in \cite{Dasgupta:2013ihk}. For self-consistency, one
should then also verify that for the optimal values one derives, the radiative
corrections to signal efficiency can indeed be considered small relative to the tree level result.

\subsection{mMDT}
Let us follow the above described procedure for the mMDT and extract
the optimal value of $\ycut$.

One needs to study the background
mistag rate in the window $M_H-\delta M< M_j<M_H+\delta M$ which corresponds to a range in $\rho$, $\rho_H-\delta \rho
<\rho<\rho_H+\delta \rho$, with $\delta \rho \approx 2 M_H \delta M/p_T^2$ where we have used $R=1$.

We then have the following expression for the signal significance:
\begin{equation}
\label{eq:sigsig}
\frac{\eps_S}{\sqrt{\eps_B}} = \frac{1-2 y_\mathrm{cut}}{\sqrt{\Sigma \left(\ycut,\rho_H+\delta\rho \right)-\Sigma\left(\ycut,\rho_H-\delta\rho \right)}},
\end{equation}
with $\rho_H=\frac{M_H^2}{p_T^2}$ and where $\Sigma(\ycut,\rho)$ is the
integrated mMDT background jet mass distribution calculated in
Ref.~\cite{Dasgupta:2013ihk}. Thus the quantity $\Sigma
\left(\ycut,\rho_H+\delta\rho \right)-\Sigma\left(\ycut,\rho_H-\delta\rho \right)$ represents the
integral of the background jet-mass distribution over the mass window
corresponding to signal tagging with mMDT. Note that we have treated
the signal efficiency at lowest order.

We can find the value of $\ycut$ that maximises signal significance by
taking the derivative of the RHS of Eq.~\eqref{eq:sigsig} wrt $\ycut$ and setting it to zero which gives:
\begin{equation}
\frac{-4}{1-2 \ycut} = \frac{\Sigma'\left(\ycut,\rho_H+\delta\rho
  \right)-\Sigma' \left(\ycut,\rho_H-\delta\rho
  \right)}{\Sigma\left(\ycut,\rho_H+\delta\rho
  \right)-\Sigma\left(\ycut,\rho_H-\delta\rho \right)},
\end{equation}
where $\Sigma'$ denotes a derivative wrt $\ycut$.
Neglecting higher order corrections in $\delta\rho$, the optimal value
for $\ycut$ satisfies
\begin{equation}
\label{eq:effmax}
\frac{-4}{1-2 y_{\mathrm{cut}}}  = \frac{\frac{d}{d\yc} \left(\frac{\partial \Sigma}{\partial 
\rho} \right)_{\rho=\rho_H}}{\left (\frac{\partial \Sigma}{\partial \rho} \right)_{\rho=\rho_H}}.
\end{equation}
Next we use the analytical expressions for $\Sigma \left(\ycut,\rho
\right)$ derived in Ref.~\cite{Dasgupta:2013ihk}. The fixed-coupling
result for the mMDT for $\rho < \ycut$ reads:

\begin{equation}
\Sigma(\ycut,\rho) = \exp \left[-\frac{C_F \alpha_s}{\pi} \left ( \ln
\frac{1}{\ycut} \ln \frac{\ycut}{\rho} -\frac{3}{4} \ln\frac{1}{\rho}+\frac{1}{2} \ln^2 \frac{1}{\ycut}\right) \right].
\end{equation}

We can use this result in Eq.~\eqref{eq:effmax}, assuming that the
optimal value lies in the region $\rho < \ycut$, \footnote{This is reasonable
at high $p_T$, since $\rho_H (1 \text{TeV} )\approx 0.015$ much
smaller than typical $\ycut$ values $\sim 0.1$ quoted in the literature.} and doing
so gives us an implicit equation for optimal $\ycut=y_{\text{max}}$:
\begin{equation}
\label{eq:maxima}
\frac{-4 y_{\mathrm{max}}}{1-2 y_{\mathrm{max}}} = C_F \frac{\alpha_s}{\pi} \ln \frac{y_{\mathrm{max}}}{\rho_H}+\frac{4}{3+4 \ln y_{\mathrm{max}}}.
\end{equation}

\begin{figure}[ht]
  \includegraphics[width=0.49 \textwidth]{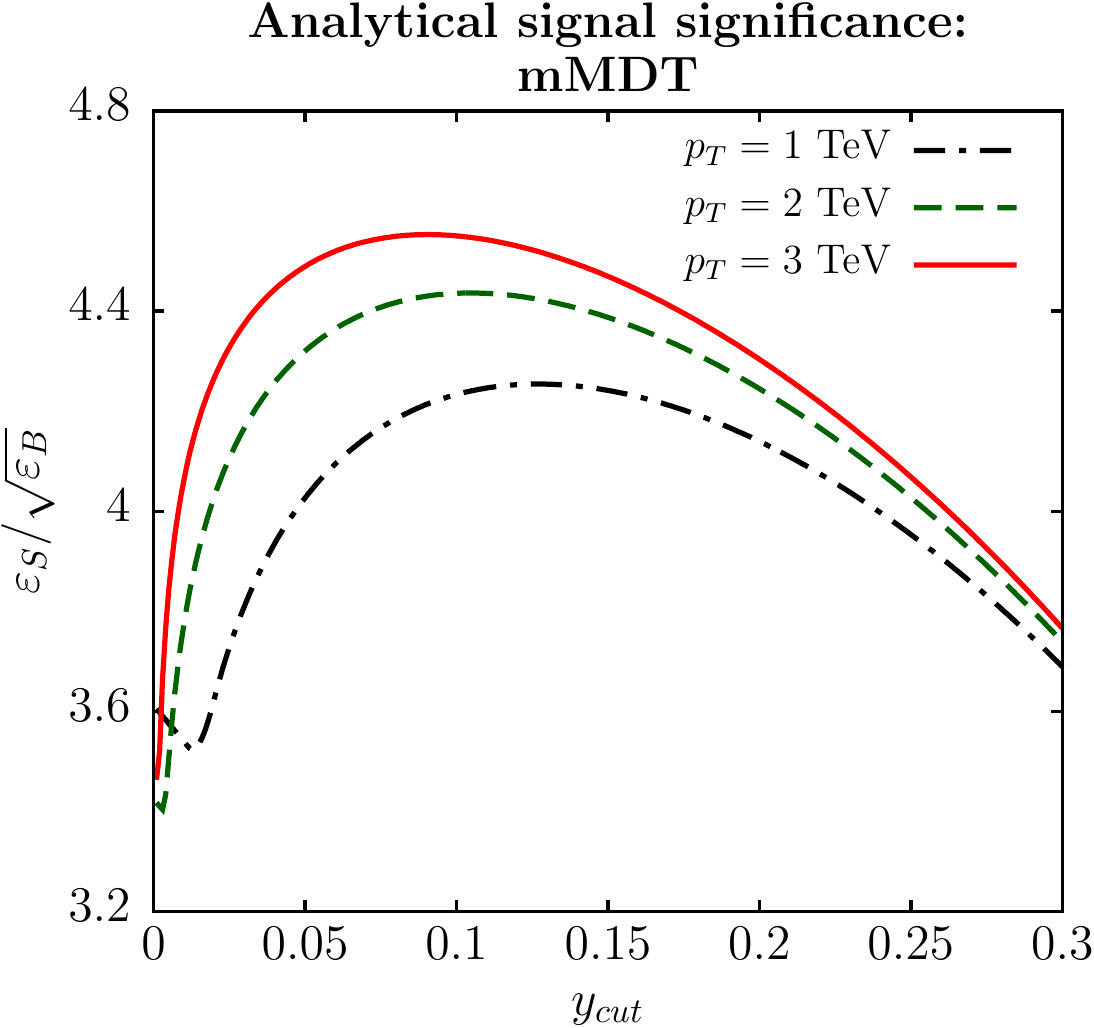}
  \includegraphics[width=0.49 \textwidth]{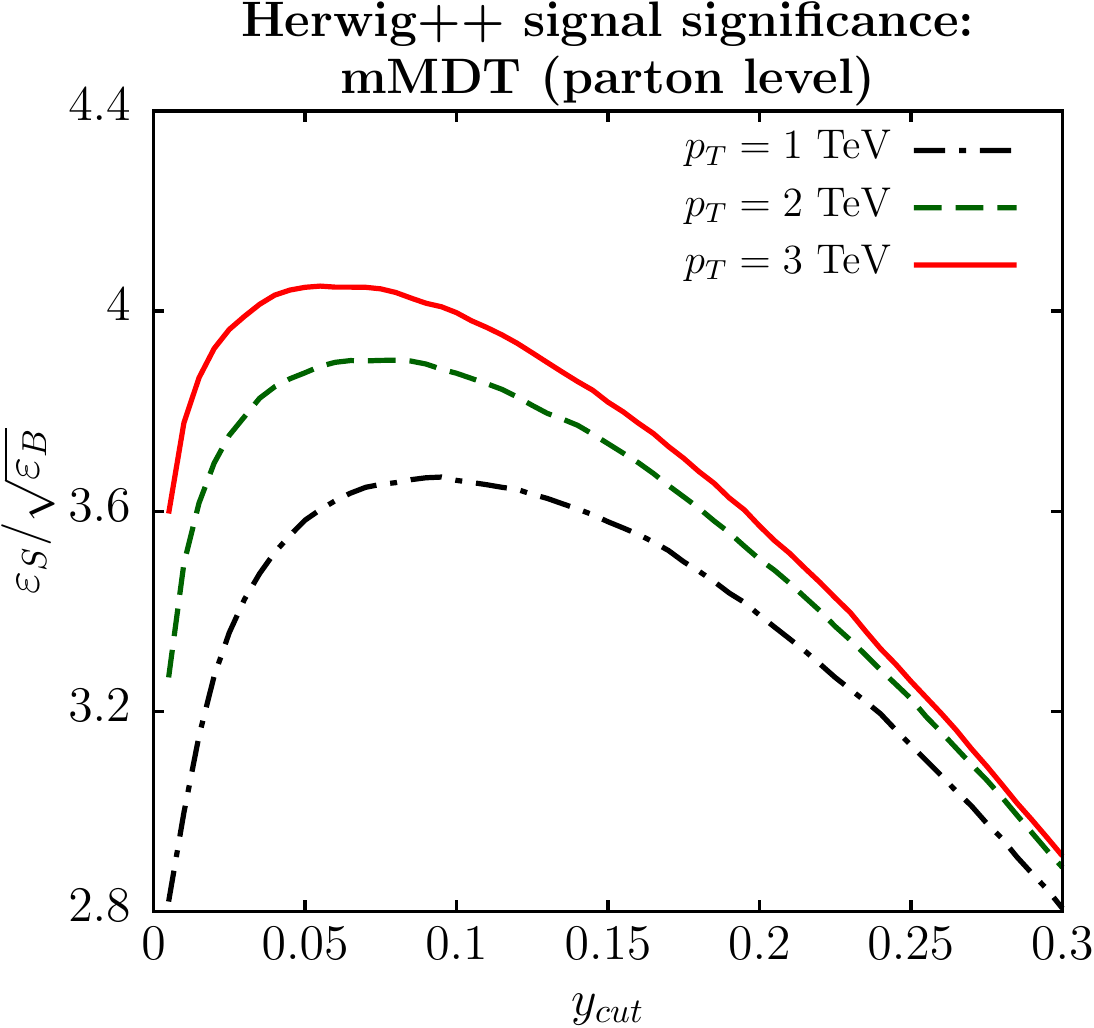}\\\\
  \includegraphics[width=0.49\textwidth]{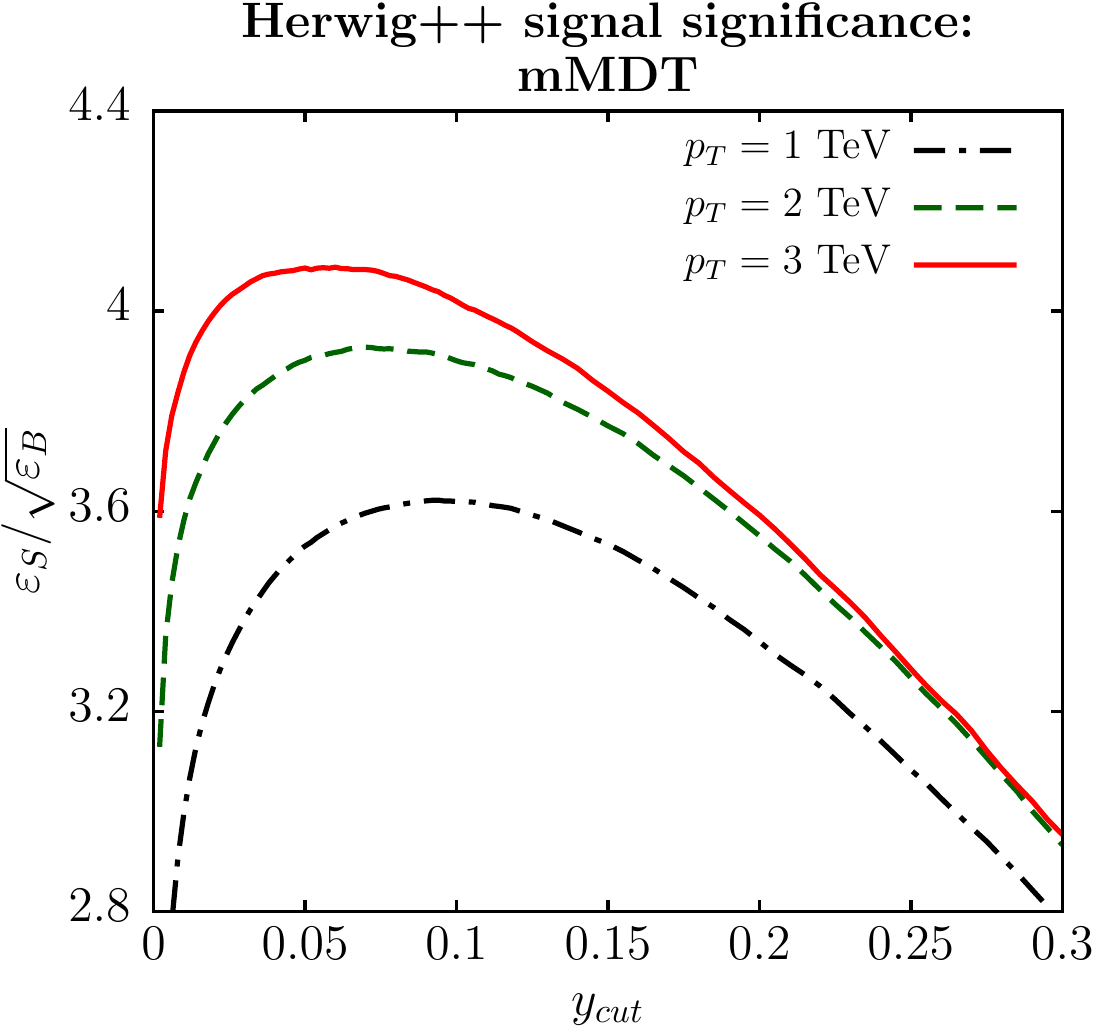}
  \includegraphics[width=0.474 \textwidth]{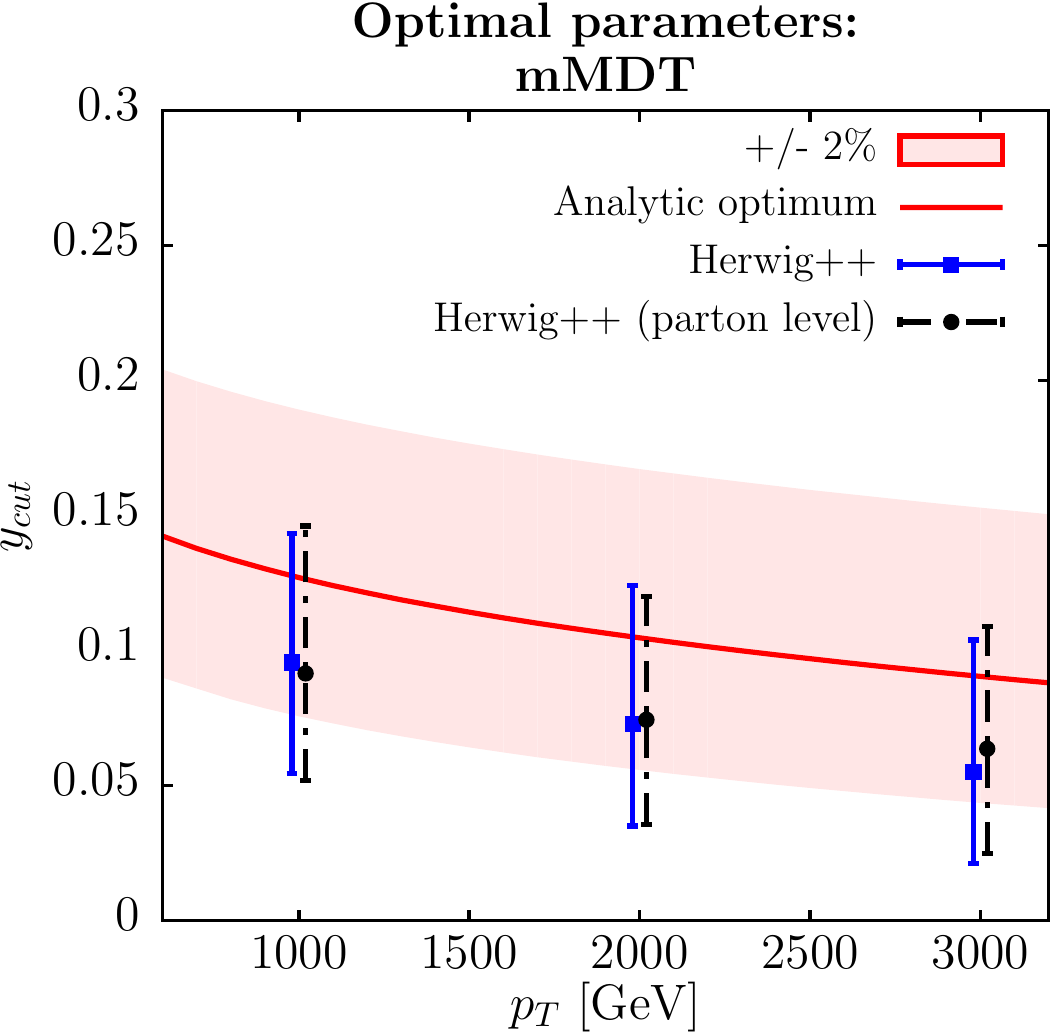}
    \caption{mMDT analytical signal significance from tree level signal and resummed background as a function of $\ycut$ (top left) compared to \herwig \cite{Bahr:2008pv} 
    at parton level (top right) and with hadronisation and MPI (bottom left). The signal process used is $pp \to ZH$ where we require the Higgs and Z to decay hadronically 
    and leptonically respectively with quark backgrounds. We place a generator level cut on the Higgs transverse momentum $p_{T}$ of 1, 2 and 3 TeV. Jets are tagged around the Higgs mass with a mass window $\delta M=16$ GeV.
    The bottom right panel shows the analytic optimal $\ycut$ values as a function of $p_T$ (red line) with a $2\%$ variation in signal significance about the peak (red shaded area).
    We overlay the optimal results for $\ycut$ obtained using \herwig with hadronisation and underlying event at 1, 2 and 3 TeV, 
    with an equivalent $2\%$ variation about the peak signal
    significance (blue bars) and at parton level (black bars).
      \label{fig:SS_MMDT}
    }
\end{figure}

One can numerically solve the above equation, which contains the
essential information about the optimal $\ycut$ and its dependence on
$p_T$. The values we obtain for $p_T = 1,2,3$ TeV with $\alpha_s=0.1$ are approximately
  $0.124$, $0.102$ and $0.088$ respectively. Whilst we have not
  included running coupling effects in the above derivation, one finds it is straightforward to do
  so. Using the full calculation of  Ref.~\cite{Dasgupta:2013ihk} for
  the background, i.e. including running coupling effects and a
  transition to the plain mass like behaviour for $\rho > \ycut$,  we
  compute the analytical signal significance plotted in
  Fig.~\ref{fig:SS_MMDT} as a function of $\ycut$. 

From Fig.~\ref{fig:SS_MMDT} we note firstly that the peak position of the
analytical signal significance is approximately in agreement with the
numbers we quoted immediately above for the fixed-coupling
calculation. A kink can be seen in the analytical result for $p_T=1$
TeV, the origin of which is the transition from a single-logarithmic
dependence on $\rho$ valid at $\rho < \ycut$, to the usual double
logarithmic result for the plain mass for $\rho>\ycut$.
We have also shown, in the same figure, results from
\herwig at both parton level and at full hadron level including UE. We
find the \herwig results at parton level in quite reasonable agreement with
the simple analytical estimates we have made, for both the peak
positions and the evolution of optimal $\ycut$ with $p_T$, though the
values of the peak signal significance itself differ somewhat. It is noteworthy also that hadronisation and UE do not
change the picture significantly  at the $p_T$ values we have studied
here. One other feature that emerges from both analytical and MC
studies is that the peaks themselves are fairly broad so that choosing
a slightly non-optimal $\ycut$ does not greatly impact the tagger
performance. We have also provided in Fig.~\ref{fig:SS_MMDT} a
direct comparison between optimal values from \herwig (including all
effects) and analytical estimates. We show the results for the range
of $\ycut$ values (denoted by the pink shaded region) that correspond
to a $\pm 2 \%$ variation around the peak signal
significance. For \herwig instead we indicate the same range of $\ycut$
values by the blue bars shown. We find a good degree of overlap within
this tolerance band between full \herwig results and analytical
estimates. 

One can draw at least a couple of inferences from our observations above. Firstly, as
we have argued, radiative corrections to the signal are clearly of minor
significance to the tagger performance for mMDT. The fact that the
analytics are generally in good agreement with \herwig points to the
importance of the background contribution in the context of the signal
significance and the success of analytical approaches in describing
this background \cite{Dasgupta:2013ihk}. The fact that non-perturbative
effects play an evidently minor role at the values of $p_T$ studied above is also
reassuring from the point of view of a robust understanding of tagger
performance. 

We end with a caveat. If one moves to still lower $p_T$ values then one has to reconsider some of the arguments above. Here
one would have a situation where say at 200-300 GeV we expect to be in
the region where $\rho> \ycut$ and so Eq.~\eqref{eq:maxima} does not
directly apply. This apart, perhaps more significantly one can expect
UE to start playing a larger role due to the larger effective radius
$\sim \frac{M_j}{p_T}$ where UE particles accumulate without being
removed by the asymmetry cut. Here one ought to consider the use of mMDT
with filtering and optimise the parameters of both methods together as in the
original analysis \cite{Butterworth:2008iy}. 

\subsection{Pruning and Y-pruning}
\begin{figure}[ht]
  \includegraphics[width=0.49 \textwidth]{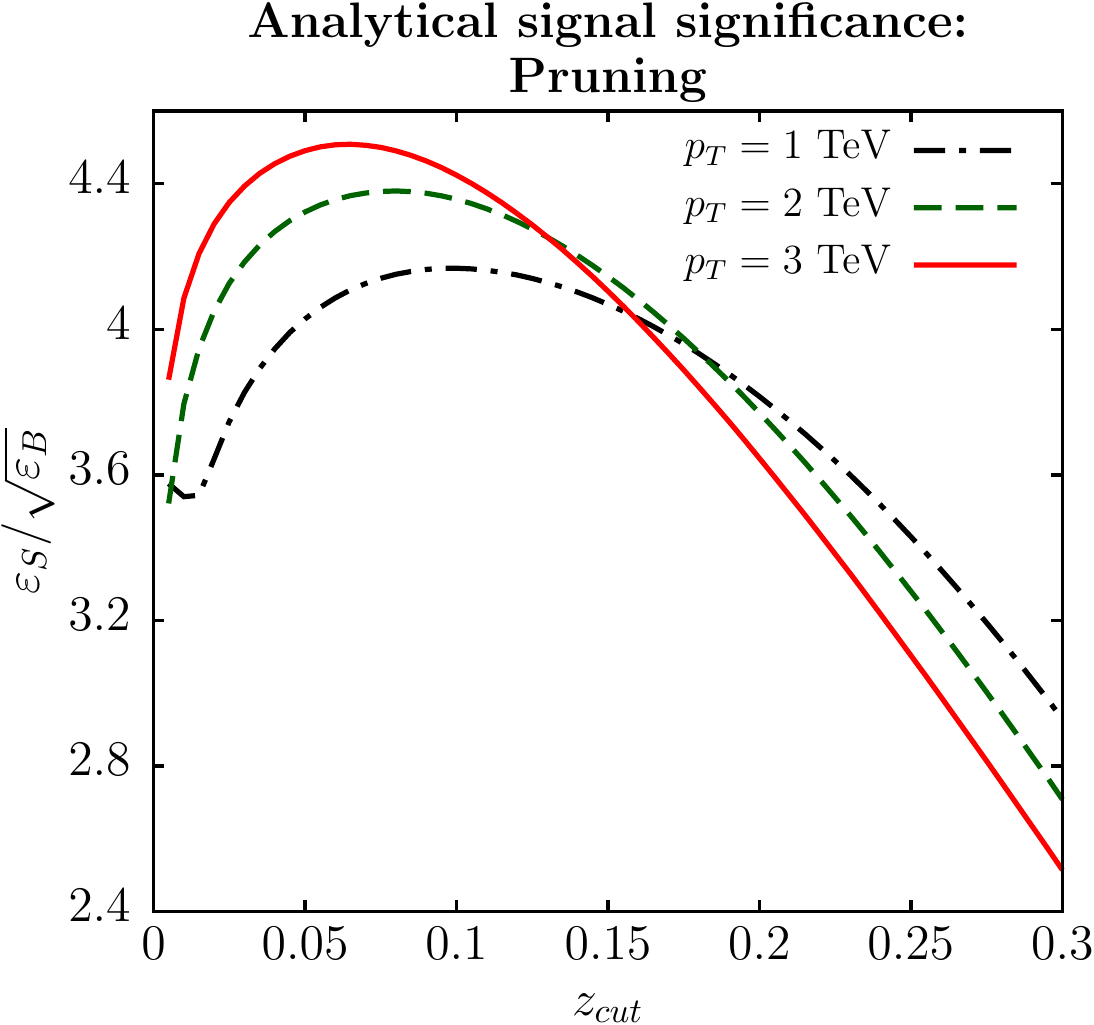}
  \includegraphics[width=0.49 \textwidth]{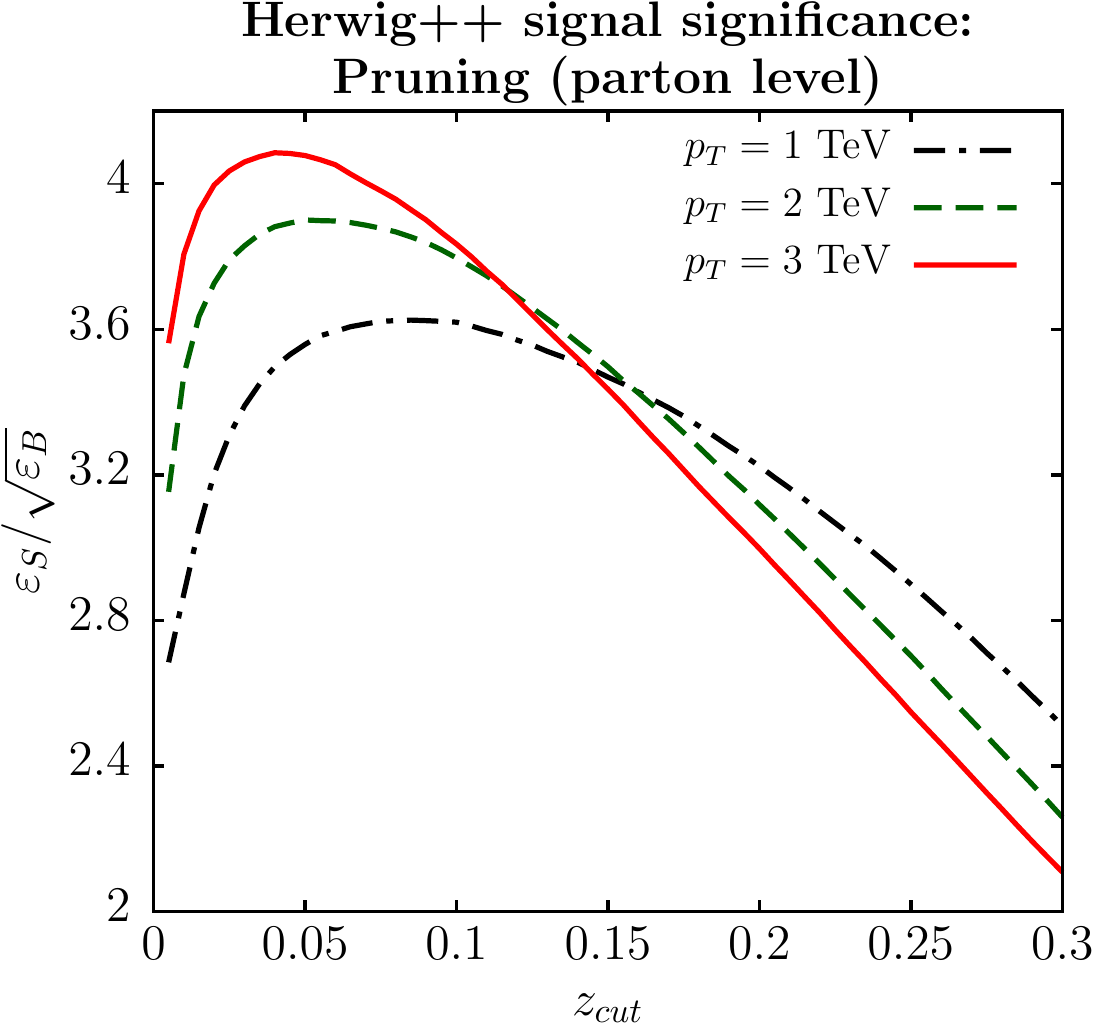}\\\\
    \includegraphics[width=0.49\textwidth]{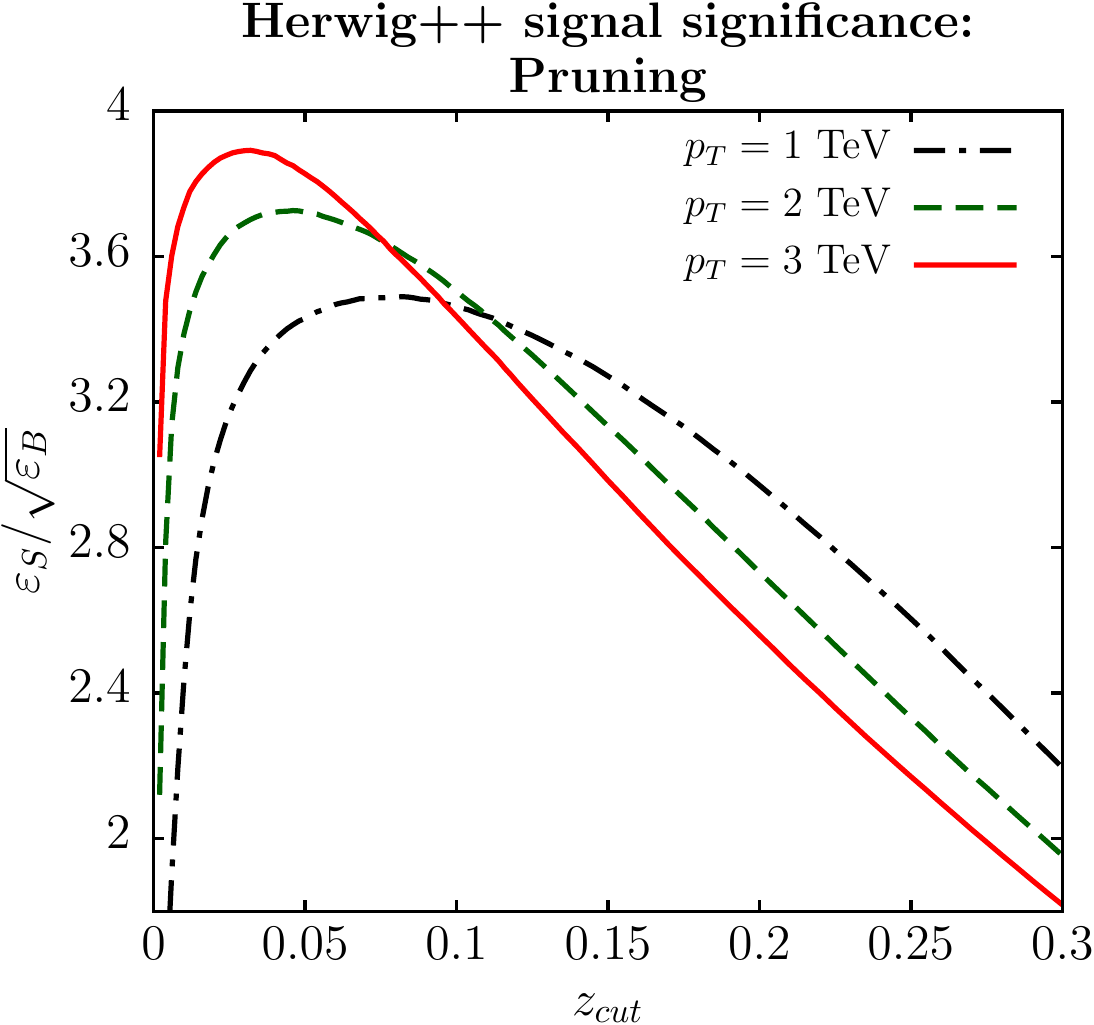}
    \includegraphics[width=0.474 \textwidth]{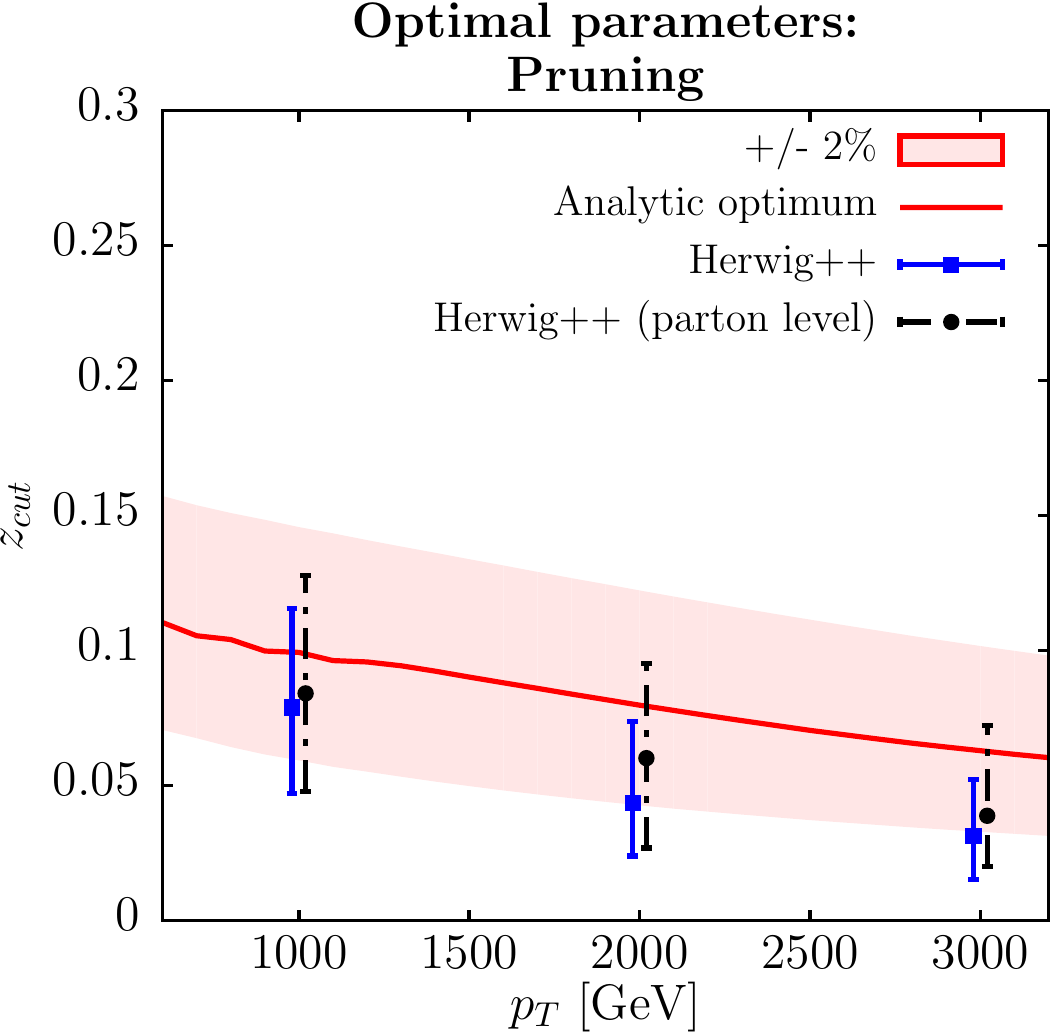}
    \caption{Pruning analytical signal significance from tree level signal and resummed background as a function of $\zcut$ compared to \herwig at parton level and with hadronisation and MPI. Details of generation given in \reffig{SS_MMDT}.
      \label{fig:SS_Prune}
    }
\end{figure}
Here we carry out a similar analysis for the case of pruning. The
resummed expression for pruning, for QCD jets, is considerably more
complicated than for mMDT. The result essentially has two components
which in Ref.~\cite{Dasgupta:2013ihk} were dubbed the Y and I
components respectively. We have already dealt with Y-pruning in
some detail in this article in the context of signal jets. For the
background, as we have also discussed in a previous section,
Y-pruning, for small jet masses, $\rho< \zcut^2$, consists of a
suppression of the leading order single-logarithmic result by a Sudakov like form factor and gives rise
to a desirable suppression of the background in the signal region, for
high $p_T$ values. The I pruning contribution, on the other hand, starts at order $\alpha_s^2$  and is as singular as the plain jet mass
i.e. double logarithmic. For the sum of Y and I components, i.e. for
pruning as a whole, one observes two transition points: for
$\rho>\zcut$ the behaviour is like the plain mass (as for mMDT), for
$\zcut^2< \rho <\zcut$ there is a single log behaviour as in the
leading-order result and as for mMDT, while for $\rho<\zcut^2$ we see
the I-pruning contribution starts to become more important which can
cause growth of the background and the appearance of a second peak for
quark jets and a shoulder like structure for gluon jets.

We do not, for brevity, present here the resummed results for pruning
for QCD jets, referring the reader instead to section 5.3 of
Ref.~\cite{Dasgupta:2013ihk}. Here we simply plot the analytical signal
significance for pruning as for mMDT, with neglect of radiative
corrections to the signal efficiency, but with the full resummed
calculation for QCD background, which we take to be quark jets alone. The resulting signal significance is
displayed in \reffig{SS_Prune} along with MC results at parton
and hadron level. One would expect the optimal $\zcut$ to lie in a
region that corresponds to the mMDT like region i.e. such that
$\rho_H>\zcut^2$. Choosing a larger $\zcut$ would push us into the
region where the background starts to grow due to onset of I-pruning
and hence the signal significance falls off.
At 1 TeV where $\rho_H \approx 0.0017$ one can expect
an optimal value of $\zcut$ to be below $\sqrt{\rho_H} \approx
0.125$ while for 3 TeV  one may expect a value closer to $0.04$ and
these expectations are roughly consistent with what one notes with
both the analytical and MC results shown. Once again we observe that
non-perturbative effects do not change the essential picture one
obtains from analytics and have only a limited impact on the signal
significance relative to parton level.

The pruning results have clear qualitative differences from the case of
mMDT. In particular at higher $p_T$ we have to be more precise about the choice of $\zcut$ due to the somewhat narrower peak in the
signal significance. We can compare, as for mMDT, analytical results
to those from \herwig, once again with a $\pm 2 \%$ tolerance band
shown in the bottom right figure of \reffig{SS_Prune}. We
observe that within this small tolerance band the results are
compatible though at higher $p_T$ perhaps less so than for mMDT. 

In the original paper \cite{Ellis:2009me}, the authors conclude that
the optimal $\zcut$ value for pruning is $0.1$ when using the C/A
algorithm to cluster the initial jet, as we do here.
This optimisation was performed at a moderate transverse momenta ($100-500$ GeV for W bosons) compared to this paper, however our results are consistent as we approach this region.
For larger boosts, we observe that the optimal value choice for
$\zcut$ tends to slightly smaller values ($\zcut \sim 0.075$).

\begin{figure}[ht]
  \includegraphics[width=0.49 \textwidth]{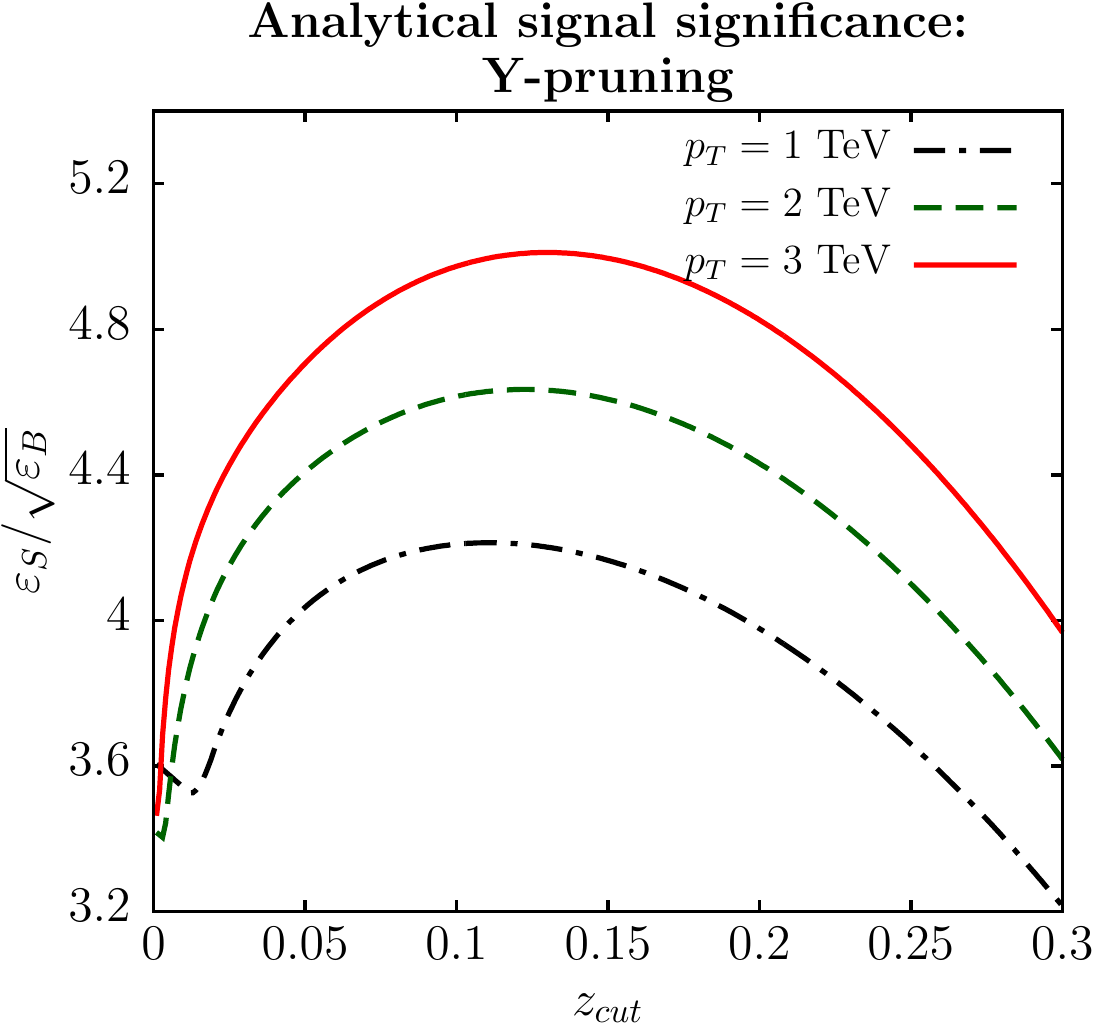}
  \includegraphics[width=0.49 \textwidth]{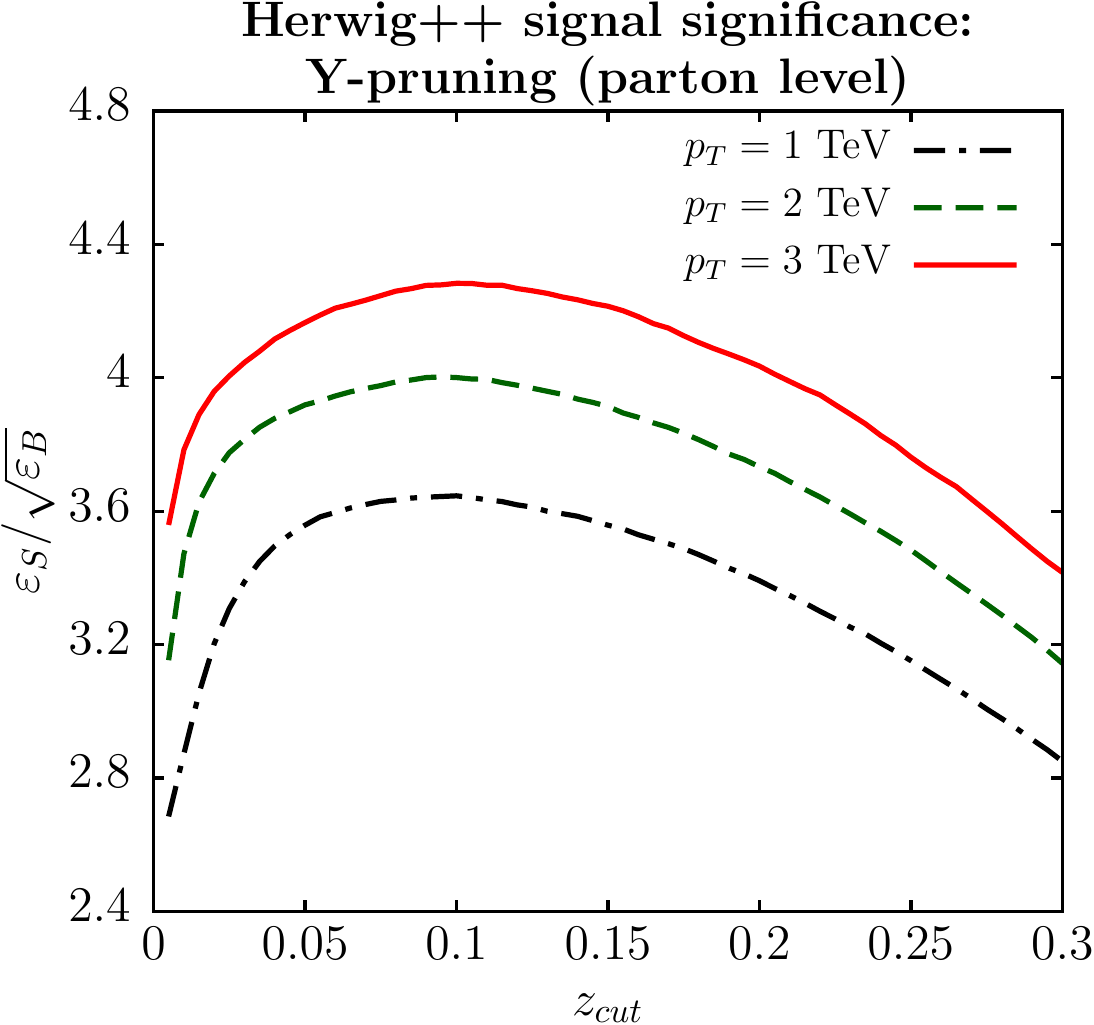}\\\\
    \includegraphics[width=0.49\textwidth]{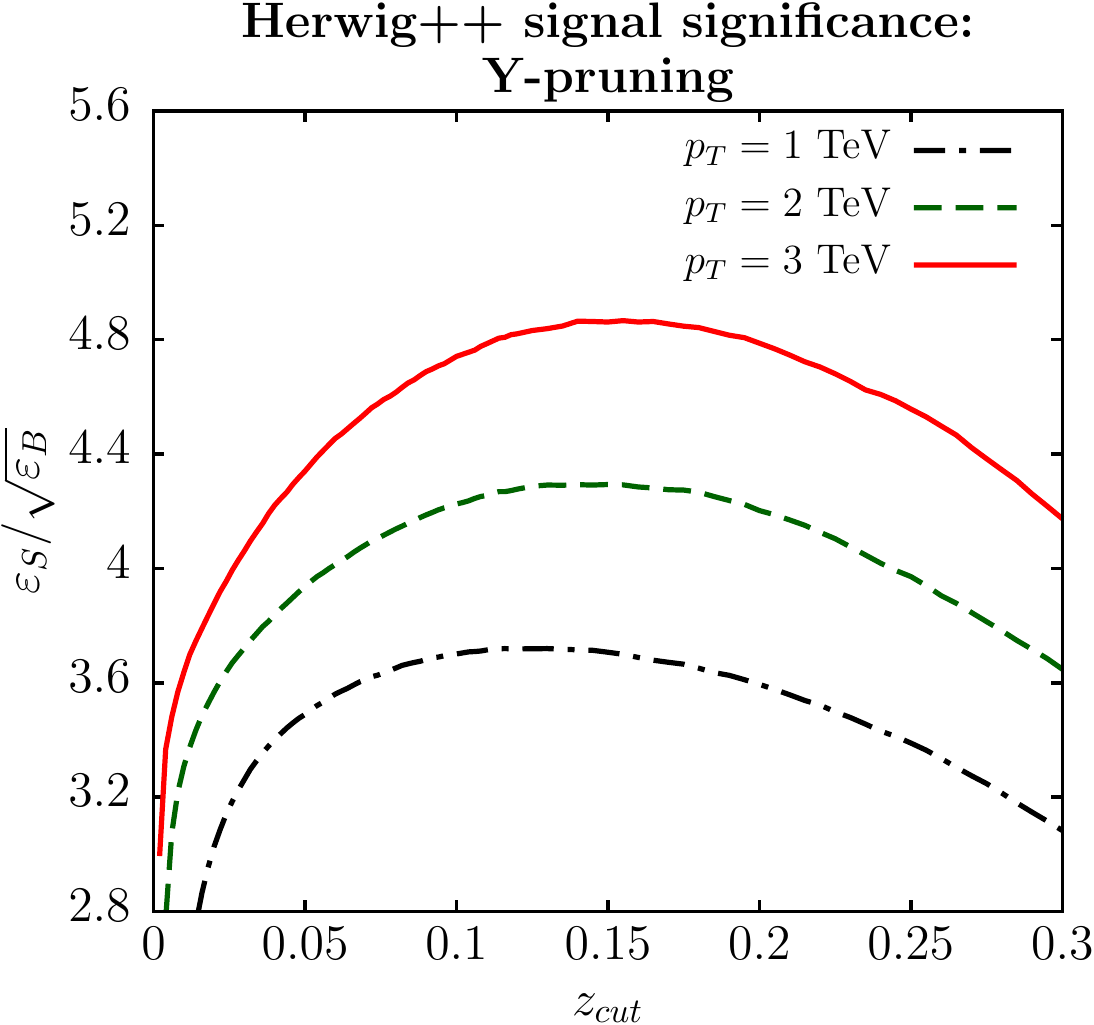}
    \includegraphics[width=0.474 \textwidth]{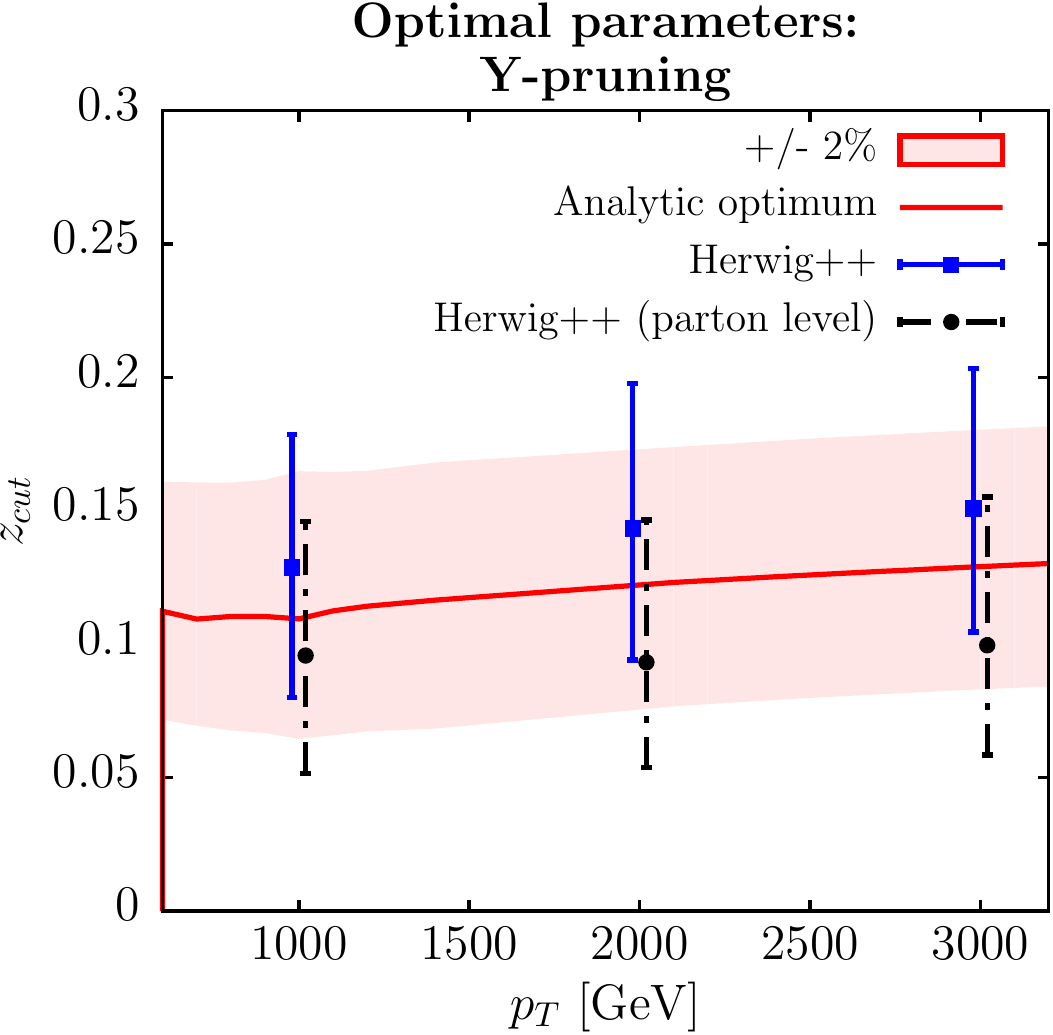}
    \caption{Y-pruning analytical signal significance from tree level signal and resummed background as a function of $\zcut$ compared to \herwig at parton level and with hadronisation and MPI.  Details of generation given in \reffig{SS_MMDT}.
      \label{fig:SS_YPrune}
    }
\end{figure}

We also present in \reffig{SS_YPrune} results for the signal
significance of Y-pruning, again taking quark jets as background.
Here we note firstly that analytics are again broadly in agreement with MC
results for the shape of the signal significance as a function of
$\zcut$. Secondly the peaks are quite broad and so choosing a somewhat
non-optimal value of $\zcut$ does not critically affect the
significance. Furthermore, the optimal $\zcut$ does not depend
strongly on $p_T$ and is virtually constant over the limited $p_T$ range studied.
Lastly within a $\pm 2$  \% tolerance band there is good
agreement between full MC results and simple analytics on the optimal
values of $\zcut$. 

Hence for mMDT and Y-pruning and to a slightly smaller degree for
pruning we find that, over the $p_T$ values we studied here,  analytical results based on resummed calculations
for QCD background and lowest order results for signals, with neglect
of non-perturbative effects, capture the essential features of tagger
performance, as reflected in the signal significance. An extension of
our studies to lower $p_T$ values would be of interest in order to
ascertain the further validity of the simple picture we have used for our
analytical results and probe in more detail the role of radiative
corrections to the signal and that of non-perturbative contributions. We
shall next examine the more involved case of optimal parameters for
trimming.

\subsection{Trimming}
\begin{figure}[t]
\begin{center}
\includegraphics[width=0.49 \textwidth]{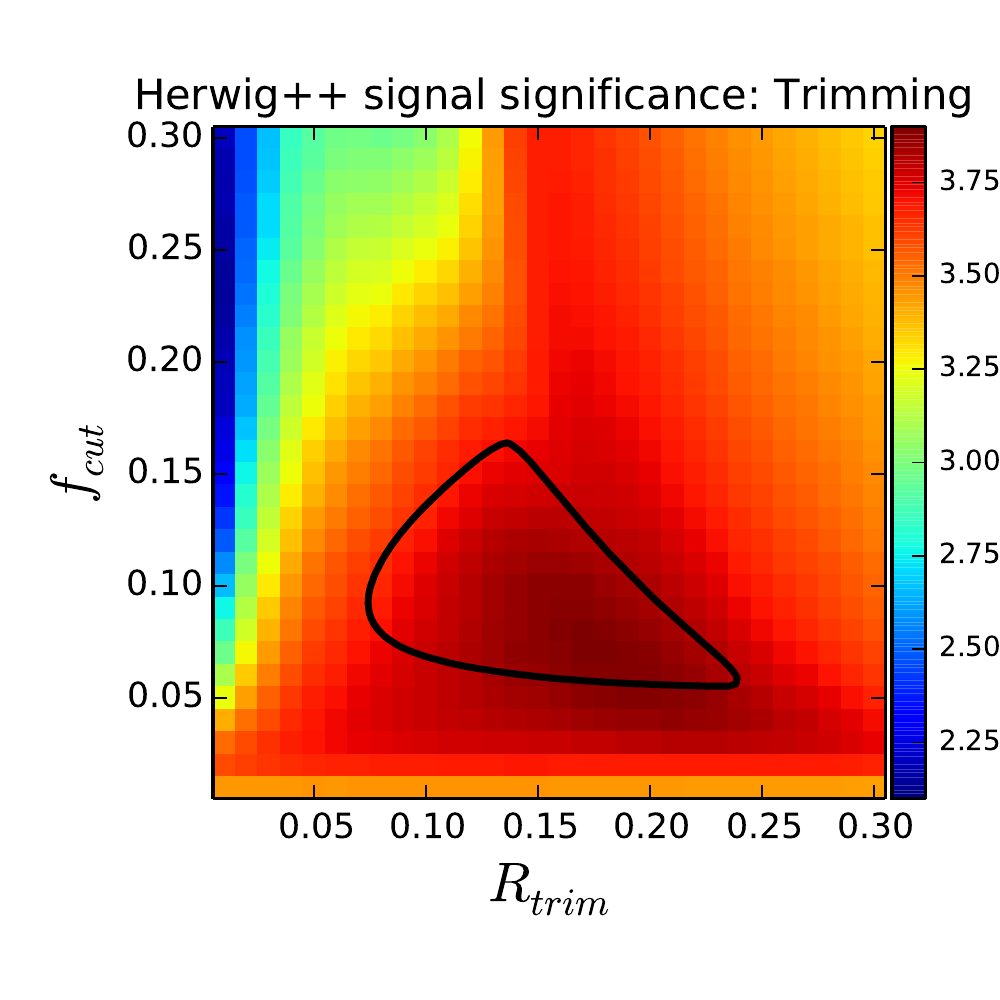}
\includegraphics[width=0.49 \textwidth]{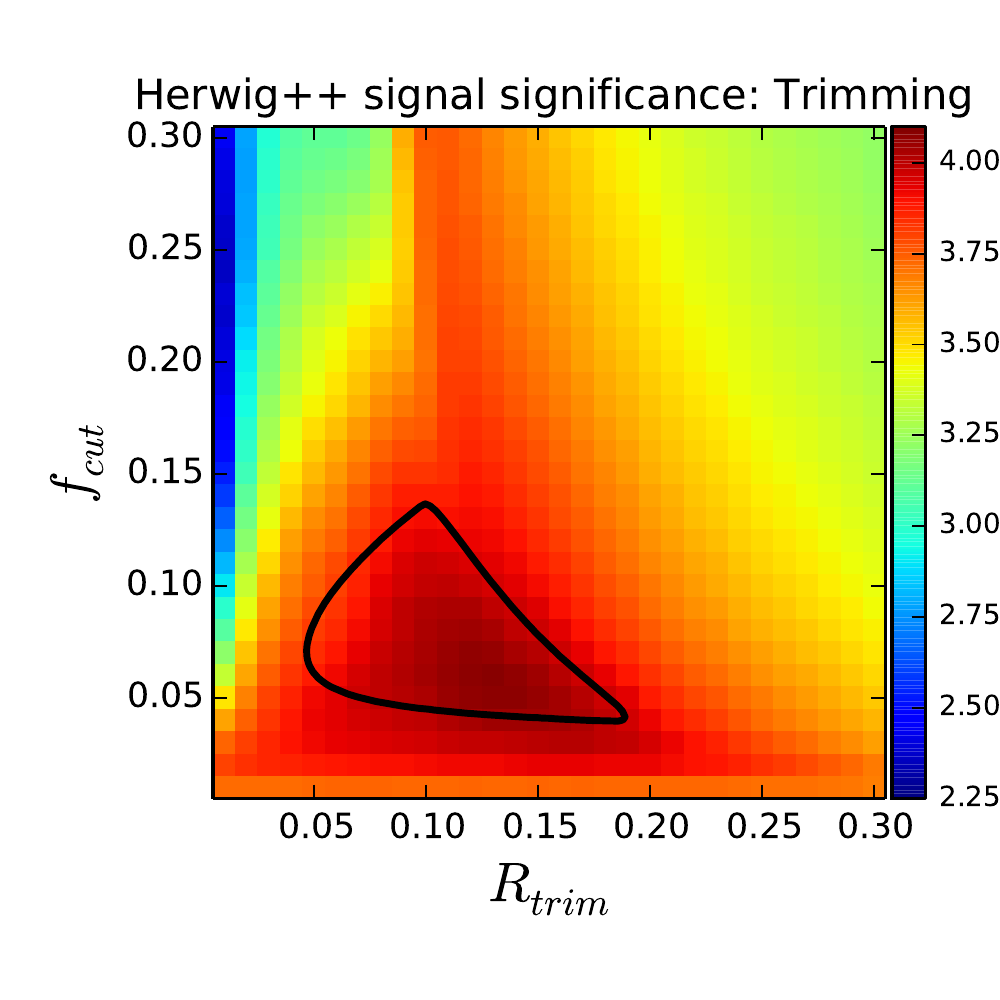}
\\
\includegraphics[width=0.49 \textwidth]{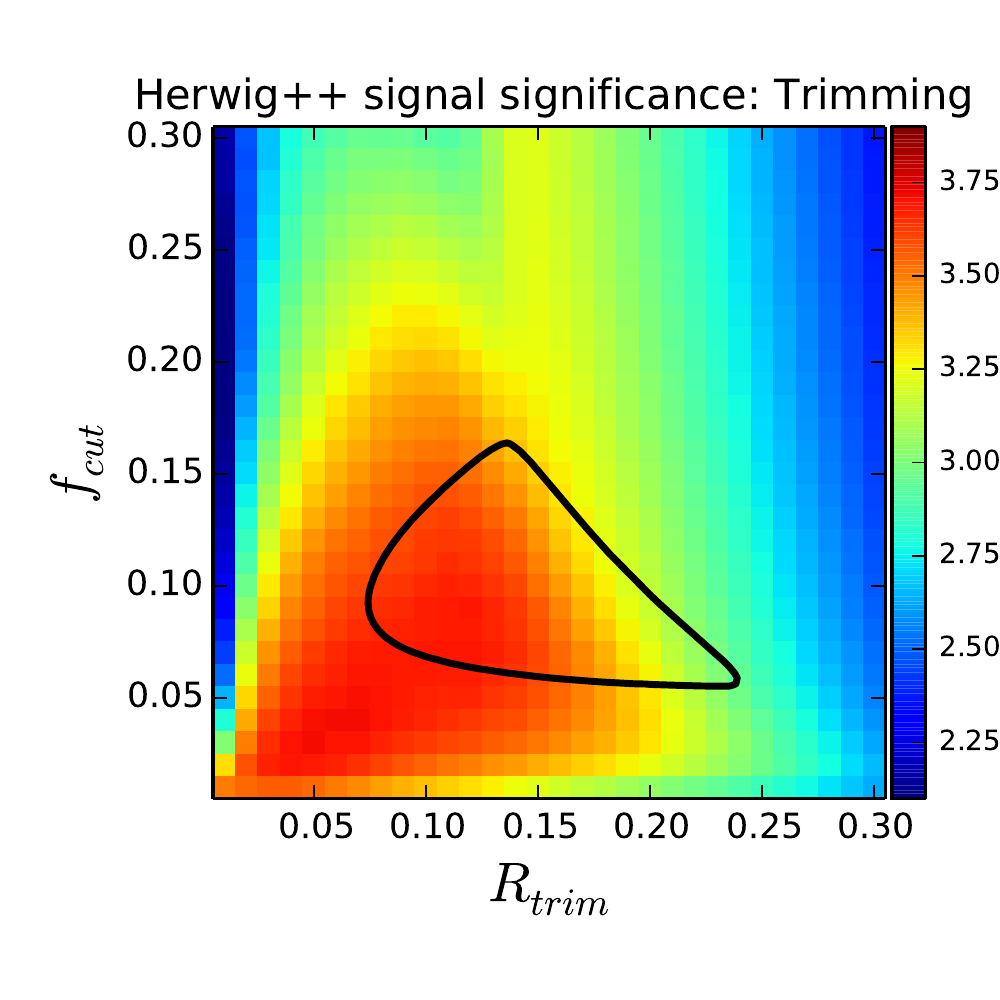}
\includegraphics[width=0.49 \textwidth]{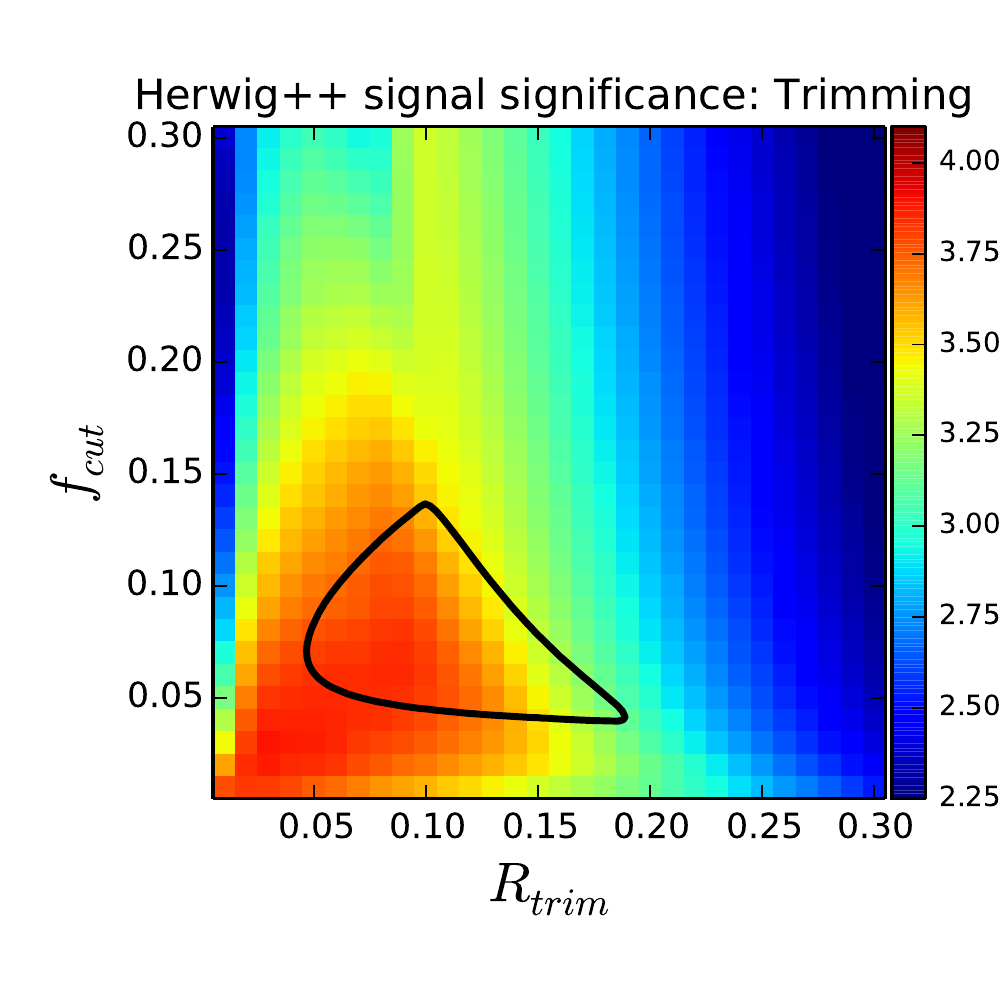}
\end{center}

\caption{2D density plot showing the trimming signal significance as a function of $\Rtrim$ and $\fcut$ using \herwig for $H \to b\bar{b}$ jets with quark backgrounds with a minimum jet transverse momentum cut.
The top panels are generated at parton level with transverse momenta 2 TeV and 3 TeV left and right and the bottom panels include hadronisation and underlying event.
The area inside the black contour represents the analytic prediction with FSR radiative corrections to the signal efficiency for optimal values within 2\% of the analytic peak signal significance.
\label{fig:Trim2D}}
\end{figure}

\begin{figure}[ht]

  \includegraphics[width=0.49 \textwidth]{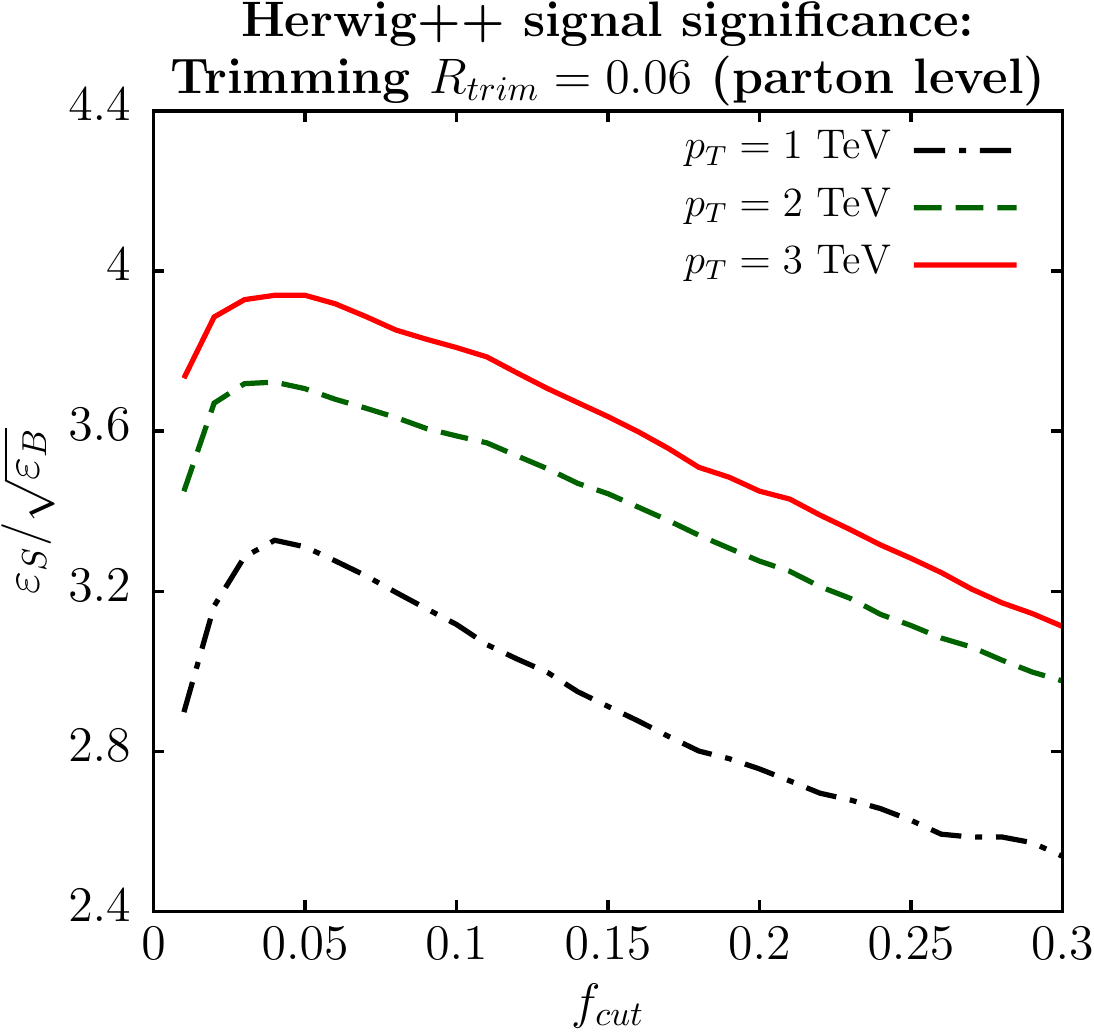}
    \includegraphics[width=0.49\textwidth]{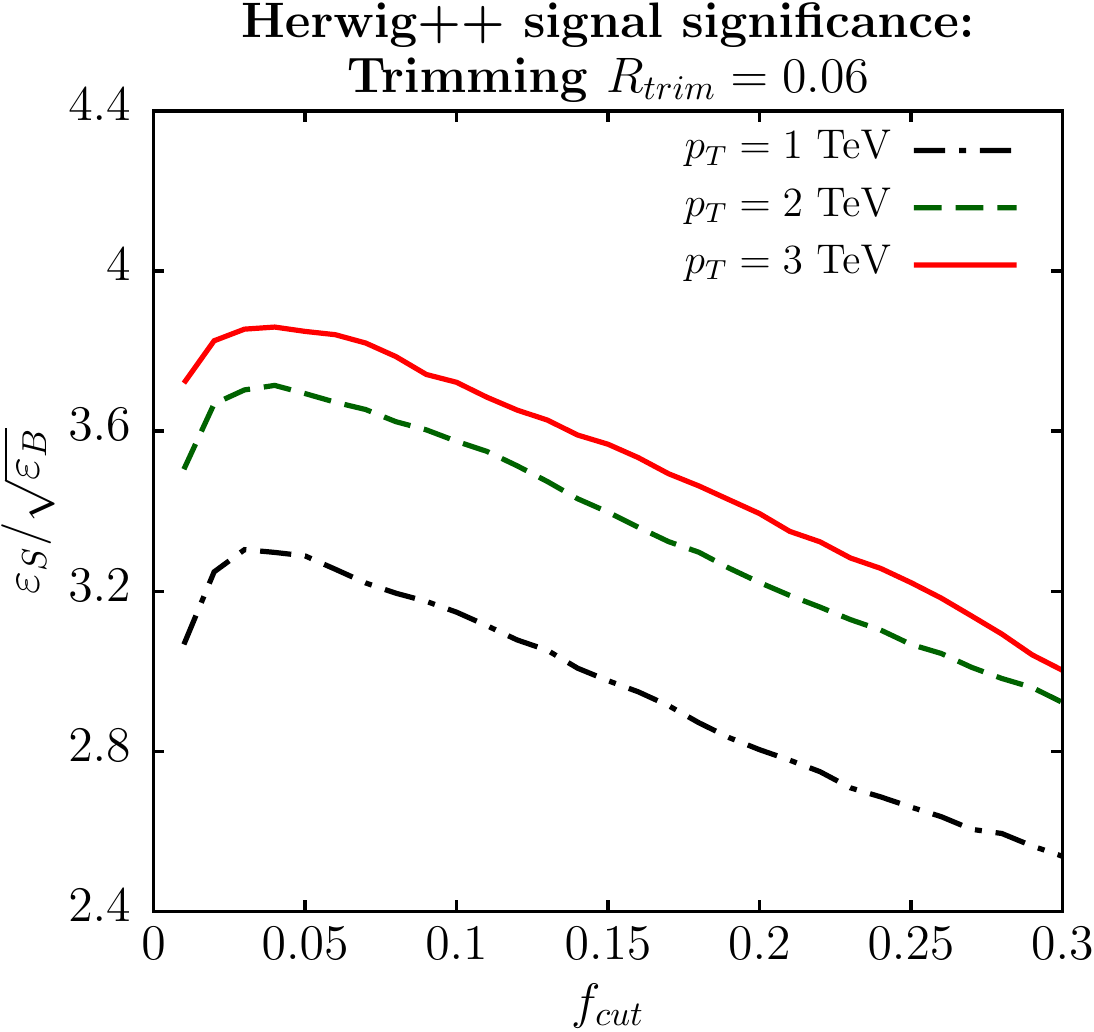}

     \caption{Trimming signal significance for $R_{trim} \sim \sqrt{\Delta} \sim 0.06$ as a function of $\fcut$ generated at parton level and with hadronisation and MPI using \herwig .
    }
          \label{fig:SS_Trimm}
\end{figure}

Here we carry out a similar analysis for trimming, but one now has to optimise two parameters, $\Rtrim$ and $\fcut$.
As performed for the pruning analysis, we use the analytic resummed expression for QCD jets given in Ref. \cite{Dasgupta:2013ihk}.
The result for the background jet mass distribution consists of a region with single log behaviour (equivalent in structure to mMDT) for $\fcut \Rtrim^2 < \rho < \fcut$ which transitions into a double logarithmic growth in the background distribution for $\rho < \fcut \Rtrim^2$.
However, in contrast to mMDT and (Y-)pruning, FSR radiative corrections to the signal efficiency are crucial for optimisation.
If one naively uses the tree level result given in \refeq{TrimBorn}, it follows that the optimum value for $\Rtrim$ tends to zero.
This is because one can ensure that signal mass window is within the single logarithmic region by simply pushing the location of the double logarithmic transition ($\rho=\fc \Rtrim^2$) to small values of the QCD jet mass, thereby avoiding the double logarithmic peak.

However, as shown in this paper, in the limit $\Rtrim \to 0$, one encounters large logarithmic corrections to the signal efficiency associated with final state radiation from the signal jet (see \refeq{epsFSRlog}).
This puts a limit on how small one can reduce the trimming radius whilst maintaining reasonable signal mass resolution.
Hence, we now include FSR radiative corrections to the signal efficiency by integrating the the expression given in \refeq{SoftTrimFSR} over $z$ and adding this term to the Born level result \refeq{TrimBorn}.
Including this radiative correction, along with the resummed QCD background, we can obtain analytical estimates for the signal significance.

In \reffig{Trim2D} we show a 2D density plot for the signal significance with trimming over a range of $\Rtrim$ and $\fcut$ values using Monte Carlo at parton level (top) and with full hadronisation and underlying event (bottom) with a transverse momentum cut at 2 and 3 TeV  left and right respectively.
We overlay a black analytical contour representing the region in which the analytical signal significance is no more than $\pm 2 \%$ away from the analytically derived peak value for $\Rtrim$ and $\fcut$.
One can see that we have reasonable agreement between the simple analytical estimates and the \herwig results at parton level.
However, when one includes non-perturbative effects, we observe that contamination from underlying event significantly reduces the signal significance as $\Rtrim$ increases.

We can use our simple analytical estimates to comment on the optimal values we observe from MC.
Firstly, for optimal values of $\fcut$ and $\Rtrim$, one would expect the signal mass window to reside in the single logarithmic mMDT-like region of the background, hence we anticipate that the optimal parameters satisfy the constraint  $\Delta > \fcut \Rtrim^2$.
This expectation is consistent with both the analytical contour (top right edge) and MC results both at parton and full hadron level.
This background driven effect is manifest as a suppression in signal significance when the product $\fcut \Rtrim^2$ becomes large (i.e top right of the contour plots).
For example, at $3$ TeV and fixed $\fcut=0.1$, one would analytically expect an optimal value for $\Rtrim \lesssim 0.13$, whilst at $\fcut=0.05$ one expects $\Rtrim \lesssim 0.19$.
These numbers are in agreement with the analytical contour and MC results.
Secondly, FSR corrections to the signal efficiency become significant in the region $\Rtrim^2 \ll \Delta$, hence one would expect the optimal trimming radius to reside in the region $\Rtrim \gtrsim \sqrt{\Delta}$.
At $3$ TeV this corresponds to $\Rtrim > 0.04$ and at $2$ TeV corresponds to $\Rtrim > 0.06$.
This is consistent with the analytical contour and MC, where we observe a reduction in signal significance in the limit $\Rtrim  \ll \sqrt{\Delta}$.

We notice that, like mMDT and pruning, the signal significance is fairly insensitive to variations in $\fcut$ provided we choose $\Rtrim$ such that $\sqrt{\Delta/\fcut} > \Rtrim$.
However, the signal significance is subject to non-perturbative corrections which increase with $\Rtrim$, and consequently one should favour the small $\Rtrim$ limit of the analytical optimal contour region $\Rtrim \approx \sqrt{\Delta}$ to minimise both signal FSR and non-perturbative corrections to the signal significance.
It is thus of interest to choose $R_{\mathrm{trim}}^2 \sim \Delta$
    and study the dependence of the signal significance on the choice
    of $f_{\mathrm{cut}}$ as in Figs.~\ref{fig:SS_MMDT}, \ref{fig:SS_Prune}, \ref{fig:SS_YPrune}. The results can be
    found in Fig.~\ref{fig:SS_Trimm} for $R_{\mathrm{trim}}=0.06$ which is identical
    to $\Delta$ at 2 TeV and of order $\Delta$ for the other $p_T$ values.
 With the given choice of $R_{\mathrm{trim}}$, reminiscent of the
 pruning radius, it is natural to compare the results to those for
 pruning reported in Fig.~\ref{fig:SS_Prune}. One notes that even with a similar
 choice of radius there are differences between the two
 techniques. While for pruning the optimal $z_{\mathrm{cut}}$
 decreases with increasing $p_T$ the optimal value for trimming stays
 more constant. The peak signal significance itself increases with
 $p_T$ in both cases. For a given $p_T$ the behaviour as a
 function of $f_{\mathrm{cut}}$ is also different, especially at larger
 $f_{\mathrm{cut}}$. These differences originate in a number of
 sources: the difference in FSR corrections and their $p_T$ dependence
 which is more pronounced for trimming, differences in the definitions
 of $f_{\mathrm{cut}}$ and $z_{\mathrm{cut}}$ and last but not least
 differences arising from QCD background jets with pruning and
 trimming (see Ref.~\cite{Dasgupta:2013ihk}). In order to better
 understand the role for example of FSR effects, in the above context, we note that for
 pruning one can simply replace the signal efficiency by $1-2
 z_{\mathrm{cut}}$ as we have done for our analytical studies of
 optimal parameter values in pruning. If one similarly uses $1-2
 f_{\mathrm{cut}}$ in computing the signal efficiency for trimming one
 observes that the result with $R_{\mathrm{trim}}^2 \sim \Delta$ is
 closer to that for pruning and optimal $f_{\mathrm{cut}}$ values show a very similar
 trend with $p_T$ to those for pruning.  However for trimming $p_T$ dependent FSR corrections
 cannot be neglected, especially at low $p_T$, and play an important role in pushing the optimal
 $f_{\mathrm{cut}}$ to smaller values than would be obtained by
 turning off FSR effects. This is the main reason behind the
 relative insensitivity of optimal $f_{\mathrm{cut}}$ values seen with
 trimming, over the $p_T$ range studied in Fig.~\ref{fig:SS_Trimm}.
\section{Conclusions}
In this article we have studied perturbative radiative corrections and
non-perturbative effects for the case of signal jets, specifically for
boosted Higgs production with $H \to b \bar{b}$, with the application
of jet substructure taggers. For the former we have carried out
relatively simple analytical calculations both to assess the impact of
ISR and FSR as well as to study its dependence on various parameters,
such as a mass window of width $\delta M$ on either side of the signal
mass, the fat jet $p_T$, the mass of the resonance $M_H$, and the
parameters of the various  taggers. To examine non-perturbative
effects we have confined ourselves to MC studies.

Our study was motivated by relatively recent calculations dedicated to
the case of QCD background jets and in particular work presented in Ref.~\cite{Dasgupta:2013ihk}. There it was noted that while taggers
should in principle discriminate against jets from QCD background, the degree to
which this happened and the impact on the background jet mass
distribution was not always as desired. While taggers such as pruning,
mMDT and trimming were essentially identical over a limited range in the normalised square jet mass
$\rho=M_j^2/\bra{p_T^2 R^2}$, significant differences in performance and behaviour were
observed at small values of $\rho$, which especially at high $p_T$
corresponded to masses in the signal mass region of interest.
Likewise taggers should, in principle, not affect significantly signal
jets, retaining them as far as possible. Additionally most taggers have a
grooming element (via the $\fcut/\ycut/\zcut$ criteria) that is responsible
for clearing the jet of contamination from ISR/UE thereby helping in
the reconstruction of sharper mass peaks.  Here our aim was to carry
out analytical and MC studies to investigate in detail the impact of taggers
on signal especially with regard to the interplay between tagger
parameters as well as kinematic cuts such as jet $p_T$, masses and
mass windows. 

Our findings on the whole indicate that tagger performance is more
robust for the case of signal jets than was apparent for QCD
background. Most taggers are quite similar in their response to ISR
and generally significantly ameliorate the loss of the signal efficiency
seen for plain jet mass cuts, without these substructure
techniques. An exception to this situation was the case of Y-splitter
where the ISR and UE contamination resulted in a loss of signal efficiency
identical to that seen for plain jets.

Likewise for FSR, the radiative losses that one sees are on the whole
modest for a reasonably wide range of tagger parameters. Here
an interesting question opens up about the potential role for
fixed-order calculations in the context of jet substructure
studies. This is because one observes an absence of genuine
logarithmic enhancements for sensibly chosen tagger parameter values. 
The signal efficiency, for the taggers studied here, ought then to be better described by exact
calculations that incorporate hard gluon radiation or by combinations
of matrix element corrections and parton showers than by the
soft/collinear  emissions encoded in pure parton showers. We carried out a
comparison between an MC description of the signal efficiency and
exact order $\alpha_s$ results for various taggers, reported in appendix B. We find that we can reasonably adjust parameters
such as the size of our mass window $\delta M$ to obtain good
agreement between the two descriptions. Such observations may also be
useful beyond the immediate context of our work, in situations where
differences in tagger performance could come from regions of phase
space that are not under the control of a soft eikonal approximation. In these situations one would ideally want to combine
resummed calculations, where necessary, with fixed-order calculations
i.e. carry out matched resummed calculations.
A summary of the results presented in this paper for the logarithmic structure of radiative corrections to the signal efficiency for each tagger are given in Table.~\ref{table:Summary}.
\begin{table}[t]
\begin{center}
\begin{tabular}{ l | c c c }
  Tagger &  ISR & FSR \\
  \hline \\         
  Plain & $R^2\ln{\frac{R^2}{\epsilon \Delta}}$ & $\Delta \ln{\Delta}$\\\\
  Trimming & $R^2\ln{\frac{1}{\fcut}}$ & $2\ln{\frac{\Delta}{\Rtrim^2}} C_2(\fcut,\epsilon)$\\\\
  Pruning & $R^2\ln{\frac{1}{\zcut}}$ & $\frac{2 \pi}{\sqrt{3}}\ln{\frac{\zcut}{\epsilon}}$\\\\
  Y-pruning & $R^2\ln{\frac{\zcut R^2}{\Delta}}$ & $\frac{2 \pi}{\sqrt{3}}\ln{\frac{\zcut}{\epsilon}}$\\\\
  mMDT & $R^2\ln{\frac{1}{\ycut}}$ & $0.646 \ln{\frac{\ycut}{\epsilon}}$\\\\
  Y-splitter & $R^2\ln{\frac{R^2}{\epsilon \Delta}}$ & $\order{\ycut}$\\\\
\end{tabular}
\caption{A table summarising the logarithmic structure of radiative corrections to the signal efficiency for each tagger.
For each tagger we show the coefficient of $-\frac{\alpha_s C_F}{\pi}$
for ISR and FSR results in the small $\Delta$ and $\zcut/\ycut/\fcut$
limit. We have defined $\epsilon =2 \delta M/M_H$ as in the main text.
The coefficient $C_2$ for the trimming FSR logarithm is given in \refeq{coefficients}.
\label{table:Summary}
}
\end{center}
\end{table}

A development we have made here is the introduction of a
combination of Y-splitter with trimming in an attempt to 
improve the response of Y-splitter to ISR/UE contamination. The main reason why we
made this effort was due to the fact that we observed that Y-splitter
was very effective at suppressing QCD background in the signal
region. The resulting improvement of signal efficiency coupled with
the fact that the background suppression from Y-splitter remained essentially intact
after the use of trimming, meant that the combination of Y-splitter+trimming actually outperforms other
taggers studied here, in particular, at high $p_T$. Our observation is in keeping with
the general idea that suitably chosen tagger combinations may prove to
be superior discovery tools compared to currently proposed individual
methods \cite{Soper:2010xk}. In fact it is now becoming increasingly common
to use combinations of techniques such as N-subjettiness
\cite{Thaler:2010tr} with for instance mMDT in an
effort to maximise tagger performance (see e.g. \cite{Soyez:boost}). There is also much effort aimed
at better understanding tagger correlations and we expect that our forthcoming analytical calculations for the case of Y-splitter with trimming will shed
further light on some of these issues \cite{PowDasSiod:2015}.

Lastly, we have carried out an analytical study of optimal parameter
values for various taggers. Having observed modest radiative corrections to the signal we neglected these effects
and found  that analytical estimates, based on lowest order results for the signal and resummed calculations for QCD
background, generally provide a good indicator of the dependence of
signal significance on the tagger parameters. The analytical formulae which also do not include
non-perturbative effects  give rise to optimal values that are fairly
compatible with those produced by full MC studies. This is encouraging
from the point of view of robustness of the various methods considered
since a dependence of optimal values on MC features (hadronisation
models or MC tunes) are potentially not ideal. 

We note in closing that for other methods, such as N-subjettiness for
example, there will also be a suppression of signal jets due to the
fact that such observables directly restrict radiation from the signal
prongs. Thus in those cases radiative corrections arising from
soft/collinear emissions by signal prongs are highly significant as
can be noted from Ref.~\cite{Feige:2012vc}. We hope that our work taken
together with studies of such observables will enable a more complete
understanding of features of signal jets in the context of jet
substructure studies and provide yet stronger foundations for future 
developments.

\section*{Acknowledgements}
We are particularly grateful to Gavin Salam and Mike Seymour for
illuminating discussions on the topic of jet substructure.
We also thank Simone Marzani and Gregory Soyez for further useful
discussions. We thank an anonymous referee for helpful remarks and suggestions which we have implemented in the current version of the article.
   We are also grateful to the Cloud Computing for Science and Economy 
project (CC1) at IFJ PAN (POIG 02.03.03-00-033/09-04) in Cracow whose resources 
were used to carry out most of the numerical calculations for this project. 
Thanks also to Mariusz Witek and Miłosz Zdybał
 for their help with CC1. 
This work was funded in part by 
the MCnetITN FP7 Marie Curie Initial Training Network PITN-GA-2012-315877.
We would like to thank the UK's STFC for financial support.
This work is supported in part by the Lancaster-Manchester-Sheffield Consortium for Fundamental Physics under STFC grant ST/L000520/1.

\appendix
\section{Angular integration for FSR}
To work out the coefficient of the soft FSR we need to perform the
angular and $z$ integrals for the antenna pattern in
Eq.~\eqref{eq:anttrim} for trimming and likewise for all taggers. 
Generally, for a single gluon emission, one has to evaluate the
contribution from FSR emission outside two cones of radius $r$ centred
on the $b$ and $\bar{b}$ quarks. The choice of $r$ depends on the
tagger in question, so after carrying out the angular integration, one
can set $r^2$ as $\Rtrim^2$ for trimming, $\Delta$ for pruning and
$\theta_{b\bar{b}}^2=\Delta/(z(1-z))$ for mMDT and lastly  integrate over
$z$. 

One has to then evaluate the integral
\begin{equation}
I = \int \frac{d\Omega}{2 \pi}\frac{1-\cos
  \theta_{b\bar{b}}}{(1-\cos\theta_{bk}) (1-\cos\theta_{\bar{b}k})}
\Theta \left(\theta_{bk}^2 -r^2 \right) \left(\theta_{\bar{b}k}^2 -r^2 \right),
\end{equation}
where we have now explicitly written the conditions for the gluon to
be at an angle $\theta^2> r^2$ wrt both hard partons. \footnote{While
  we have retained, at this stage, the full angular antenna pattern
  for ease of comparison to standard formulae, we shall later take the
  small angle approximation to compute the final answer.} 
The simplest way to evaluate the integral above is to first consider
an integration over the entire solid angle and then to remove the
contribution from inside two cones around the hard parton
directions. We shall assume that the cones do not overlap, so shall
consider $r < \theta_{b \bar{b}}/2$. For larger $r$, as appropriate for
mMDT where $r=\theta_{b\bar{b}}$, we perform a numerical calculation
and find that our results agree with those of Rubin~\cite{Rubin:2010fc}.

Therefore we write
\begin{equation}
I = I_{\mathrm{all}} - I_{\mathcal{C}_b}  - I_{\mathcal{C}_{\bar{b}}}
\end{equation}
where $I_{\mathrm{all}}$ is the integration over the full solid angle
and $I_{\mathcal{C}_{b,\bar{b}}}$ are the integrals inside the region
corresponding to cones around $b$ and $\bar{b}$ directions respectively.
$I_{\mathrm{all}}$ can be evaluated by standard techniques and yields, after
azimuthal averaging, the textbook result corresponding to angular ordering of soft emission.
\begin{equation}
I_{\mathrm{all}} = \int d\left(\cos \theta_{bk}\right) \frac{\Theta \left(\theta^2_{b \bar{b}}-\theta^2_{bk}\right)}{1-\cos\theta_{bk}}+ \int d\left(\cos\theta_{\bar{b}k} \right)\frac{\Theta \left(\theta^2_{b \bar{b}}-\theta^2_{\bar{b}k}\right)}{1-\cos\theta_{\bar{b}k}}.
\end{equation}

The contribution inside the cone around $b$, $I_{\mathcal{C}_b}$, can
be evaluated as follows. Taking the $b$ direction as the ``$z$'' axis we define the parton directions by the unit vectors:
\begin{eqnarray}
\vec{n}_b &=& \left (0,0,1 \right), \nonumber\\
\vec{n}_{\bar{b}} &=& \left( 0,\sin \theta_{b \bar{b}},\cos \theta_{b \bar{b}}\right), \nonumber\\
\vec{n}_k &=& \left(\sin \theta_{bk} \sin \phi, \sin \theta_{bk}\cos \phi, \cos \theta_{bk}\right).
\end{eqnarray}
The in-cone subtraction term for $\mathcal{C}_1$ can then be written as 
\begin{equation}
I_{\mathcal{C}_b}=\int \frac{d\phi}{2\pi} d\left(\cos \theta_{bk} \right) 
\frac{1-\cos\theta_{b \bar{b}}}{\left(1-\cos\theta_{bk}\right)\left(1-\cos\theta_{b \bar{b}} \cos \theta_{bk}-\sin\theta_{bk} \sin\theta_{b \bar{b}} \cos \phi\right)} \Theta \left(r^2-\theta^2_{bk}\right).
\end{equation}
Integrating over the azimuthal angle $\phi$ gives 
\begin{equation}
I_{\mathcal{C}_b} = \int d\left( \cos\theta_{bk} \right) \frac{1-\cos \theta_{b \bar{b}}}{\left (1-\cos \theta_{bk}\right) |\cos\theta_{bk}-\cos \theta_{b \bar{b}}|} \Theta \left (r^2-\theta_{bk}^2 \right).
\end{equation}
This term can be combined with the corresponding contribution (the
first term) in $I_{\mathrm{all}}$, and taking the small-angle approximation for $\cos \theta \approx 1-\theta^2/2!$ one obtains 
\begin{equation} 
I = \int_0^{r^2} d\theta^2_{bk}
\left(\frac{1}{\theta^2_{bk}}-\frac{\theta^2_{b
      \bar{b}}}{\theta^2_{bk} \left(\theta^2_{b
        \bar{b}}-\theta^2_{bk}\right)}\right) +
\int_{r^2}^{\theta_{b\bar{b}}^2}\frac{d \theta^2_{bk}}{\theta^2_{bk}}
+ b \leftrightarrow \bar{b},
\end{equation}
where we have also included $I_{\mathcal{C}_{\bar{b}}}$ via the interchange $b \leftrightarrow \bar{b}$.
The collinear divergence along each hard parton direction is cancelled
by the in-cone contributions, leaving only a wide-angle contribution.
Carrying out the angular integrations we get 
\begin{equation}
I= 2 \log \left(\frac{\theta_{b \bar{b}}^2-r^2}{r^2}
\right),
\end{equation}
which agrees with the result found by Rubin \cite{Rubin:2010fc}
written in terms of the variable $\eta=\frac{r}{\theta_{b\bar{b}}}$, for $\eta <\frac{1}{2}$. In the collinear limit, $r \ll
\theta_{b\bar{b}}$, we get the result for trimming quoted in the main text and used in
Eq.~\eqref{eq:SoftTrimFSR}. To obtain the result for pruning we
substitute $\theta_{b\bar{b}}^2 =\Delta/(z(1-z))$ and $r^2=\Delta$,
then carry out the $z$ integral, which gives:
\begin{equation}
I=2 \int_{\zcut}^{1-\zcut}  dz \ln \left(\frac{1-z(1-z)}{z(1-z)}
\right)=\frac{2 \pi}{\sqrt{3}}+\mathcal{O}\left(\zcut\right),
\end{equation}
which corresponds to the result quoted for pruning in Eq.~\eqref{eq:pruningfsr}.

For mMDT where $r =\theta_{b \bar{b}}$ our calculation above, which
assumed non-overlapping cones around the $b$ and $\bar{b}$, does not
apply. For this purpose we have evaluated the angular integration
numerically and for $M_H/p_T \ll 1$  i.e. when one can use the
small-angle approximation, the result is $\approx 0.646$ as found by Rubin for
the corresponding quantity $J(1)$. 

\section{Fixed-order results vs parton showers for FSR corrections}
\begin{figure}[h]
\begin{center}
\includegraphics[width=0.49 \textwidth]{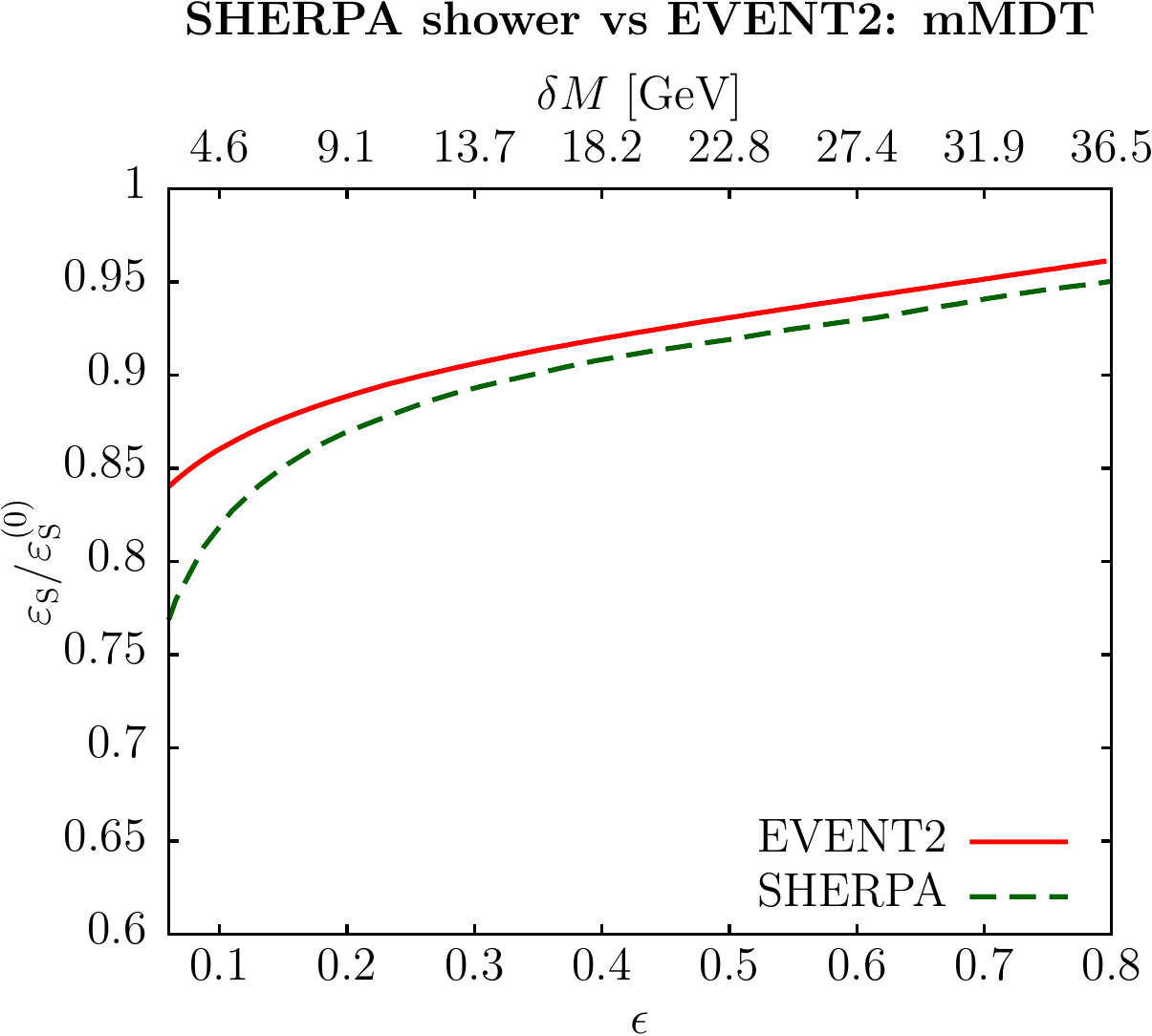}
\includegraphics[width=0.49 \textwidth]{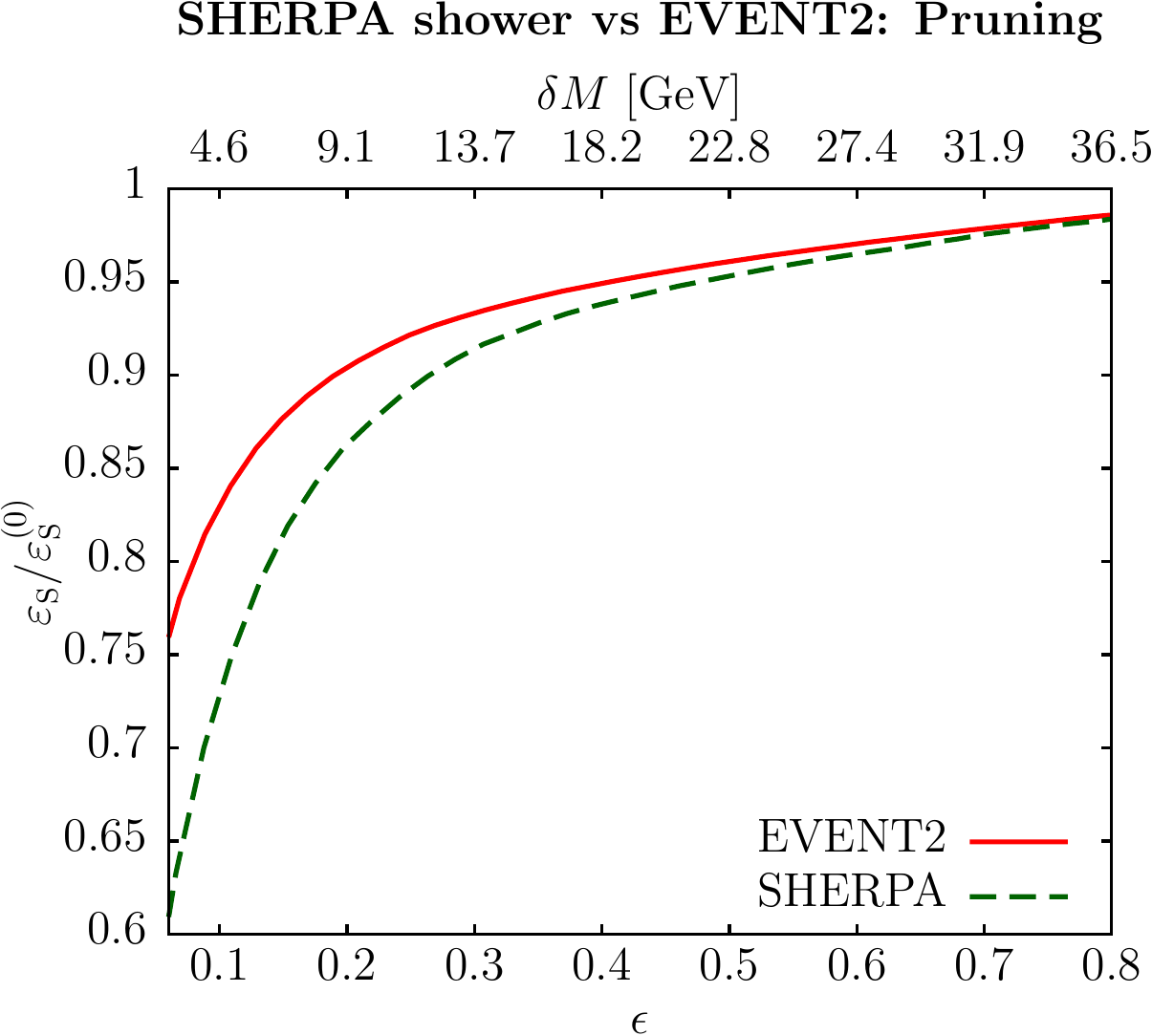}
\end{center}
\caption{Ratio for signal efficiency normalised to lowest-order result, with
  \eventtwo and \textsf{Sherpa 2.0.0},  for  $e^{+}e^{-}$ annihilation with virtual Z
  production and hadronic  decay, where we consider a  Z boson with a transverse boost
  to $p_T =3$ TeV.
\label{fig:ev2vsherpa}}
\end{figure}

We have noted that FSR computed  using the soft approximation
gives numerically very small corrections to the leading-order results,
for sensible choices of the mass window $\delta M$, and the tagger
parameters $\ycut,\zcut,\fcut$ and $\Rtrim$. This of course means that such
calculations are not a good guide to the actual tagger performance
i.e. the signal efficiency, since they do not produce genuine
logarithmic enhancements. One can expect instead that fixed-order
calculations, with correct treatment of hard non-collinear radiation
at order $\alpha_s$ and beyond, will provide a better picture of the
behaviour of taggers. Given that resummation effects are not likely to
be significant it becomes of interest to compare signal efficiencies
obtained with pure fixed-order calculations to those from MC
generators. One may anticipate that precise order $\alpha_s$
calculations give quite similar results to full MC parton showers,
owing to the dominance of hard radiation and the consequent  lack of importance of
multiple soft/collinear emissions. 

To test this we ideally need to carry out an exact order $\alpha_s$
calculation for the process $H \to b\bar{b}g$. Such a calculation can be
straightforwardly performed by taking the exact $H \to b\bar{b}g$ matrix
element and integrating over phase space after application of cuts
corresponding to jet finding and tagging in various algorithms. While
straightforward this exercise proves cumbersome and has in any case to be carried
out with numerical integration. One may instead try to obtain the same
information more economically by exploiting existing fixed-order codes.

One of the most reliable and long-standing fixed-order programs
available to us is the code \eventtwo \cite{Catani:1996jh} for $e^{+}e^{-}$
annihilation. We can exploit this program by considering the process
$e^{+}e^{-} \to Z^0 \to q \bar{q}$ at lowest order and with an
extra gluon emission i.e. up to order $\alpha_s$. One can perform a
boost such that the $Z^0$ is produced with a large momentum along a
given direction and then its decay products will,  a significant
fraction of the time, form a single fat jet. One can then apply the
boosted object taggers to tag the $Z$ boson  imposing a mass window
requirement $\delta M$, around $M_Z$ as we have done throughout this
paper, for the Higgs boson. The situation is similar but not identical
to the case of the Higgs we have thus far considered, due to the
polarisation of the $Z$ boson so that the matrix element for $Z$
decay to quarks differs from Higgs case and efficiencies at tree-level
and beyond are affected, giving for example a different dependence on
$\zcut,\ycut$ at lowest-order. Nevertheless all of our conclusions
about radiative corrections apply to this case as well, including our
findings about the logarithmic structure of FSR contributions, since
these results follow from the radiation of a gluon from the $q
\bar{q}$ pair, which is given by a process independent antenna
pattern, that factorises from the process dependent lowest order decay
of a scalar (i.e. Higgs) or a $Z$ boson.

Therefore in order to test our basic notion that fixed-order
calculations should give a comparable FSR contribution to tagging
efficiency, to that from MC event generators, it should suffice to study
results for boosted $Z$ bosons from \eventtwo on the one hand and
MC on the other. In order to minimise any process dependence one
should choose precisely the same hard process for both the fixed-order
and MC and hence we choose to study the virtual $Z$ boson contribution in
$e^{+}e^{-} \to q \bar{q}$ events, with the hadronically decaying
$Z$ boosted to 3 TeV (as in the \eventtwo case) with the MC generator \sherpa shower \cite{Gleisberg:2008ta}.

We first study the signal efficiencies, normalised to the lowest order
result, that are obtained with \eventtwo
and \sherpa for mMDT and pruning for $\ycut=\zcut=0.1$. These are shown in
Fig.~\ref{fig:ev2vsherpa} as a function of the mass window, where $\epsilon=2\delta M/M_Z$ as in the main text. 

A first observation is that there is a reasonable degree of
qualitative and quantitative similarity between LO and shower
estimates, over a wide range of mass windows, which establishes
further our point about the essential perturbative stability of
taggers against FSR
corrections.
The difference between the normalised signal efficiencies for SHERPA and \eventtwo are $2\%$ or less when $\delta M$ is greater than $\sim8$ GeV or $\sim13$ GeV for mMDT and pruning respectively. One should not in any case
consider mass windows significantly lower than these values at high $p_T$, in order to
minimise NP hadronisation corrections from ISR. Differences start to
become more marked for very low mass windows in particular for
pruning, signalling the need for resummation and hadronisation
corrections. We have also verified that hadronisation corrections have
a minimal impact above the $\delta M$ values stated above and hence
basically preserve the picture one obtains already at leading-order.
\begin{figure}[h]
\begin{center}
\includegraphics[width=0.49 \textwidth]{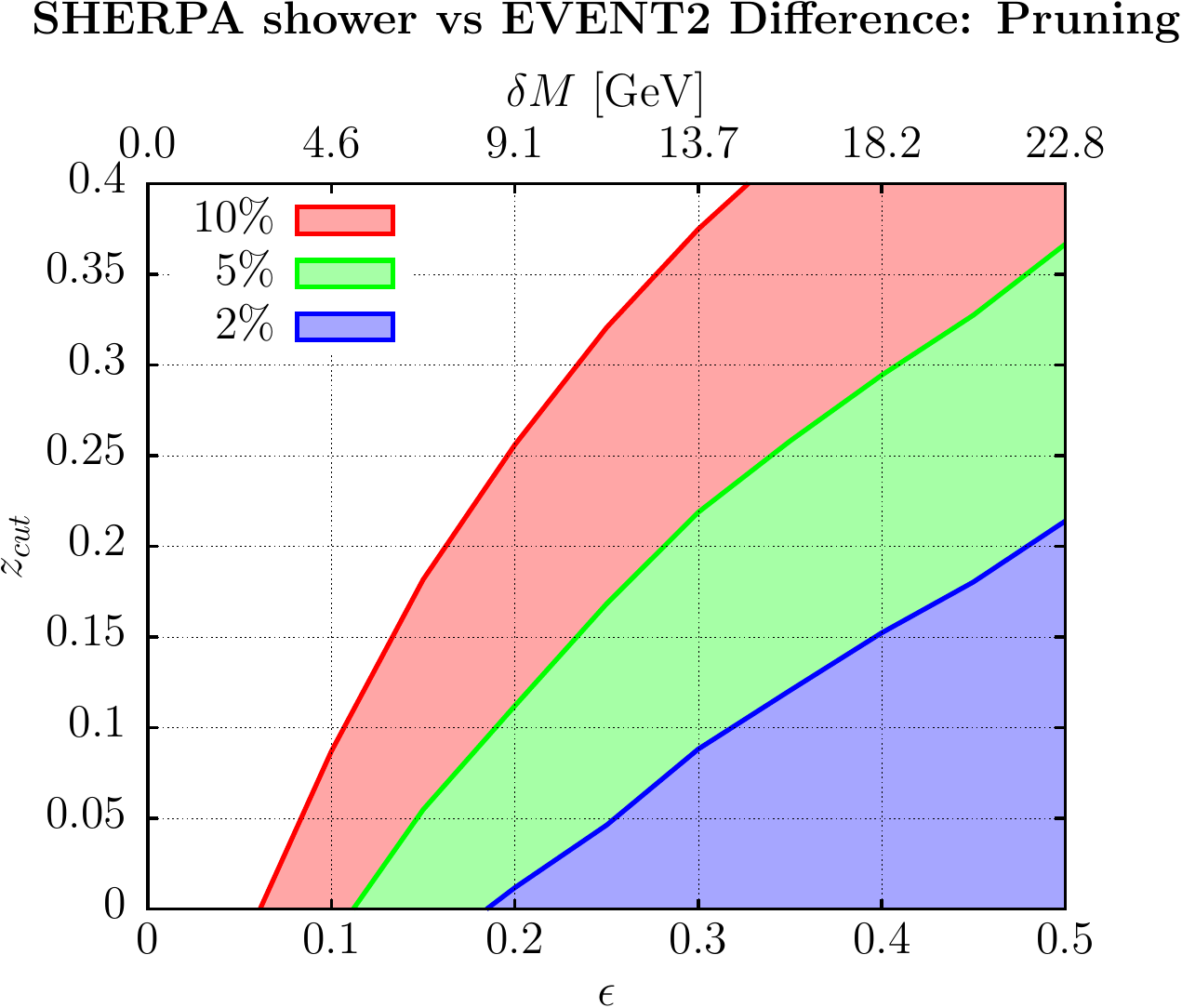}
\includegraphics[width=0.49 \textwidth]{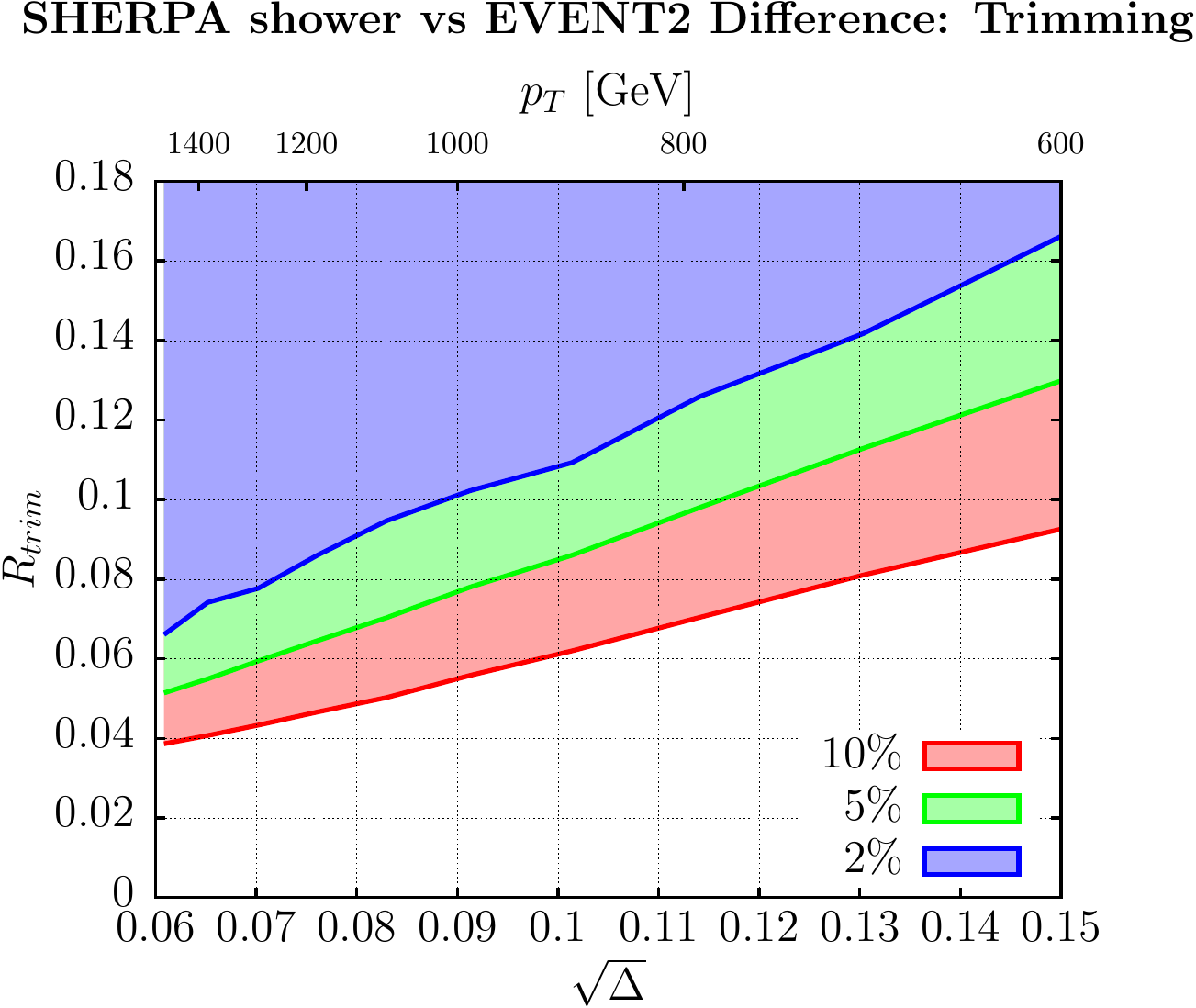}
\end{center}
\caption{Contour plot showing the maximum percentage difference in signal efficiency between \eventtwo at order $\alpha_s$ and \sherpa final state shower both normalised to the lowest order result.
In the left hand panel we apply pruning with different values for $\epsilon$ and $\zcut$ with $p_T=3$ TeV.
In the right hand panel we apply trimming with different values for $\sqrt{\Delta}$ and $\Rtrim$ with $\fc=0.1$.
\label{fig:ev2vsherpacontours}}
\end{figure}
One can also similarly study trimming where the choice of $\Rtrim$ is
additionally crucial to ensure that radiative corrections are minimised  so that
signal efficiency is maintained.  Another way of making this
comparison is provided in Fig.~\ref{fig:ev2vsherpacontours} where we
show the difference between \eventtwo and \sherpa efficiencies (normalised
to the lowest order result) as a function of $\zcut$ and
$\epsilon$ for pruning and as a function of $\Rtrim$ and $\Delta$ for
trimming. The values of $\delta M$, corresponding to the $\epsilon =2
\delta M/M_Z$ values, are shown on the upper axis for pruning and values for $p_T$
corresponding to $\Delta$ are shown for trimming. The blue shaded
region in each case represents parameter values where the difference between the normalised signal efficiencies for \sherpa and \eventtwo
are less than two percent while the green and pink regions correspond to
less than five and ten percent respectively.   
From the plot for pruning  one notes that  there is a correlation
between values of $\epsilon$ and $\zcut$ needed to minimise radiative
corrections. As one goes up in $\zcut$, to stay within say the five
percent zone, one has to correspondingly increase the size of the window. 
This is also  in accordance with expectations from our simple analytics
where one can expect large radiative corrections for $\epsilon \ll
\zcut$. 

Also, for trimming, one may expect a correlation between the value of
$\Delta$ and the value of $\Rtrim$ required to minimise radiative
degradation of mass. This is also reflected in \reffig{ev2vsherpacontours} where once again the green and pink
shaded regions represent differences of 5 percent and 10 percent
respectively, for the normalised signal efficiencies. For $p_T =600$
GeV, for example, choosing $\Rtrim \approx 0.13$ or larger gives less than 5
percent difference between leading order and shower descriptions. As
one lowers $\Rtrim$ radiative losses get progressively larger.

\end{document}